\documentclass[fleqn,usenatbib]{mnras}

\usepackage{newtxtext,newtxmath}
\usepackage[T1]{fontenc}

\DeclareRobustCommand{\VAN}[3]{#2}
\let\VANthebibliography\thebibliography
\def\thebibliography{\DeclareRobustCommand{\VAN}[3]{##3}\VANthebibliography}

\usepackage{graphicx}	% Including figure files
\usepackage{amsmath}	% Advanced maths commands

\title[The Formation of Li-rich Giants]{Convective Mixing: The Formation Channel of Li-rich Giants}
\author[Li et al.]{
Xue-Feng Li$^{1,2}$ \thanks{E-mail: lixuefeng@ynao.ac.cn},
Jian-Rong Shi$^{3,4}$,
Yan Li$^{1,2,5,6}$,
Hong-Liang Yan$^{3,4,7}$,
Jing-Hua Zhang$^{8}$
\\
$^{1}$Yunnan Observatories, Chinese Academy of Sciences, P.O.Box 110, Kunming 650216, China\\
$^{2}$University of Chinese Academy of Sciences, Beijing 100049, China\\
$^{3}$CAS Key Laboratory of Optical Astronomy, National Astronomical Observatories, Beijing 100101, China\\
$^{4}$School of Astronomy and Space Science, University of Chinese Academy of Sciences, Beijing 100049, China\\
$^{5}$Key Laboratory for Structure and Evolution of Celestial Objects, Chinese   Academy of Sciences, P.O.Box 110, Kunming 650216, China\\
$^{6}$Center for Astronomical Mega-Science, Chinese Academy of Sciences, Beijing 100012, China\\
$^{7}$Institute for Frontiers in Astronomy and Astrophysics, Beijing Normal University,  Beijing 102206, China\\
$^{8}$South-Western Institute for Astronomy Research, Yunnan University, Chenggong District, Kunming 650500, China
}

\date{Accepted XXX. Received YYY; in original form ZZZ}

\pubyear{2024}

\begin{document}
\label{firstpage}
\pagerange{\pageref{firstpage}--\pageref{lastpage}}
\maketitle

\begin{abstract}
Increasing observed data indicate that part of giants has abnormally high lithium (Li) inside their surface, and their proportion is around $1\%$.
Instead of pursuing the feasible mechanisms for extra Li enrichment, we focus on how to inhibit Li depletion from the main sequence (MS) phase of giants. With this in mind, we find that convective mixing is capable of achieving this goal and forming Li-rich giants, which is mainly attributed to the convection model with the convective boundary defined by the Ledoux criterion. Another factor in the formation of Li-rich giants in our convection models is related to the Li abundances of their progenitors. If the Li abundances of the progenitors exceed the meteoritic value ($\rm 3.3\,dex$), then the majority of giants will be rich in Li. This is the general pattern of stellar Li abundance evolution without factoring in extra Li depletion. We propose that other Li depletion processes should also be adopted in the future, as they may be the key to the $1\%$ puzzle.
\end{abstract}


%%%%%%%%%%%%%%%%%%%%%%%%%%%%%%%%%%%%%%%%%%%%%%%%%%

%%%%%%%%%%%%%%%%% BODY OF PAPER %%%%%%%%%%%%%%%%%%

\section{Introduction}

For the past few years, the GALAH \citep{2018MNRAS.478.4513B, 2021MNRAS.506..150B} and the LAMOST surveys \citep{2012RAA....12.1197C} have provided a large amount of observed samples on the Li abundances ($A(\rm Li)$\footnote{
$A(\rm Li) = $$\rm log$$(N_{\rm Li}/N_{\rm H})+12$, where $ N_{\rm X}$ is the atomic number densities of X.}) of giants. The samples show the same problem that the Li abundances of the giants emerge a large-scale, more than four orders of magnitude, distribution \citep{2019MNRAS.484.2000D, 2020NatAs...4.1059K,  2021ApJ...914..116G,  2021MNRAS.505.5340M}. For the normal giants with the Li abundance within the range of $\rm -1.0$ to $\rm 1.5\,dex$, they have been already many extra mixing models to try to explain \citep{2020ApJ...901L..18S, 2021MNRAS.503.2746M, 2023ApJ...943..115L}. However, there are also some abnormal giants with the Li abundance larger than $\rm 1.5\,dex$, i.e., so-called Li-rich giants \citep[e.g.,][]{1982ApJ...255..577W, 1989ApJS...71..293B, 1995ApJ...448L..41D, 2005AJ....129.2831R,  2011ApJ...730L..12K, 2016MNRAS.461.3336C, 2018NatAs...2..790Y, 2019MNRAS.482.3822S, 2020MNRAS.498...77H, 2021RAA....21...20Z, 2022arXiv220902184K}. This definition originates from the upper limit value obtained by the Li abundance of a star with $1.5\,M_{\odot}$ and solar metallicity, after assuming its convective envelope dilute the Li abundance by about 60 times from $\rm 3.3\,dex$ \citep[e.g.,][]{2000A&A...359..563C, 2023A&A...674A.157T}. Among them, \citet{2022arXiv220902184K} had discovered the most Li-rich red giant branch (RGB) star at present, 2MASS J05241392-0336543, its $A(\rm Li)$ is as high as $\rm 5.6\,dex$. In addition, \citet{2019MNRAS.482.3822S} also reported two red clump (RC) stars with a $A(\rm Li)$ of $\rm \sim 4.0\,dex$. In general, the proportion of the Li-rich giants is very low, only $\sim 1\%$ \citep{1989ApJS...71..293B, 2019ApJS..245...33G, 2021MNRAS.505.5340M}. Among the Li-rich giants, the RC stars are majority \citep[e.g.,][]{2021MNRAS.505.5340M, 2021ApJ...913L...4S, 2021NatAs...5...86Y}. 

How the Li-rich giants form is unclear, while some attempts have been made in this field, e.g., the element diffusion effects \citep{2022A&A...668A.126G}. It can be considered from two aspects: one is to increase Li content on the stellar surface, such as the external action \citep[e.g.,][]{1999MNRAS.308.1133S, 2012A&A...538A..36L, 2019ApJ...880..125C, 2020ApJ...889...33Z} and the internal process \citep{1955ApJ...121..144C, 1971ApJ...164..111C}, and the other is to inhibit or weaken Li depletion. In this article, we will focus on the weakening on the Li depletion and explore whether it is possible to form the Li-rich giants.

Stars undergo the Li depletion at various stages, such as the pre-main sequence (PMS) \citep[see e.g.][]{1965ApJ...141..993I}, the main sequence (MS) \citep[see e.g.][]{2005A&A...442..615S}, and the first dredge-up \citep[see e.g.][]{1967ARA&A...5..571I, 1967ApJ...147..624I}. However, the standard convection model almost does not exist the MS Li depletion processes \citep[e.g.,][]{1984A&A...138..431D, 1990ApJS...73...21D, 2023ApJ...943..115L}. At the first dredge-up, the stellar convective envelope extends inward, diluting the surface Li by carrying it to hotter regions. Therefore, the calibration of the bottom boundary of the convective envelope will greatly affect the retained Li content. With regard to the two criteria of convective boundary, i.e., the Schwarzschild and the Ledoux criterion,  the main difference is whether the mean molecular weight gradient is zero. Nevertheless, the computation of the convective boundary is still an open topic in stellar evolution code \citep{2014A&A...569A..63G, 2022ApJ...928L..10A, 2023Galax..11...56A}. Recently, \cite{2022ApJ...933...58C} explained the normal giant Li abundance distribution with the help of the standard convection model but had been unable to do so for the Li-rich giants. The convective zone during the first dredge-up will penetrate into the region of chemical composition changes caused by the central hydrogen-burning, which led us to explore from the convection models how weakening Li depletion enables the formation of the Li-rich giants. 

In this paper, we will investigate the impact of the convective mixing on the surface Li of the giants and the possibility of it in the formation of the Li-rich giants. The input physics and the model setting are presented in $\rm Sect.\,\ref{sec:Method}$. Then, in $\rm Sect.\,\ref{sec:Result}$, we analyze the effect of the convective boundary selection, the stellar mass and metallicity parameter, and the input Li abundance on the formation of the Li-rich giants. Then $\rm Sect.\,\ref{sec:Discussion}$ is the discussions about the convective boundary, the thermohaline mixing, the $1\%$ puzzle, and the Li enrichment. Finally, major conclusions are situated in $\rm Sect.\,\ref{sec:Conclusion}$.

\section{Method}\label{sec:Method}
\subsection{Inputs} \label{sec:input}
Using the Modules for Experiments in Stellar Astrophysics (MESA; r11701 \citep{2011ApJS..192....3P, 2013ApJS..208....4P, 2015ApJS..220...15P, 2018ApJS..234...34P, 2019ApJS..243...10P}) to construct our stellar models.
Our models are based on the template provided by the MESA, and its path is `.../.../mesa-r11701/star/test\_suite/7M\_prems\_to\_AGB'. We do not take into account rotation, element diffusion, mass loss, radiative levitation, etc. We add a strong surface convective overshooting to all models to ensure that the convective zone is uniformly mixed, which can avoid the abundance differences caused by the selection of atmospheric boundaries. The settings of the convective overshooting are as follow:
\[
\begin{array}{lp{0.8\linewidth}}
\texttt{overshoot\_f0\_above\_nonburn\_shell = 0.004}  \\ 
\texttt{overshoot\_f\_above\_nonburn\_shell = 0.80}   \\
\end{array}
\]
In order to simplify the calculation, we choose `simple\_photosphere', which is the place where the optical depth is equal to 2/3 as the atmosphere boundary. Time step and spatial resolution follow the default settings of the MESA. The basic input physics for the stellar models have been listed in $\rm Table\,\ref{tab:t1}$. The nuclear reaction network we selected is \textit{pp\_extras.net}, which includes 12 isotopes: $\rm ^{1,2}H$, $\rm ^{3,4}He$, $\rm ^{7}Be$, $\rm ^{7}Li$, $\rm ^{8}B$, $\rm ^{12}C$, $\rm ^{14}N$, $\rm ^{16}O$, $\rm ^{20}Ne$, and $\rm ^{24}Mg$. The mixing length parameter follows the default value of the MESA code: 2.0. %\citep{1968pss..book.....C}. 

For the overshooting model in $\rm Sect.\,\ref{sec:3.1.1}$, we solely consider the overshooting from the convective envelope into the interior of the star and do not involve core overshooting, and the overshooting coefficient $f_{\rm ov}$ is 0.016 \citep{2000A&A...360..952H}. In  $\rm Sect.\,\ref{sec:3.1.2}$, we introduce thermohaline mixing and semiconvection. About the thermohaline mixing model, we use the thermohaline mixing module of the MESA code \citep{2013ApJS..208....4P}, which includes three options: `$\rm Kippenhahn$' \citep{1980A&A....91..175K}, `$\rm Traxler\_Garaud\_Stellmach\_11$' \citep{2011ApJ...728L..29T}, and `$\rm Brown\_Garaud\_Stellmach\_13$' \citep{2013ApJ...768...34B}. The option 1 is the physical method, and the options 2 and 3 are the results of numerical simulation. In this work, we use the `$\rm Kippenhahn$' method to calculate the diffusion coefficient of the thermohaline mixing, and the mixing coefficient $\alpha_{\rm th}$ is 2 \citep{2010ApJ...723..563D,  2011ApJ...728L..29T, 2013ApJ...768...34B}. With reference to template `.../.../mesa-r11701/star/test\_suite/semiconvection', we construct the semiconvection model. %We refer to the rotation setup of template  `.../.../mesa-r11701/star/test\_suite/25M\_z2m2\_high\_rotation' for the rotation model in  $\rm Sect.\,\ref{sec:3.2}$. 

In this work, we do not investigate the evolutionary behavior of surface Li at the PMS. However, we still use it as a reference, specifically in terms of input Li abundance (see $\rm Sect.\,\ref{sec: ini ALi}$). Our models have all evolved from the PMS, but our attention is on the stage from zero-age main sequence (ZAMS) to the RGB tip, which ensures that our low-mass star model has experienced the RGB bump. In addition, we have additionally included RC stars in order to compare our model results with observations, so we evolve our basic models (see $\rm Sect.\,\ref{sec:3.1}$) to the RC stage appropriately in $\rm Sect.\,\ref{sec:3.2}$. When the mass fraction of the central hydrogen drops to $10^{-9}$, we mark this moment as the MS turnoff.
%\begin{verbatim}	
%\end{verbatim}
\begin{table*}
\centering
\caption{Input Physics}
\label{tab:t1}
\begin{tabular}{lll} % three columns, alignment for each
\hline
Items & Values & Descriptions/References\\
\hline
$\rm Mass$ & $0.8-1.8M_{\odot}$& $\Delta=0.1$\\
$Y_{\rm ini}$ & $0.24+2Z_{\rm ini}$ & $-$\\
$Z_{\rm ini}$ & $0.001-0.039$  & $\Delta=0.002$\\
The equation of state & $-$ &\cite{2002ApJ...576.1064R} \\
The OPAL opacity tables & $-$ & \citet{1993ApJ...412..752I, 1996ApJ...464..943I} \\
The chemical composition & $-$ & \citet{1998SSRv...85..161G} \\
The treatment of convection & $-$ & \citet{1968pss..book.....C}\\
\hline
\end{tabular}
\end{table*}

\subsection{The Setting for Initial Li Abundance} \label{sec: ini ALi}

In our MESA models, initial Li abundance follows the following rule: $A(\rm Li)_{\rm ini}=[Fe/H]+3.4$, and the $A(\rm Li)_{ini}$ is $\rm 3.4\,dex$ for solar metallicity ($Z=0.02$, $\rm [Fe/H]=0\,dex$). $A(\rm Li)_{ini}$ is the Li abundance at the beginning of the stellar model evolution, which will experience the Li depletion process during the PMS. On the one hand, we do not intend to evaluate the behaviour of the PMS Li evolution in the current work, but only to serve as a reference when investigating the effect of the stellar input Li abundance on the Li abundance in their giant phase, and so we instead set the ZAMS Li abundance to a certain value to allow the models to evolve.

On the other hands, given the differences in the actual MS Li distribution and the results of convection model prediction (e.g., \citet{2005A&A...442..615S} vs. \citet{1990ApJS...73...21D}). Therefore, we artificially supplement the Li abundance information on the ZAMS ($A(\rm Li)_{\rm ZAMS}$). In short, using controllable inputs to replace and simulate the surface Li abundance in different low-mass stars%%% induced by extra mixing
. By changing slightly the proportion of Li in $Z$, we can get disparate $A(\rm Li)_{ZAMS}$. We apply the two Li abundance inputs to our stellar models in the form of a cross-reference in $\rm Sect.\,\ref{sec:Result}$.

\subsection{Convective Boundary} \label{sec:CB}

Our aim is to investigate the impact of convective mixing on Li enrichment; therefore, we will construct two types of convective models in a comparative manner. Starting from limiting the convective boundary, we set up two models, one by considering the Schwarzschild boundary and the other by choosing the Ledoux boundary. In the uniform region of chemical composition, the formation of the convective instability needs to conform to the Schwarzschild criterion:
\begin{equation}\label{equ:e1}
\bigtriangledown_{\rm ad} < \bigtriangledown_{\rm rad},
\end{equation}
 while for the nonuniform region of chemical composition, the Ledoux criterion needs to be meet: 
\begin{equation}\label{equ:e2}
\bigtriangledown_\mu + \bigtriangledown_{\rm ad} < \bigtriangledown_{\rm rad}.
\end{equation}
Here, the mean molecular weight gradient $\bigtriangledown_{\rm \mu} = (d \ln \mu / d \ln P) $, the adiabatic temperature gradient $\bigtriangledown_{\rm ad} = (\partial \ln T / \partial \ln P)_{\rm ad}  $, and the radiative temperature gradient $\bigtriangledown_{\rm rad} = (\partial \ln T / \partial \ln P)_{\rm rad}$.

\section{The Formation of Li-rich Giants}\label{sec:Result}

\subsection{The Exploration of Convective Mixing on Li Enrichment}\label{sec:3.1}

Recently, the study of Li abundances in low-mass giants has received more and more attention from survey projects, with the peak values for mass and [Fe/H] are approximately $1.2\, M_{\odot}$ and $\rm -0.15\,dex$ ($Z\sim 0.014$) respectively \citep[e.g.,][]{2018ApJS..239...32P, 2020MNRAS.494.1348D, 2022ApJ...931..136Z}. Thus, we take the stellar parameter of our models as $1.2\, M_{\odot}$ and $Z=0.014$, and the input Li abundance is $A(\rm Li)_{ZAMS}=3.3\,dex$, i.e., the meteoritic value \citep{1989GeCoA..53..197A, 1998SSRv...85..161G, 2009ARA&A..47..481A}. 

Here, the models we built are convection models (there is no other mixing, and convection is the only mixing process), including the $\rm Conv.\,\&\,Schwarzschild$ model (using the Schwarzschild criterion) and the $\rm Conv.\,\&\,Ledoux$ model (using the Ledoux criterion).

\subsubsection{Convective Boundary Effect}\label{sec:3.1.1}

\begin{figure*}
	\includegraphics[scale=0.29]{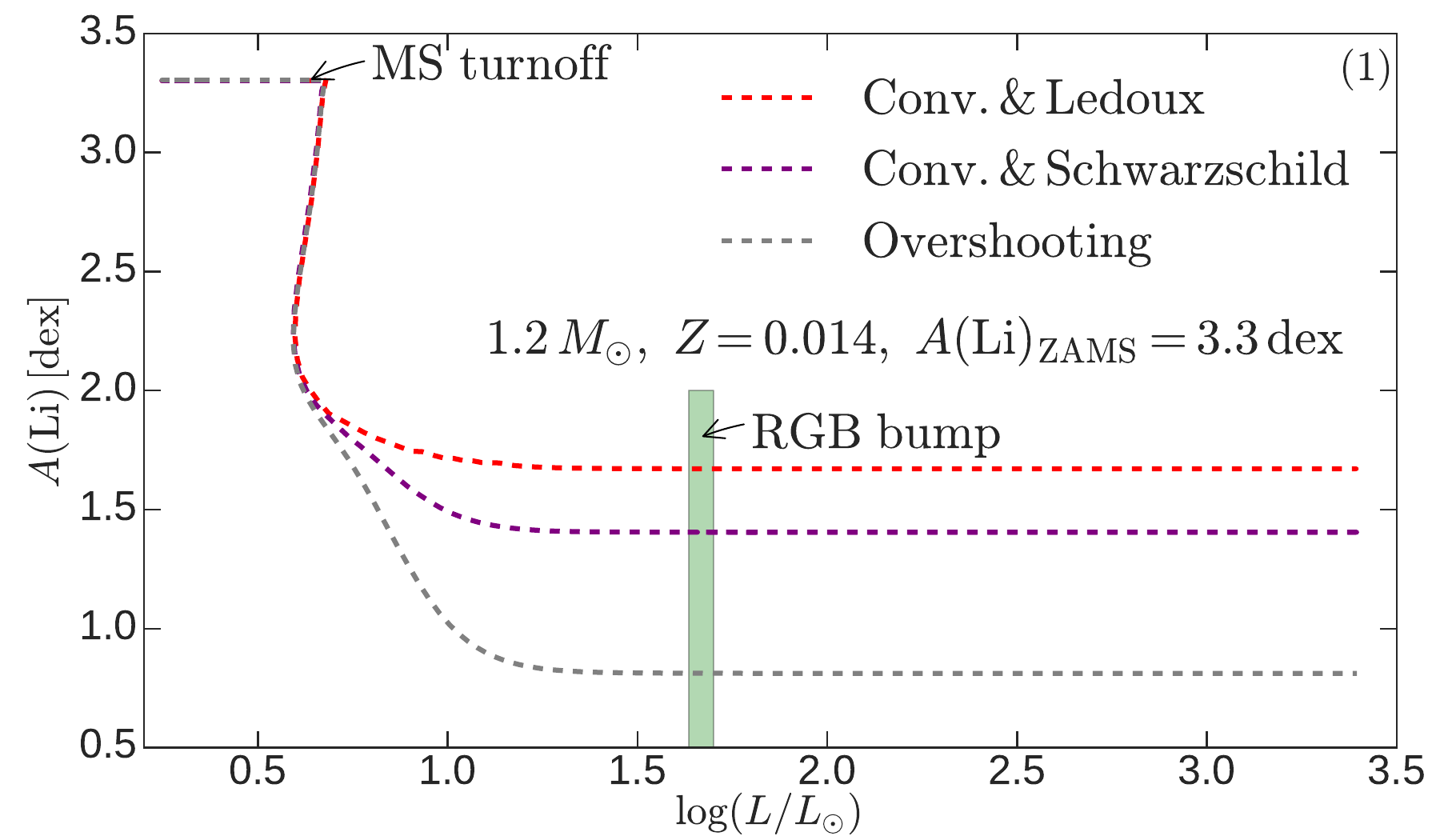}
	\includegraphics[scale=0.29]{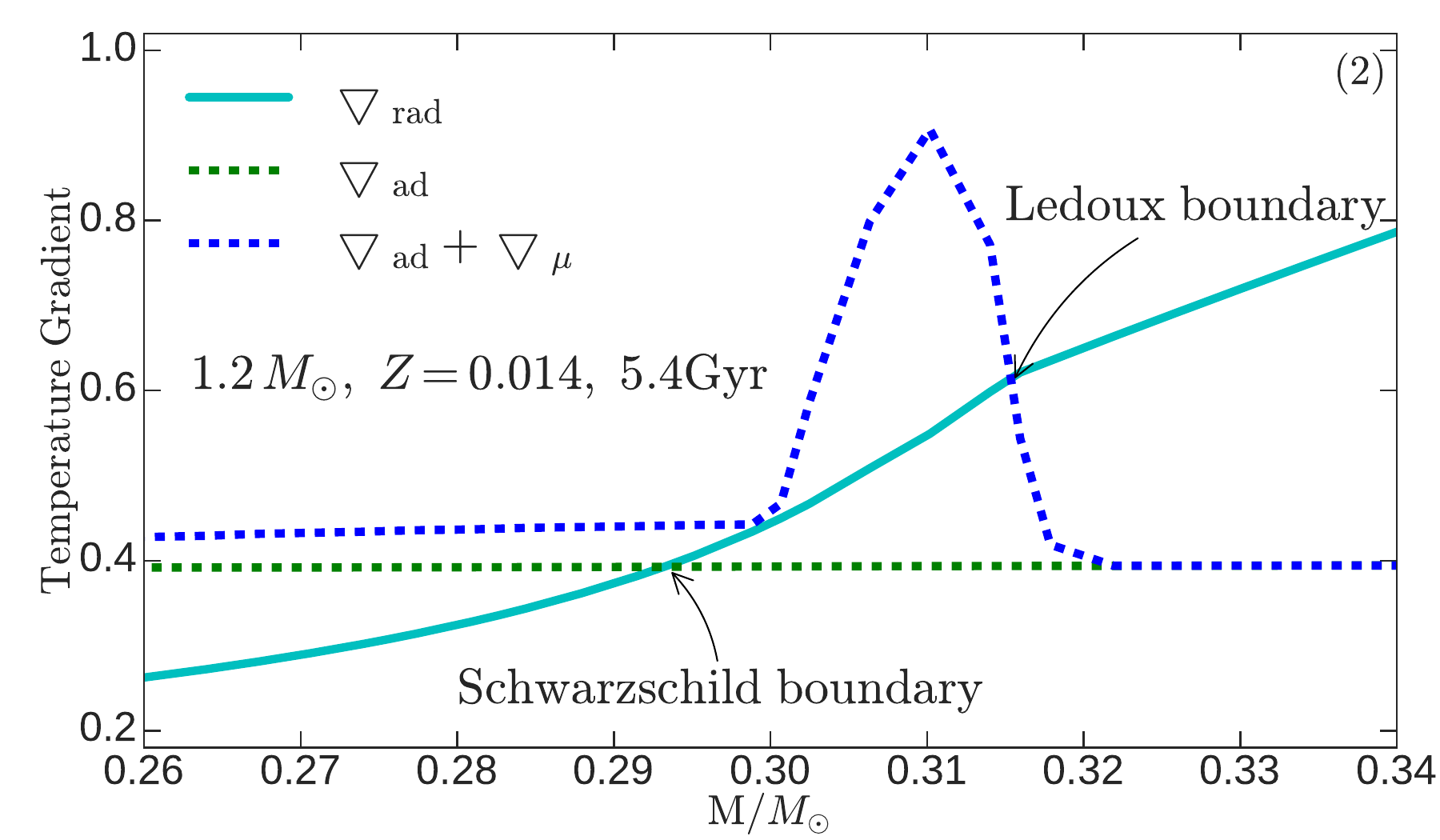}
	\includegraphics[scale=0.29]{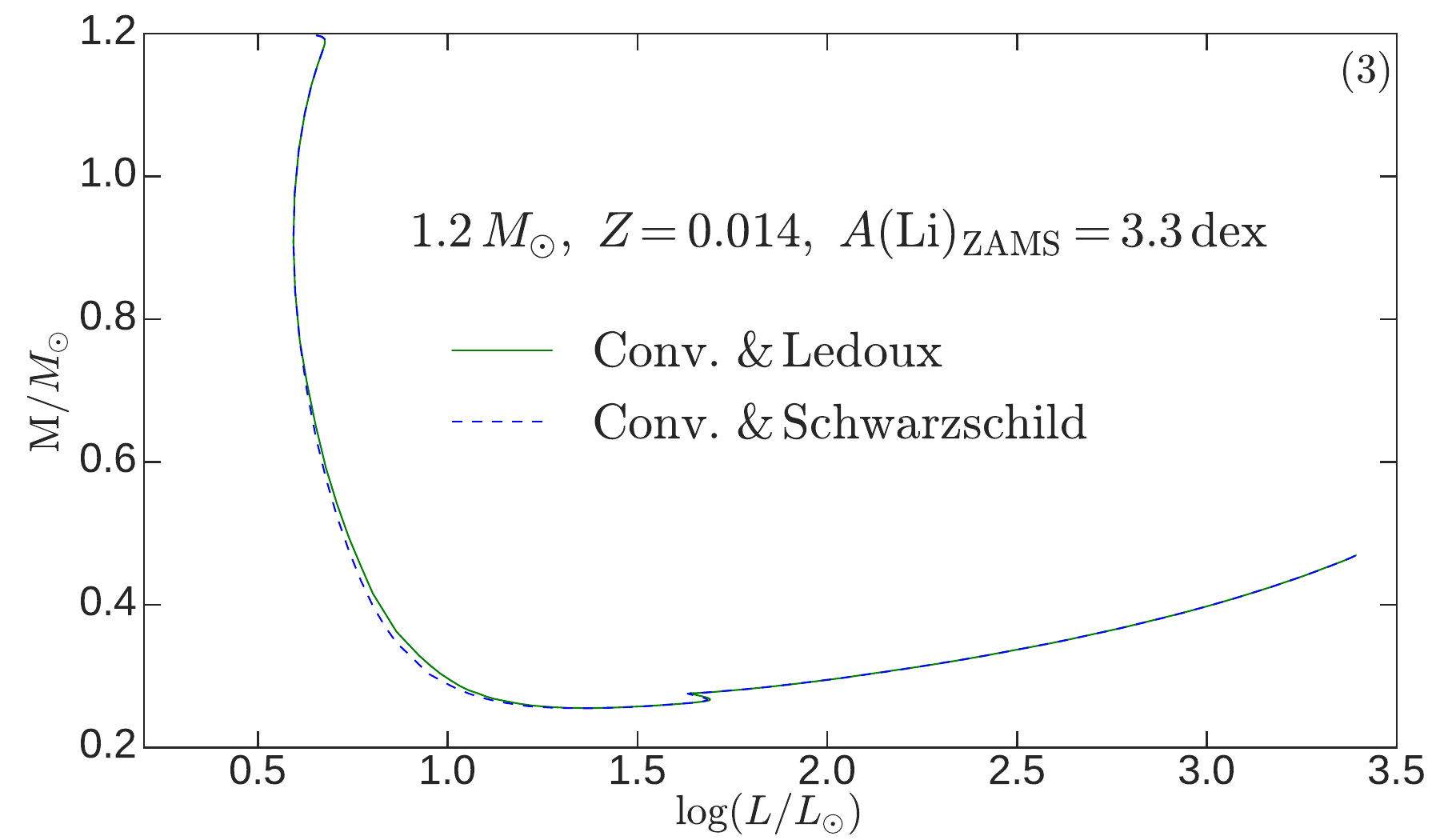}
	\includegraphics[scale=0.29]{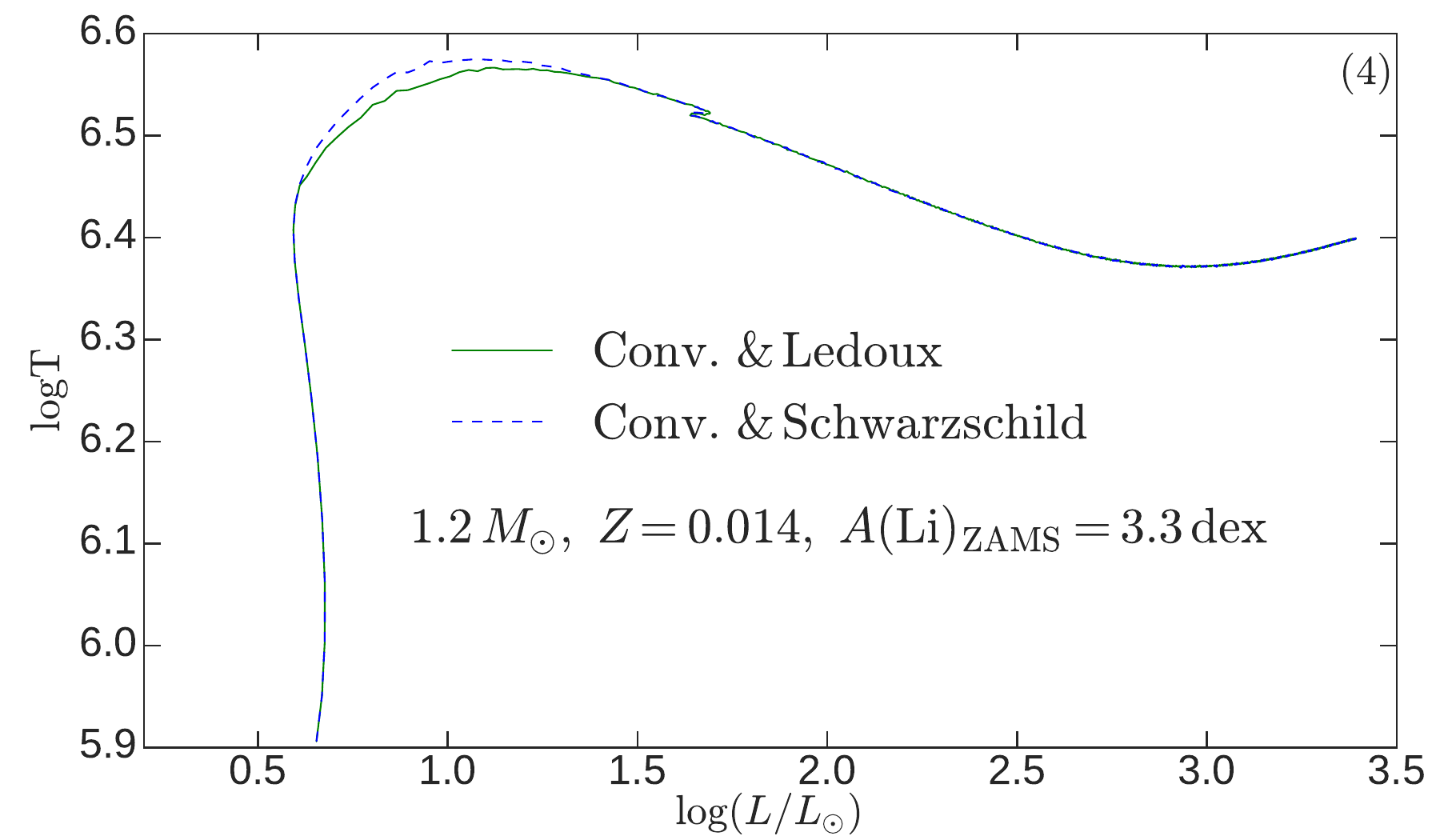}
	\caption{Li abundance evolution and structural information of stellar models. Panel (1): $A(\rm Li)$ vs. Luminosity, the $\rm Conv.\,\&\,Schwarzschild$ model (using the Schwarzschild criterion);  the $\rm Conv.\,\&\,Ledoux$ model (using the Ledoux criterion); the overshooting model considers overshooting from the bottom of the convective envelope towards the interior of the star based on the convection model, and the overshooting coefficient $f_{\rm ov}$ is 0.016 \citep{2000A&A...360..952H}. Panel (2) shows the temperature gradient profile in mass coordinates inside a star with 5.4 Gyr (i.e., $\rm log$$(L/L_{\odot}) \sim 1.0$). Panel (3) presents the change of the bottom position of the convective envelope from the MS turnoff to the RGB tip, and we use the mass label (i.e., the embedded mass of the convective envelope). In addition, we show the evolution trajectory of the temperature at the bottom of the convective envelope in panel (4).} \label{fig:boundary}
\end{figure*}

 In $\rm Fig.\,\ref{fig:boundary}$, we show the Li abundance evolution and structure information for the two convection models. In $\rm Fig.\,\ref{fig:boundary}\, (1)$, we present the evolution trajectory of the Li abundance from the ZAMS to the RGB tip for various models. Additionally, we supplement the overshooting model (convection model + overshooting of the convective envelope into the interior of the star) for reference. It should be noted that the variance in Li abundances across these models is mainly due to the inconsistent degree of Li depletion during the first dredge-up. At this stage, in comparison with $\rm Conv.\,\&\,Schwarzschild$, the $\rm Conv.\,\&\,Ledoux$ models have an effect on the Li abundance by shrinking the convective zone size, while the overshooting leads to the opposite case. Compared to the convection models, the overshooting model based on the Schwarzschild boundary accelerates the Li depletion. This is because the overshooting brings Li from the convective envelope to deeper and hotter areas inside. In all of the convection models constructed, the decrease of Li abundances only presents during the first dredge-up, while the difference in the degree of Li depletion for the two convection models is solely related to the choice of the convective boundary. In addition, the $\rm Conv.\,\&\,Ledoux$ model indicates that the Li abundance during the RGB exceeds $\rm 1.5\,dex$, i.e., Li-rich giants are formed.
 
The structural information of the relevant models is presented in $\rm Figs.\,\ref{fig:boundary}\, (2),\ (3),\ and\ (4)$. It can be seen from $\rm Fig.\,\ref{fig:boundary}\, (1)$ that the Li abundance difference between the two convection models remains unchanged in the case of luminosity is over $10\, L_{\odot}$. In order to present the location discrepancy in the convective boundary between the two models, we show the temperature gradient profile in mass coordinates at a luminosity of $10\, L_{\odot}$  in $\rm Fig.\,\ref{fig:boundary}\, (2)$.

It can be observed from $\rm Figs.\,\ref{fig:boundary}\, (2)\ and (3)$ that throughout the entire first dredge-up phase, the $\rm Conv.\&\,Ledoux$ model has a more outward position at the bottom of the convective envelope than the $\rm Conv.\&\,Schwarzschild$ model, and this difference is eliminated after the RGB bump. As $\rm Fig.\,\ref{fig:boundary}\, (4)$ indicates, this difference is the key to reducing the Li depletion.
Above $2.6\times10^6\,K$, Li is destroyed. As the convective envelope deepens during the first dredge-up, its bottom gets hotter, exacerbating the destruction of Li near the convective boundary. However, Li in the convective envelope is uniformly mixed, resulting in a decrease in surface Li as the convective envelope deepens. When selecting the Ledoux criterion for the convective boundary, as opposed to the Schwarzschild boundary, its convective boundary is nearer to the lower temperature region. Consequently, the reaction rate of Li in the $\rm Conv.\&\,Ledoux$ model is less high. Thus, the manifested results are that the degree of Li depletion is weakened.

The choice of the convective boundary is related to whether the local chemical composition is uniform. During the first dredge-up, the convective envelope can erode inward to approach the hydrogen-burning shell, where a discontinuity of element abundance exists. Then,  employing the Ledoux criterion becomes a feasible option. 

\subsubsection{Effect of Metallicity and Mass}\label{sec:3.1.2}

\begin{figure*}
	\centering
	\includegraphics[scale=0.29]{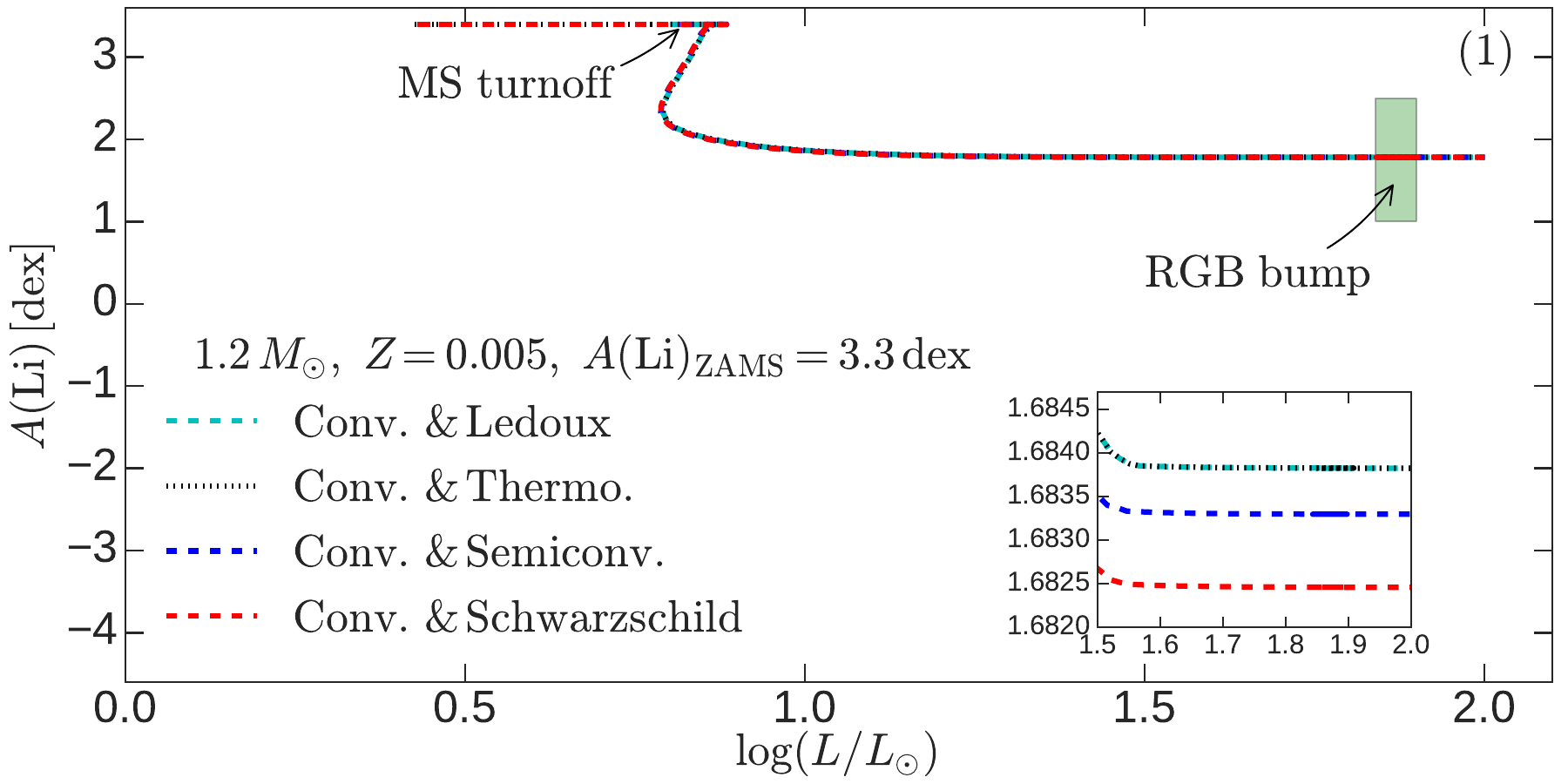}
	\includegraphics[scale=0.29]{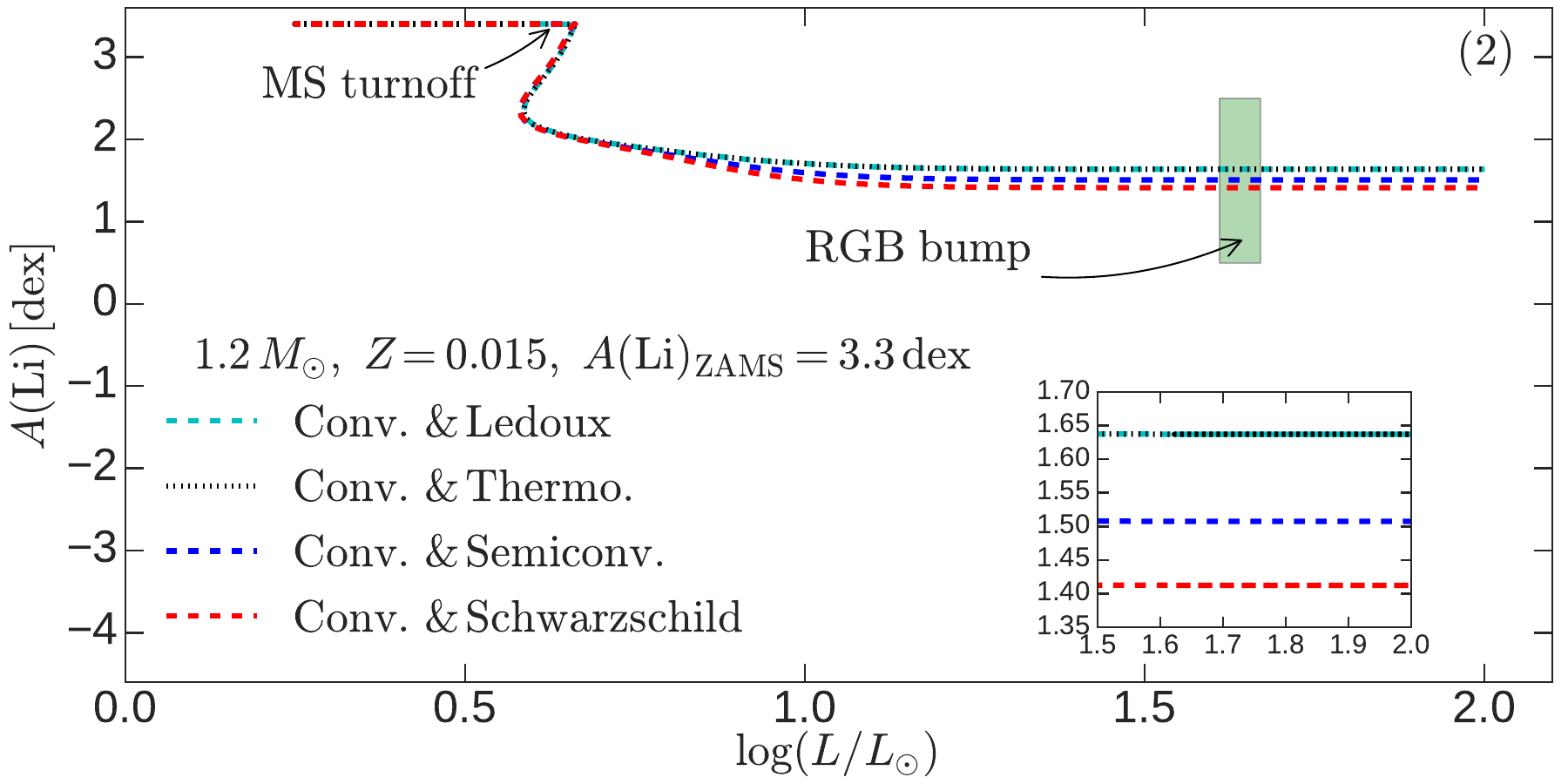}
	\includegraphics[scale=0.29]{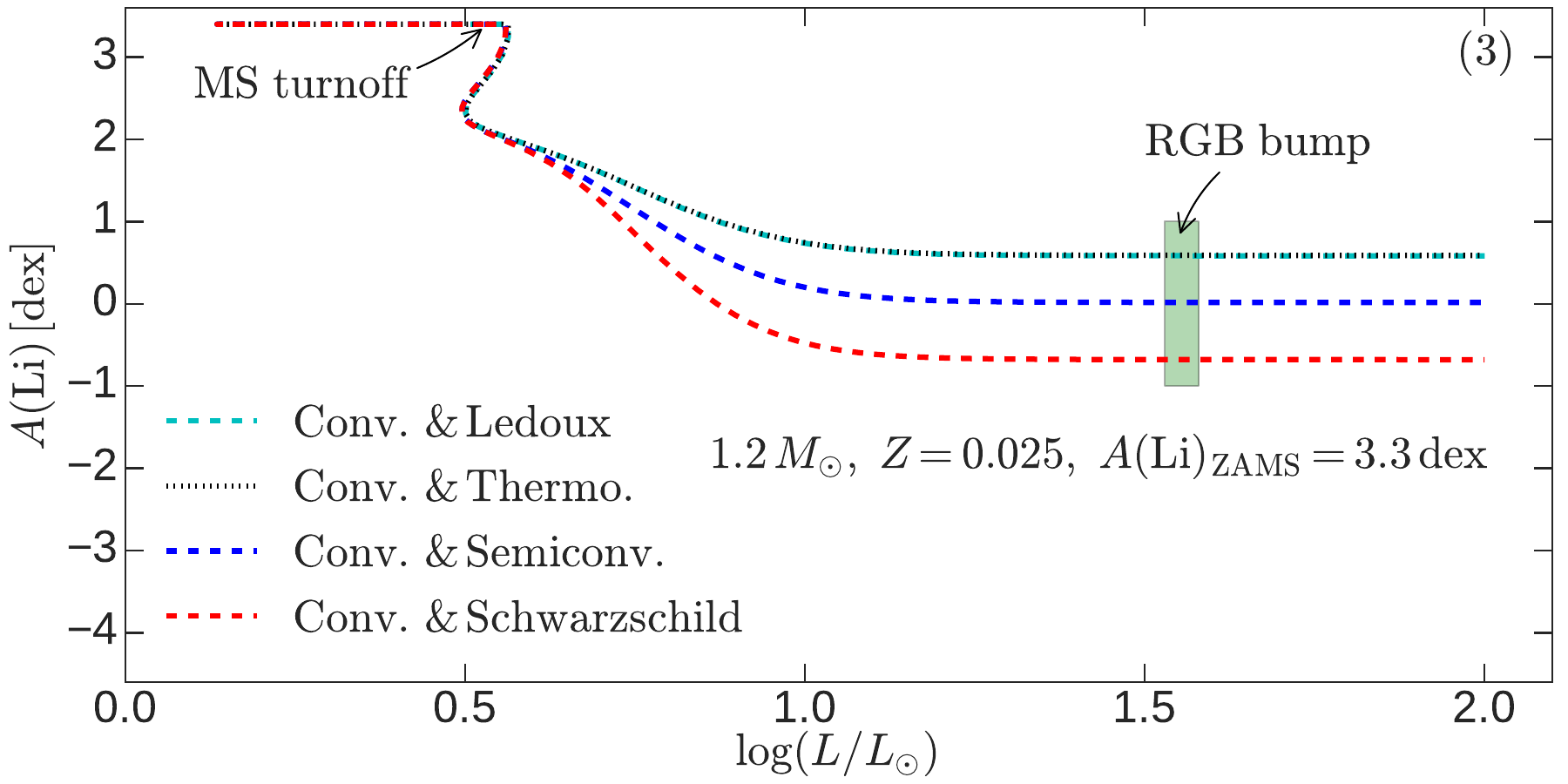}
	\includegraphics[scale=0.29]{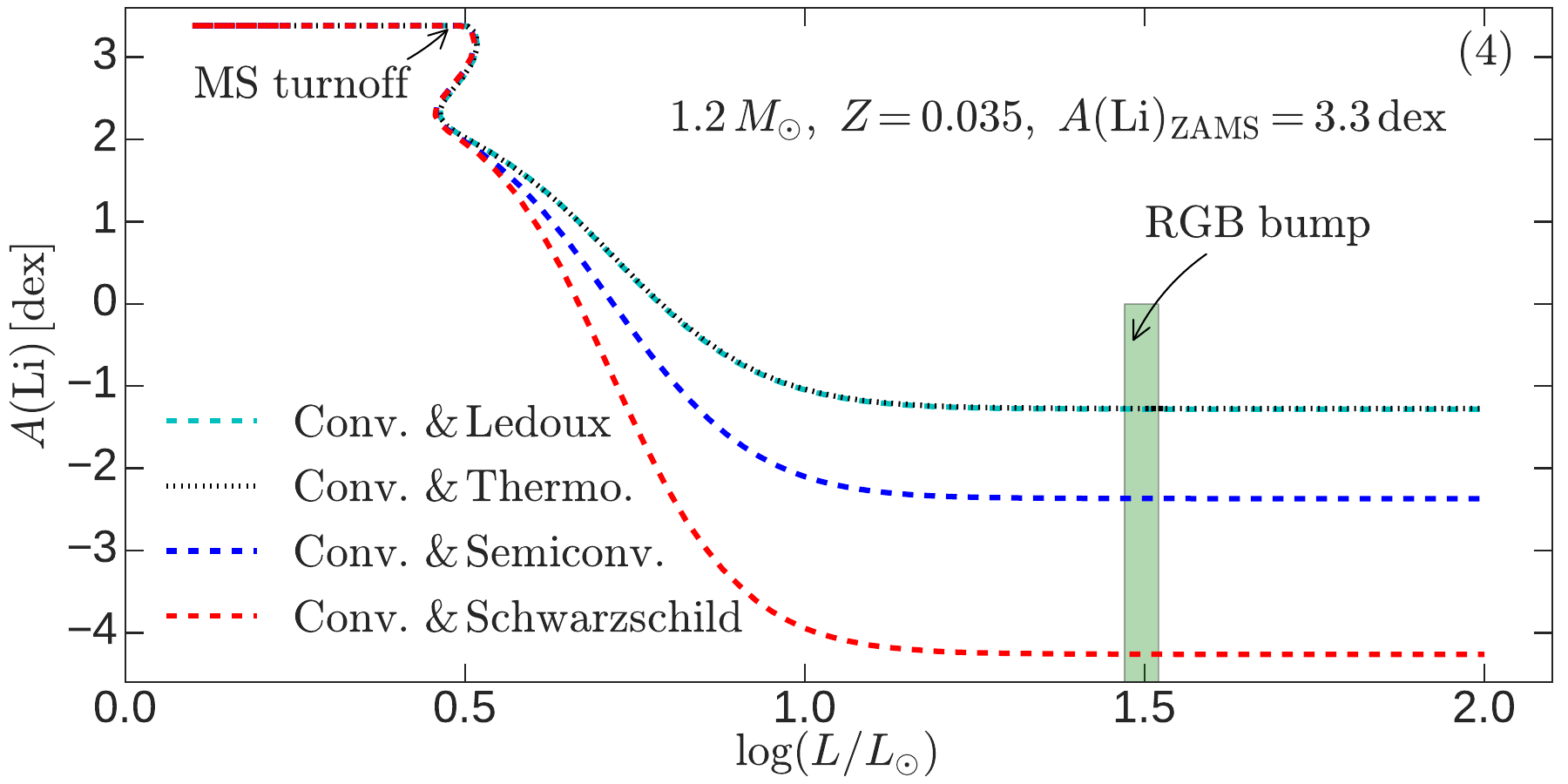}
\caption{Similar to $\rm Fig.\,\ref{fig:boundary}\,(1)$, but the evolution stage is from the MS turnoff to luminosity rise to $100\,L_{\odot}$. We choose four metallicities, i.e, $Z$=0.005, 0.015, 0.025, and 0.035. We introduce four models. The $\rm Conv.\, \&\, Schwarzschild$ and the $\rm Conv.\, \&\, Ledoux$ models are the convection model in $\rm Fig.\,\ref{fig:boundary}$.  $\rm Conv.\, \&\, Thermo.$: the thermohaline mixing model, and the mixing parameter $\alpha_{\rm th}$ is 2 \citep{2010ApJ...723..563D,  2011ApJ...728L..29T, 2013ApJ...768...34B}; $\rm Conv.\, \&\, Semiconv.$: convection + semiconvection. Here, we take the semiconvection coefficient $\alpha_{\rm sc}$ as 0.03 \citep{1991A&A...252..669L, 2006A&A...460..199Y}. The subgraphes in the panels (1) and (2) are partial enlargements. Refer to $\rm Fig.\,\ref{fig:appendix1}$ for the related Kippenhahn diagram.}
\label{fig:structure}. 
\end{figure*}

\begin{figure*}
\centering
\includegraphics[scale=0.29]{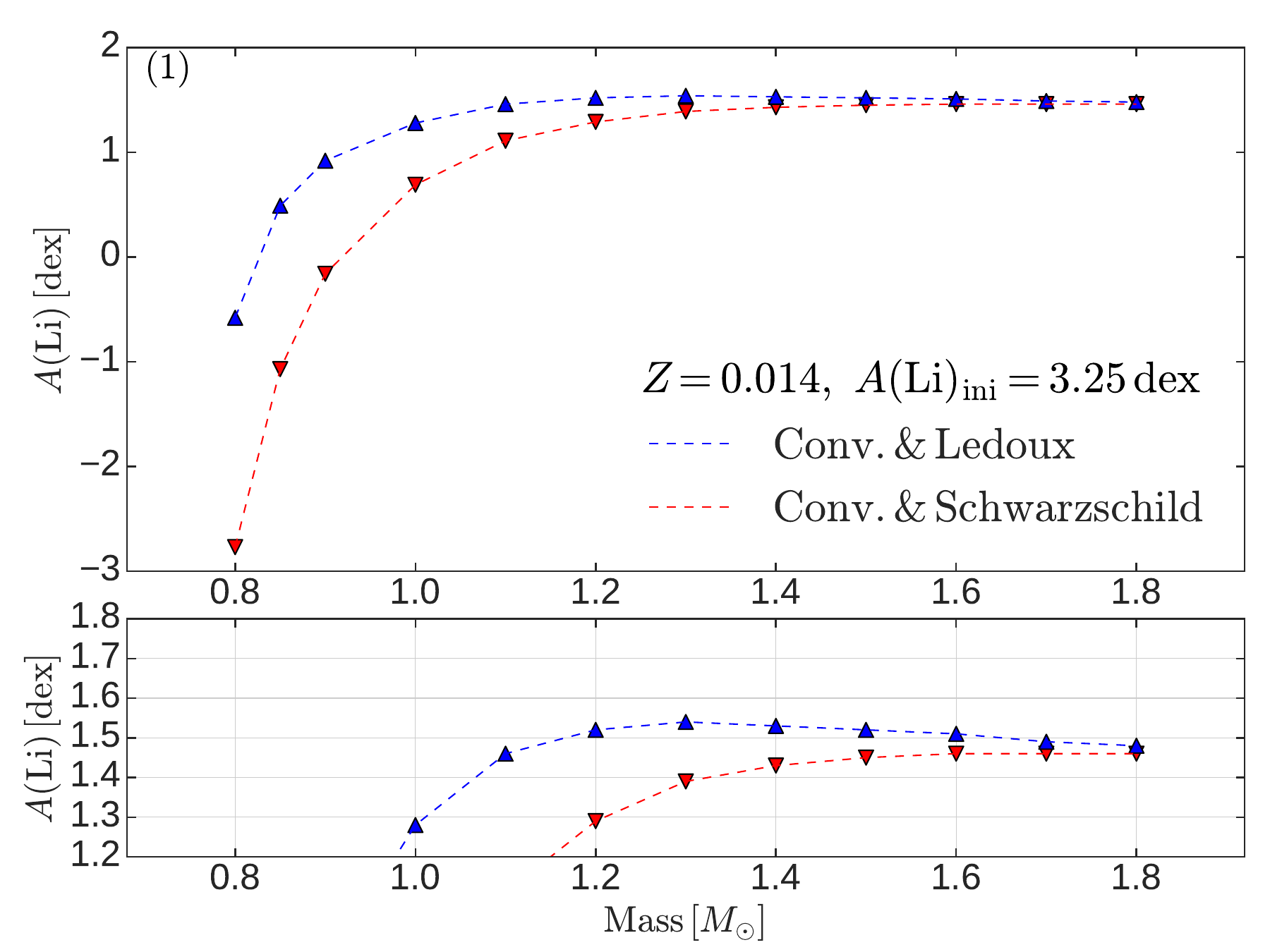}
\includegraphics[scale=0.29]{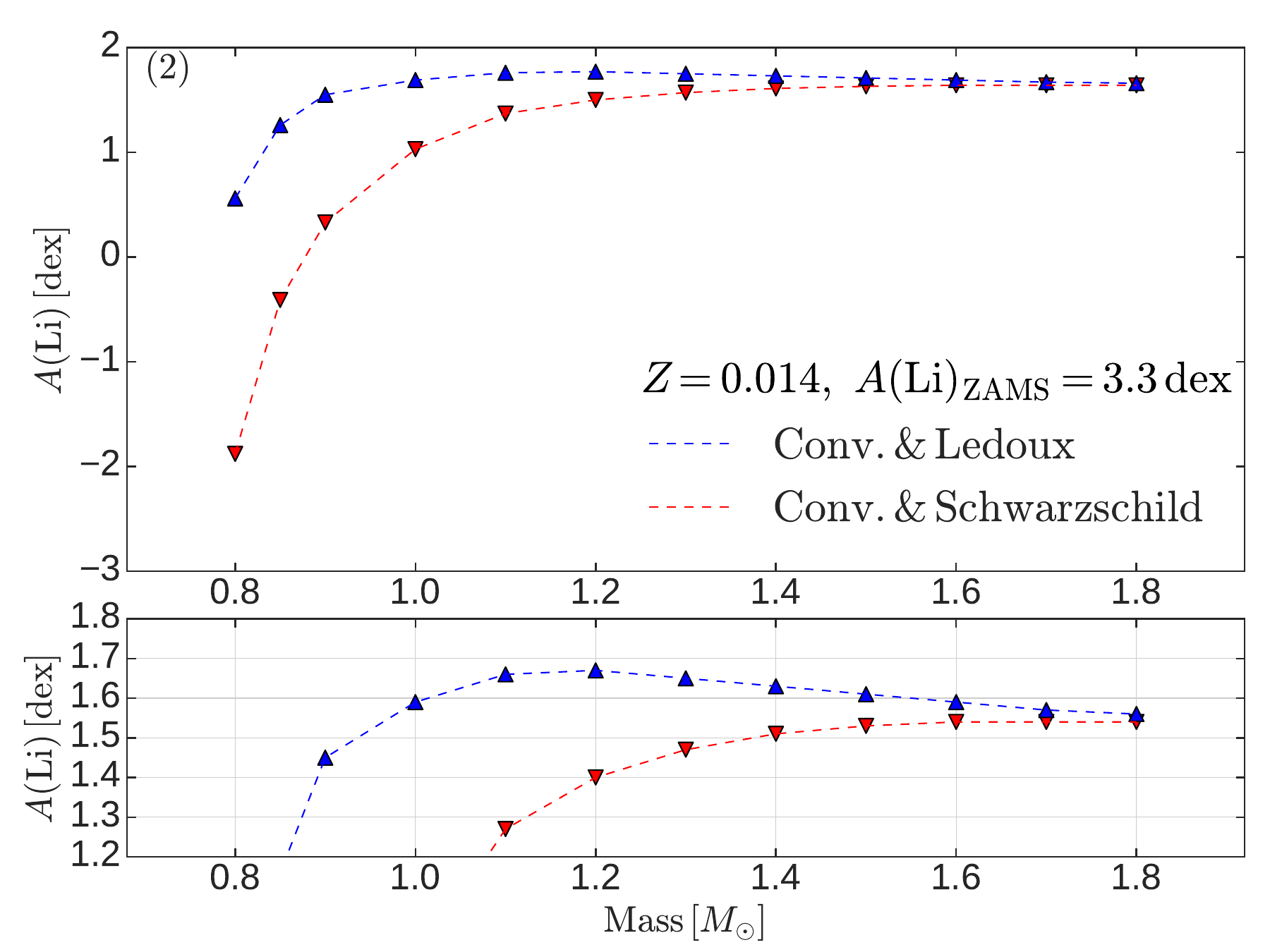}
\caption{The Li abundances of giants vs. stellar mass. We take the Li abundance of the RGB tip to plot. The lower panels are the local enlarged image corresponding to the upper panels. The $\rm Conv.\, \&\, Schwarzschild$ and the $\rm Conv.\, \&\, Ledoux$ models are marked in red and blue, respectively. Refer to $\rm Figs.\,\ref{fig:appendix3}\ and\ \ref{fig:appendix2}$ for the related Kippenhahn diagrams with panels (1) and (2).} \label{fig:convmass} 
\end{figure*}

The temperature gradient and the opacity are closely related, and higher metallicities mean higher opacities. Consequently, we need to examine the impact of metallicity on the models. Additionally, we add thermohaline mixing and semiconvection, as both necessitate attention to the molecular weight gradient. The formation of the thermohaline mixing requires the existence of an inverse molecular weight gradient, i.e.,$\bigtriangledown_\mu < 0$ \citep[e.g.,][]{1972ApJ...172..165U, 2007A&A...467L..15C}, while the semiconvection meets $\bigtriangledown_{\rm ad} < \bigtriangledown_{\rm rad} < \bigtriangledown_\mu + \bigtriangledown_{\rm ad}$ \citep{1983A&A...126..207L}. The Li abundance over luminosity for the $\rm Conv.\&\,Schwarzschild$, the $\rm Conv.\&\,Ledoux$, the thermohaline mixing, and the semiconvection models are displayed in $\rm Fig.\,\ref{fig:structure}$. Since the Li abundances of our models output are almost constant after the RGB bump, in order to show more detail on a limited scale, we show the results from the MS turnoff to luminosity $=100\,L_{\odot}$ in $\rm Fig.\,\ref{fig:structure}$. This choice ensures that the models with different metallicities pass through the RGB bump.
The figure indicates a positive correlation between the difference in Li abundance and metallicity caused by the selection of the convective boundary. At lower metallicities, the difference of the convective boundary has minimal effect on the giant Li abundances; however, this effect grows with rising metallicity. The introduction of the Ledoux boundary can significantly improve the extreme Li depletion caused by the Schwarzschild criterion defined boundary model in the case of high metallicity. Such as, in $\rm Fig.\,\ref{fig:structure}\, (4)$, the $\rm Conv.\&\,Ledoux$ model improves the Li abundance by about three orders of magnitude compared to the $\rm Conv.\&\,Schwarzschild$ model. In higher metallicity models, the opacity is correspondingly higher, where higher temperature gradients are required for effective energy transfer, resulting in the convective boundary being closer to higher temperature interior regions. As a result, the Li depletion is more pronounced than in the lower metallicity models. On the other hand, in the higher metallicity models, both the Schwarzschild and the Ledoux boundaries are in the higher temperature region, which makes their Li abundance depletion obvious, but in this case, a slight discrepancy in the boundary can bring about a significant disparity in Li abundance. Therefore, the obvious differences between their Li abundances occur at higher metallicities. Above results are precisely to compensate for the shortcomings of the standard convection model of \cite{2023ApJ...943..115L}. 

Since the convective boundary lies between the Schwarzschild and Ledoux boundaries for the semiconvection models, the influence of the semiconvection effect on the Li depletion is between the two convection models. While the thermohaline mixing model is in line with the $\rm Conv.\, \&\, Ledoux$ model.
There are two possible reasons. One is the numerical value of the thermohaline mixing coefficient $\alpha_{\rm th}$. In our work, we choose the recommended value, $\alpha_{\rm th}=2$, of the numerical simulation results \citep{2010ApJ...723..563D,  2011ApJ...728L..29T, 2013ApJ...768...34B}. Obviously, it is inconsistent with the evolution behavior of Li abundance at higher coefficient selection, such as the thermohaline mixing showing significant Li depletion after the RGB bump in the case of $\alpha_{\rm th}=100$ \citep[e.g.,][]{2020NatAs...4.1059K, 2023ApJ...943..115L}. Since the diffusion coefficient and the mixing coefficients meet $D_{\rm th}\,\propto\, $$\alpha_{\rm th}$ \citep[e.g.,][]{2007A&A...467L..15C,2010A&A...522A..10C,2013ApJS..208....4P}, it can be seen that the model with $\alpha_{\rm th}=2$ can only provide a very low diffusion coefficient compared to $\alpha_{\rm th}=100$, which will make it difficult to achieve the transportation of Li and beryllium, so the behavior of Li is as shown in $\rm Fig.\,\ref{fig:structure}$. The other is also worth thinking whether the thermohaline mixing zone and the convective zone are connected. The diverse descriptions for the thermohaline mixing will give different results, see \cite{2011A&A...533A.139W} for more details.

The convection models seem to be limited by mass effects in explaining the giant Li abundances \citep[e.g.,][]{2022ApJ...933...58C, 2023ApJ...943..115L}, with higher masses generally holding more surface Li \citep{1989ApJ...347..835G, 2023AJ....166...60T}. Then the impact of mass is also brought into our scope of concern. In addition, we also take two different Li abundance inputs into account, i.e., $A(\rm Li)_{ini}$ and $A(\rm Li)_{ZAMS}$. $\rm Fig.\,\ref{fig:convmass}$ shows the Li abundances at the RGB tip predicted by two convection models in the mass range of $0.8-1.8\,M_{\odot}$. Contrary to the result of obtained by considering the metallicity parameter in $\rm Fig.\,\ref{fig:structure}$, the influence of the two convection models on the Li abundance becomes weaker with the increase of mass parameters. Compared with $\rm Figs.\,\ref{fig:convmass}\, (1)$ and (2), the initial Li abundance is also a critical factor affecting the Li content of the giants. Our models show a higher chance of forming a Li-rich giant when the unified initial Li abundance is $\rm 3.3\,dex$. Moreover, the convection model with the convective boundary defined by the Ledoux criterion also increases this probability.

As above and $\rm Sect.\,\ref{sec:3.1.1}$, we find that the treatment for the convective boundary significantly affects the evolution of the giant Li abundance, indicating a possible formation channel of the Li-rich giants.

\subsection{Effect of MS Li Abundances on the Li Abundances of Giants}\label{sec:3.2}
\begin{figure}
	\centering
	\includegraphics[scale=0.29]{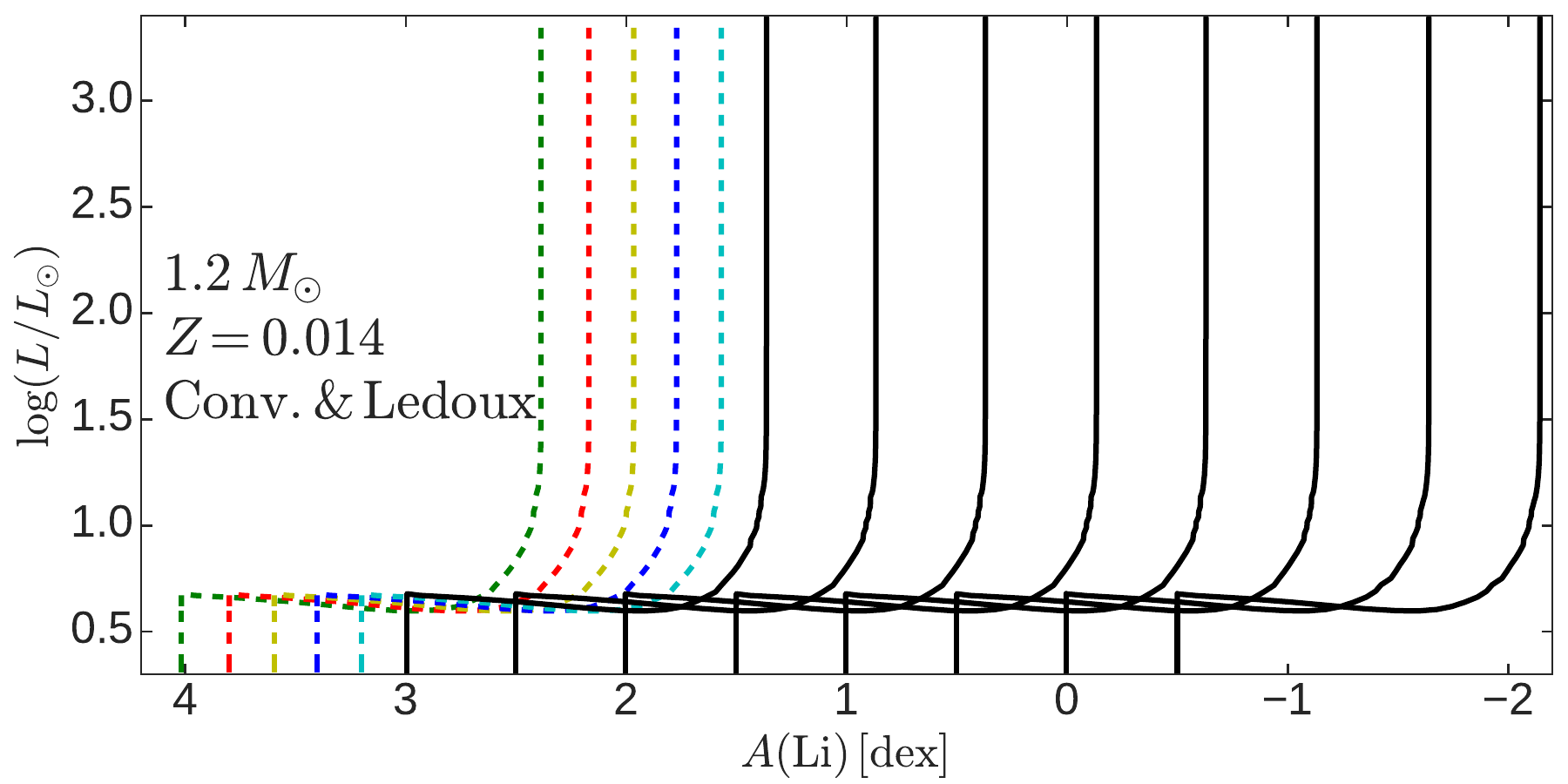}
	\caption{Luminosity vs. $A(\rm Li)$. The evolution of Li abundance from the  ZAMS to the RGB tip for stellar models with different $A(\rm Li)_{ZAMS}$ with $4.0$, $3.8$, $3.6$, $3.4$, $3.2$, $3.0$, $2.5$, $2.0$, $1.5$, $1.0$, $0.5$, $0.0$, and $\rm -0.5\,dex$. Non-black lines indicate an initial Li abundance of $\rm 3.2\,dex$ or greater.}
	\label{fig:dLi}
\end{figure}

\begin{figure*}
	\centering
	\includegraphics[scale=0.29]{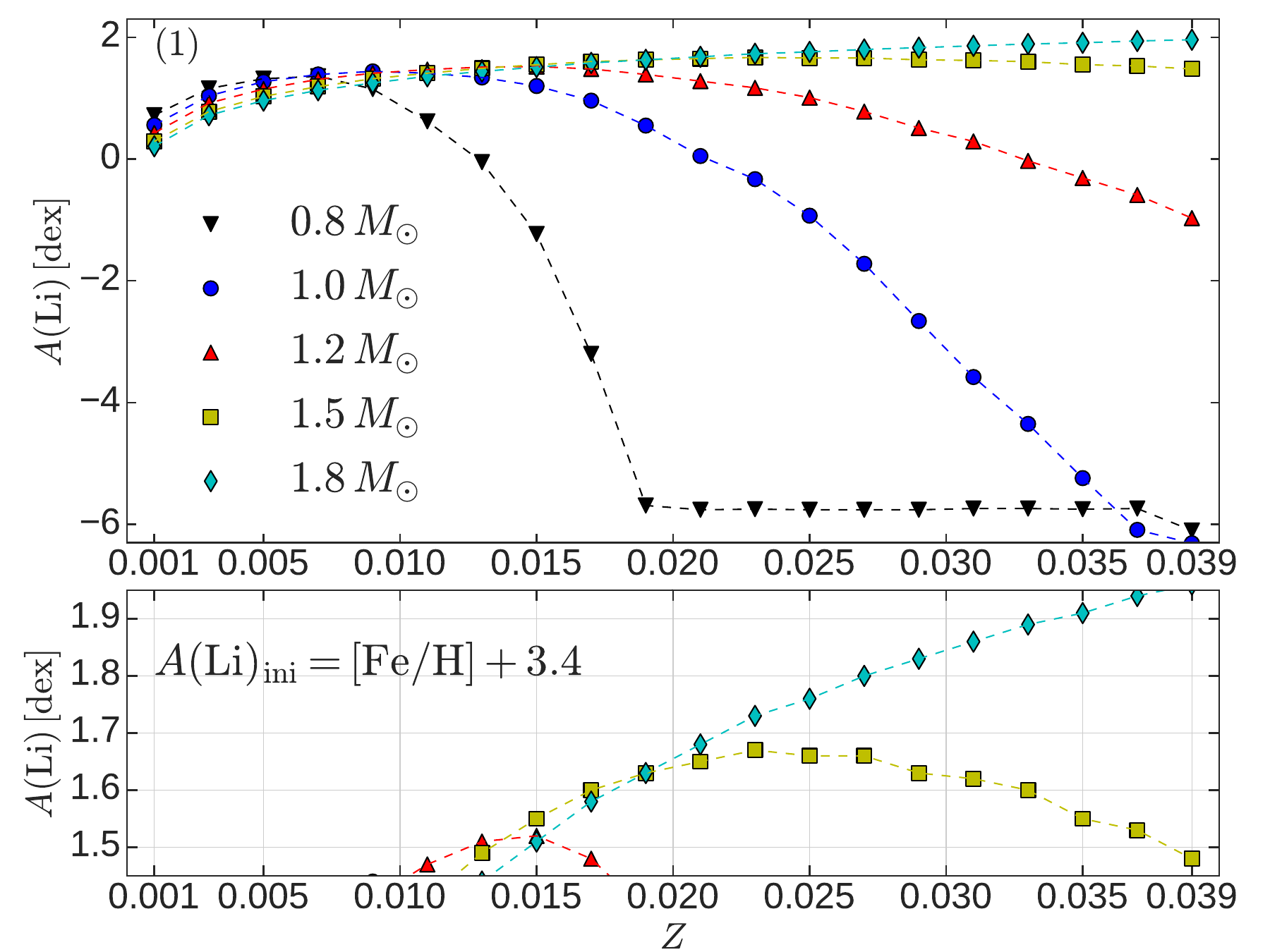}
	\includegraphics[scale=0.29]{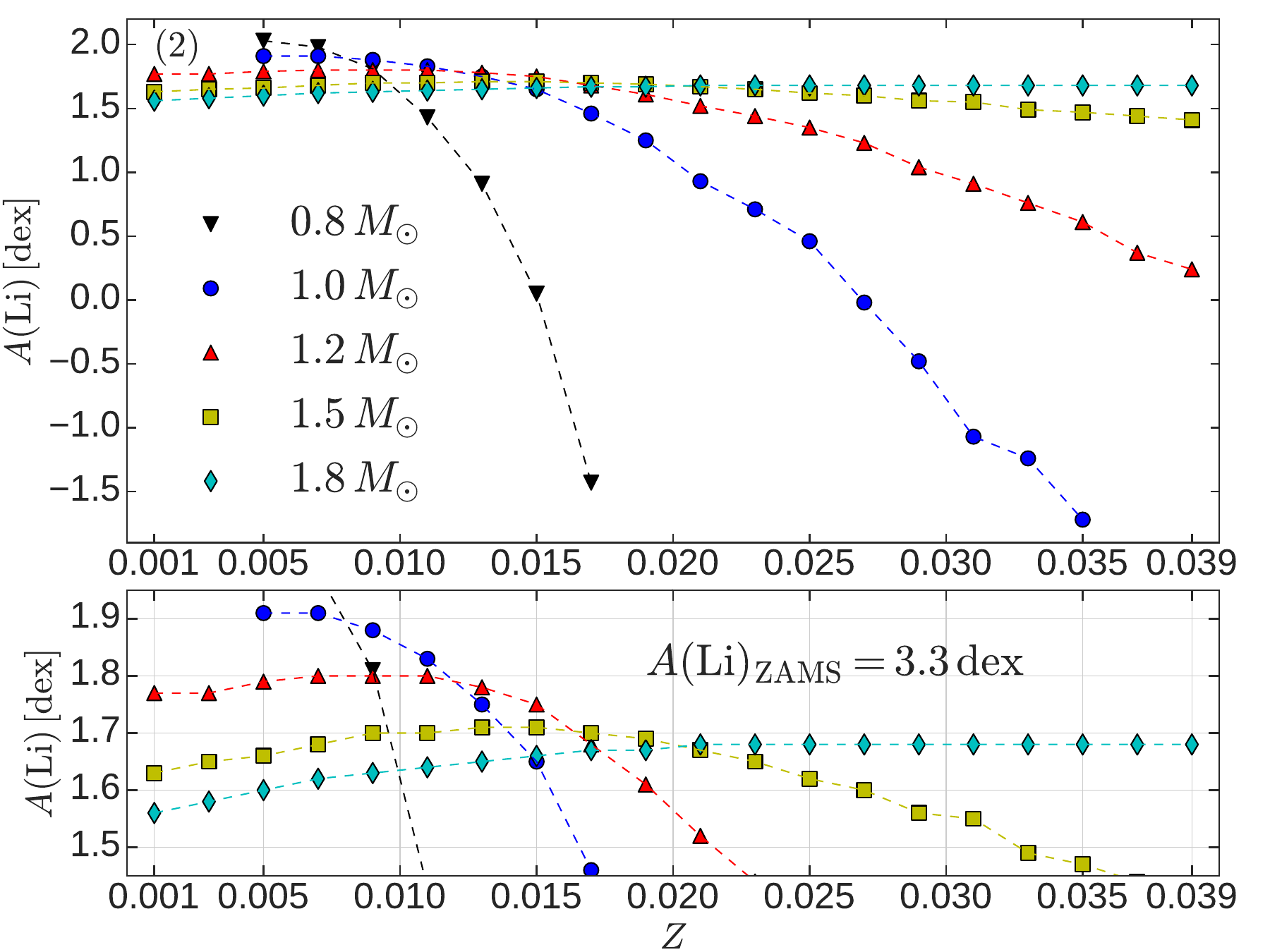}
	\caption{The Li abundances at the RGB tip with different stellar parameters. The difference between upper panels is the initial Li abundance input. The lower panels are the local enlarged image corresponding to the upper panels.} \label{fig:miz}
\end{figure*}

\begin{figure}
	\centering
	\includegraphics[scale=0.29]{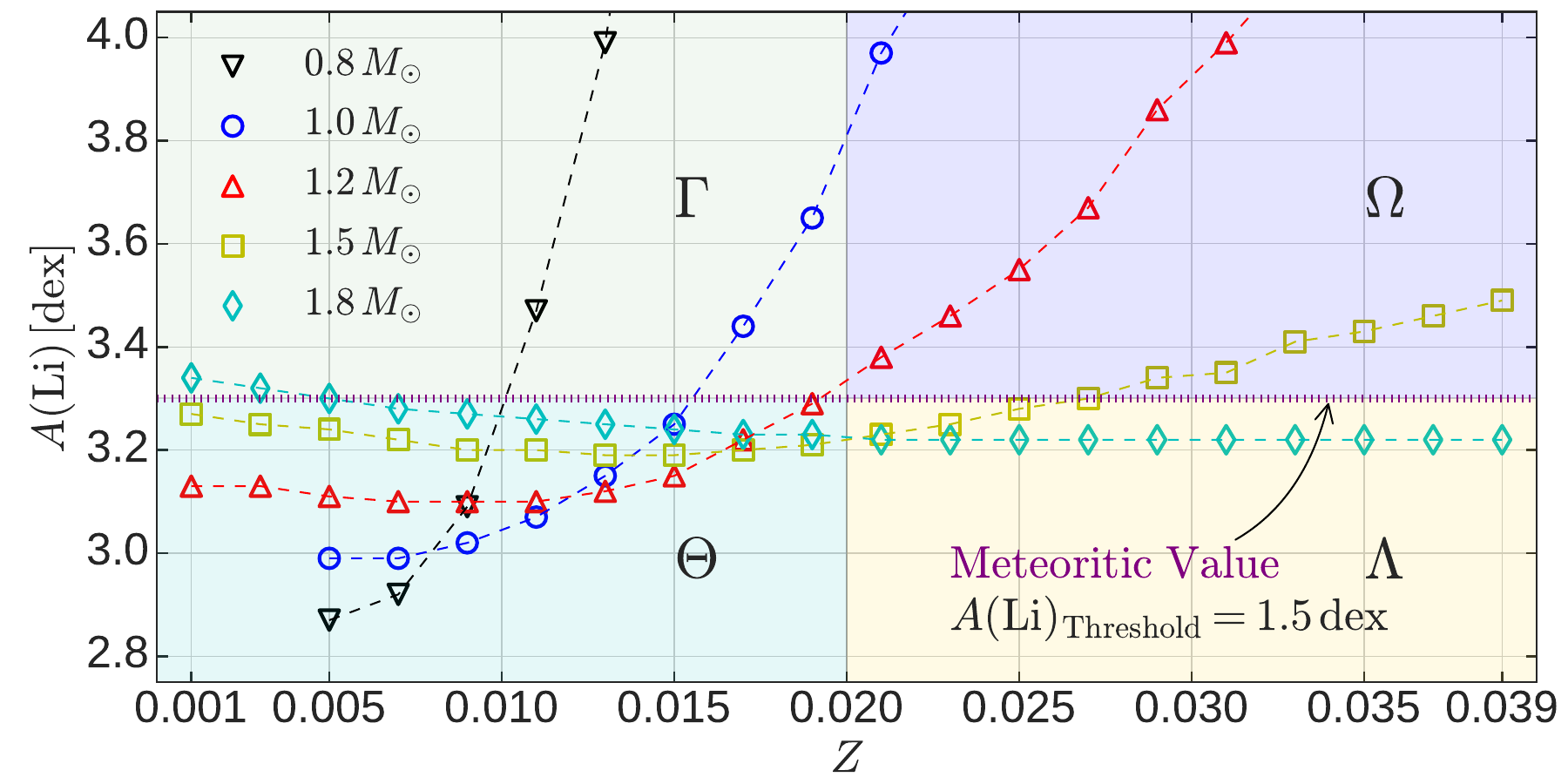}
	\caption{$A(\rm Li)_{\rm ZAMS}$ vs. $Z$. The Li abundance low limit of the ZAMS stars when the Li abundance of giants are $\rm 1.5\,dex$ (i.e., $A(\rm Li)_{\rm Threshold}$). We mark the meteoritic value with purple. We divide the figure into four areas $\Gamma$, $\Omega$, $\Theta$, and $\Lambda$, each marked with a different color.} \label{fig:mt}
\end{figure}

\begin{figure}
	\centering
\includegraphics[scale=0.29]{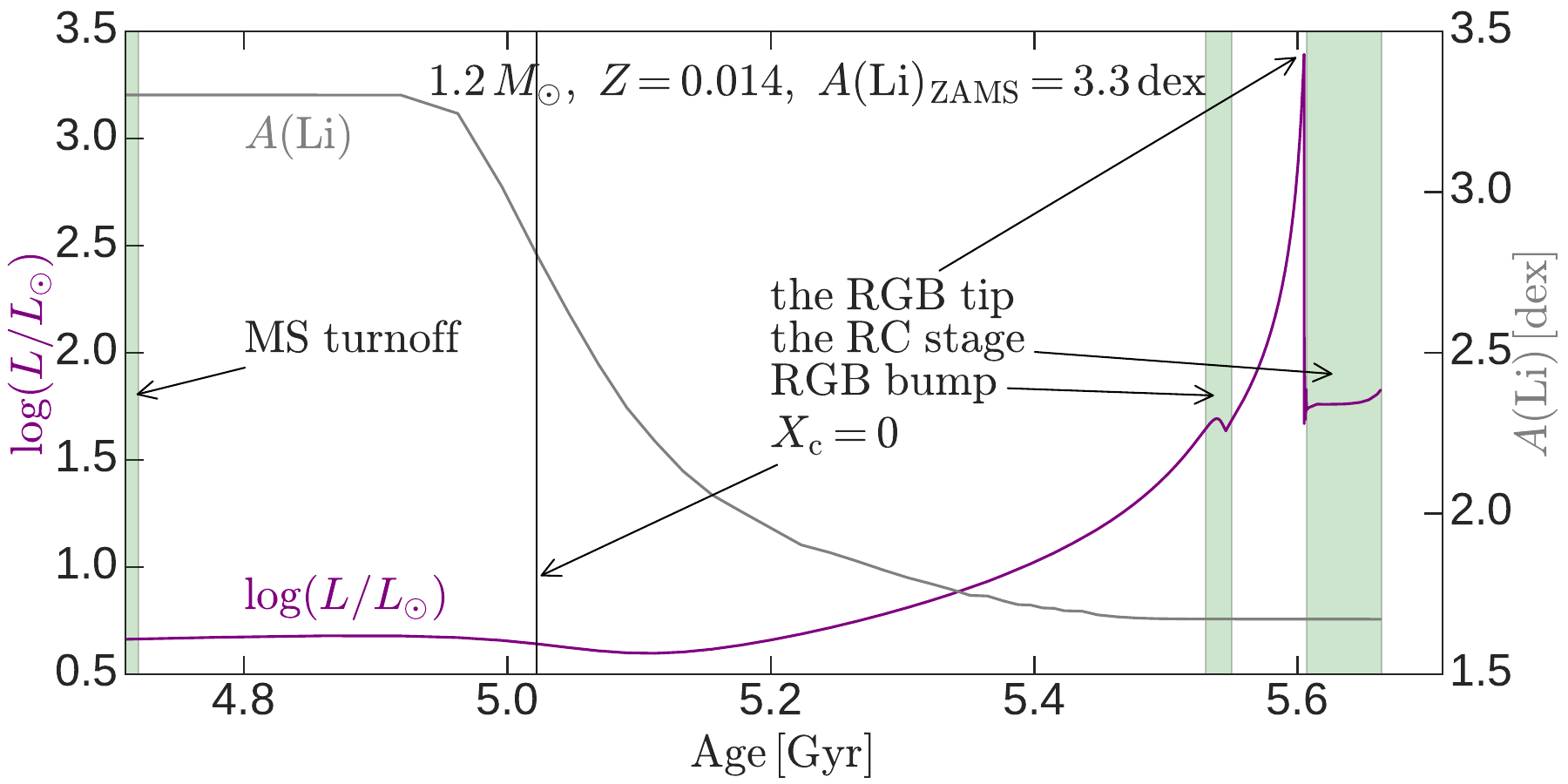}
	\caption{$\rm Luminosity/$$A(\rm Li)$ vs. $\rm Age$. The left y-axis: luminosity, and we mark the evolution curve in purple. The right y-axis: $A(\rm Li)$, we mark the evolution curve in grey. We mark four moments or time periods with shaded areas, namely the MS turnoff, the RGB bump, the RGB tip, and the RC stage. In addition, we use a solid vertical line in the diagram to mark the time when the mass fraction of the central hydrogen is 0 (i.e., $X_{\rm c}=0$).} \label{fig:hr}
\end{figure}

It is known from $\rm Fig.\,\ref{fig:convmass}$ that another factor contributing to the formation of Li-rich giants in our convection models is the Li abundance of giant progenitor. Here, the giant progenitor refers to a star in the foregone stage of the RGB, e.g., unevolved stars, MS stars, and MS turnoff stars.
Since low-mass stars will spend most of their lives during the MS, we take the MS stars as the main giant progenitors to conduct the follow-up analysis. In this section, we will explore the effect of the Li abundance of the MS stars on the formation of the Li-rich giant with the $\rm Conv.\,\&\,Ledoux$ model. Although the thermohaline mixing and the $\rm Conv.\,\&\,Ledoux$ models are consistent in the results, the main focus of this paper is to explore the role of convective boundaries on the Li abundance of giants. Therefore, $\rm Sect.\,\ref{sec:3.2}$ does not cover the thermohaline mixing process. Here, we would like to clarify that the thermohaline mixing model is used only as a tool for theoretical exploration (refer to $\rm Sect.\,\ref{sec:4.2}$), and not as a basis for drawing specific conclusions about Li abundance in all giant phases.

Just like the work of \citet{1984A&A...138..431D} and \citet{1990ApJS...73...21D}, our models with the Ledoux boundary and input $A(\rm Li)_{ZMAS}$ do not experience the Li depletion during the MS. Furthermore, these models do not take the evolution of surface Li into account during the PMS. Considering the limitations of the models, we use a distribution with $\rm -0.5-+4.0\,dex$ as the input for Li abundance. $\rm Fig.\,\ref{fig:dLi}$ presents the evolution of stellar surface Li for varying $A(\rm Li)_{\rm ZAMS}$. Due to the inward migration of the convective envelope during the first dredge-up, the Li depletion concentrates in this phase. The depletion is about $\rm 1.5\,dex$ for the stellar models with $1.2\,M_{\odot}$ and $Z=0.014$. If the Li abundance is $\rm \sim 3.0\,dex$ at the MS turnoff, a giant star with a Li abundance of $\rm 1.5\,dex$ will be finally formed. Our models predict one possible outcome there is a strong correlation between the Li abundance distribution of the giants and that of their progenitors. Starting from our convection models, only the degree of the Li depletion caused by the first dredge-up can affect the Li abundance in the later period of the RGB.  Therefore, for the given initial Li abundance distribution input, whether it be our models or the post-MS models, the final result will also obtain similar Li abundance distribution.  Since there is no Li enrichment process in our models, the final result will be simple elementary arithmetic for the initial Li abundance distribution. As a result, the distribution of the obtained giants Li abundance is directly determined by the input.  However, the immediate consequence of this is that the weight of the initial Li abundance distribution is the same as that of the Li abundance distribution of the giants, which needs to be verified further by exploring the correlation between the two by matching observations in the future.

At present, the observation of giant Li abundance has shifted to large-scale data surveys, which can provide a large amount of sample data. The mass of the low-mass giant is mainly distributed in $\sim 0.8-1.8\,M_{\odot}$ \citep[e.g.,][]{2019ApJ...880..125C,2022ApJ...931..136Z}, while the metallicity $\rm [Fe/H]$ is concentrated in the range of $\rm \sim -1.2 - +0.4\,dex$ \citep[e.g.,][]{2019ApJS..245...33G,2021ApJ...914..116G,2021MNRAS.505.5340M}. Therefore, we extend $\rm Fig.\,\ref{fig:convmass}$ to a wider range of stellar parameters (i.e., the Mass and $Z$ ranges in $\rm Table\,\ref{tab:t1}$), and the results are presented in $\rm Fig.\,\ref{fig:miz}$.  In  $\rm Fig.\,\ref{fig:miz}\,(1)$, only 1.5 and $1.8\, M_{\rm \odot}$ stars have more opportunities to form the Li-rich giants, while $\rm Fig.\,\ref{fig:miz}\,(2)$ expands the scope of such opportunities. Their differences once again confirm that the Li abundances of the giant progenitors are pivotal factor for the formation of the Li-rich giants. As can be seen from $\rm Fig.\,\ref{fig:miz}\, (2)$, the stars with lower masses and metallicities can maintain higher surface Li during the RGB instead. However, the problem also arises in the stars with lower masses, especially the higher metallicity stars, which appears to be difficult or impossible to become the Li-rich giants through natural Li depletion of their progenitors. But more importantly, our models propose a possible formation channel for the Li-rich giants, i.e., the origin of the vast majority of the Li-rich giants is their super Li-rich progenitors ($A(\rm Li)>3.3\,dex$).

%%Based on the big bang nucleosynthesis theory, the primordial Li abundance is usually $\rm 2.7\,dex$ \citep{2003PhLB..567..227C}. From $\rm Fig.\,\ref{fig:dLi}$, it can be noticed that the Li depletion level in our convection models is independent of the initial value. Therefore, if the input is reduced from $\rm 3.3$ to $\rm 2.7\,dex$, the Li abundances of the giants will be concentrated around $\rm 1.0\,dex$, which is consistent with the observed Li abundance peak \citep{2021ApJ...914..116G}.

In $\rm Fig.\,\ref{fig:miz}\, (2)$, we find that the same Li abundance input will contribute to disparate Li abundance values of the Li-rich giants with different masses and metallicities. Therefore, we present in $\rm Fig.\,\ref{fig:mt}$ the required $A(\rm Li)_{ZAMS}$ value for the giants with a Li abundance of exactly $\rm 1.5\,dex$. Interestingly, if the Li abundance input is close to the meteoritic value \citep[$\sim \rm 3.3\,dex$, e.g.,][]{2009ARA&A..47..481A}, most of the low-mass stars will naturally form the Li-rich giants with our $\rm Conv.\,\&\, Ledoux$ model (i.e., the  $\Theta$ and $\Lambda$ areas). 
Our models predict that the stars with lower metallicities are more likely to form the Li-rich giants for the $\rm Fig.\,\ref{fig:mt}$ area $\Theta$. The metallicity of the Li-rich giants is concentrated in the range of $\rm -0.5<[Fe/H]<+0.3\, dex$ \citep{2020MNRAS.494.1348D, 2021MNRAS.505.5340M}, with a peak of $\rm \sim -0.15\,dex$ \citep{2019ApJS..245...33G}. The mass distribution of the Li-rich giants is more concentrated in the range of $1.0-1.5\,M_{\odot}$ \citep{2019ApJ...880..125C}. Stars in the range of $\rm [Fe/H]<0\,dex$, i.e., the areas $\rm \Gamma$ and $\rm \Theta$, have a higher probability of becoming Li-rich stars, which is in good agreement with \citet{2021MNRAS.505.5340M}. However, when $\rm [Fe/H]>0\,dex$ (i.e., the areas $\rm \Omega$ and $\rm \Lambda$), our model works for stars with higher masses (the area $\rm \Lambda$), whereas for stars with lower masses (the area $\rm \Omega$), higher $A(\rm Li)_{ZAMS}$ is needed for the formation of the Li-rich giants. As shown in $\rm Fig.\,\ref{fig:miz}$, lower mass stars are more sensitive to the opacity and, as a result, the Li depletion during the first dredge-up is very significant, a defect that our convective models find difficult to compensate for.

The Li-rich giants here can be divided into RGB and RC stars. We present the evolution trajectory of luminosity and the Li abundance with age from the MS turnoff to the end of the RC stage in $\rm Fig.\,\ref{fig:hr}$. Our models predict the same Li abundances in the late RGB and the RC phases. In Li-rich giants, however, the proportion of the RC stars is higher \citep[e.g.,][]{2021MNRAS.505.5340M, 2021ApJ...913L...4S, 2021NatAs...5...86Y}. Perhaps the observed differences between the RGB and the RC stars are due to some physical processes during the helium flash, such as internal gravity waves \citep{2020ApJ...901L..18S}. It can be seen from \citet{2021MNRAS.505.5340M} that there is basically no special difference between the Li-rich RGB and RC stars in the range of $\rm  -1.2 < [Fe/H] < +0.4\,dex$, and the result of \citet{2022ApJ...931..136Z} shows that the mass peaks of the two are about $\rm 1.5$ and $1.0\,M_{\odot}$, respectively. For the Li-rich RGB stars, \citet{2021MNRAS.505.5340M} and \citet{2022ApJ...931..136Z} shown opposite results, with \citet{2021MNRAS.505.5340M} predicting has a higher proportion in the range of $\rm [Fe/H]>0\,dex$, while \citet{2022ApJ...931..136Z} present the opposite result. For the Li-rich RC stars, they predicted the same result, with the higher the metallicity, the higher the proportion of Li-rich RC stars. While in the $\rm [Fe/H]<0\,dex$ range, the Li-rich RGB and RC stars are basically the same ratio. Therefore, the results of the $\rm Conv.\,\&\,Ledoux$ models are suitable for the most of Li-rich RGB stars because our results show that more massive stars in our metallicity range are more likely to form Li-rich giants. Due to the peak mass of the Li-rich RC stars is around $1.0\,M_{\odot}$, however, only area $\Theta$ is suitable for the Li-rich RC stars. It can be seen that extra Li enrichment processes in the stages after the RGB tip are more necessary for the formation of the Li-rich RC stars.

\section{Discussion}\label{sec:Discussion}

\begin{figure*}
	\centering
	\includegraphics[scale=0.29]{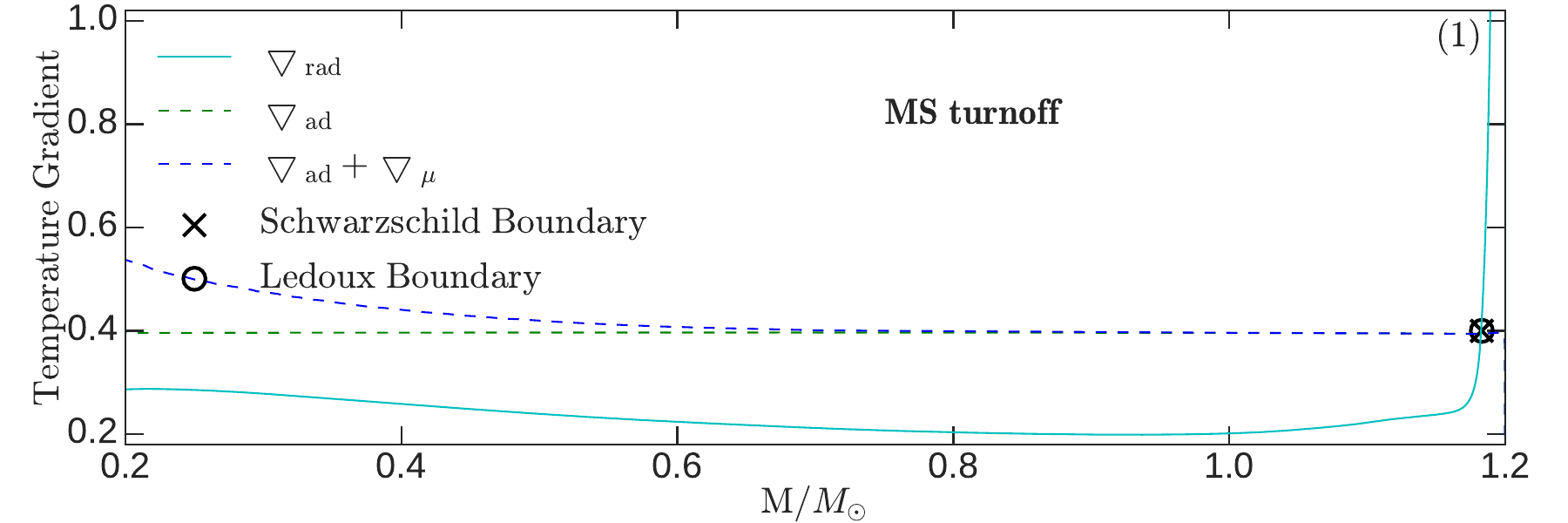}
	\includegraphics[scale=0.29]{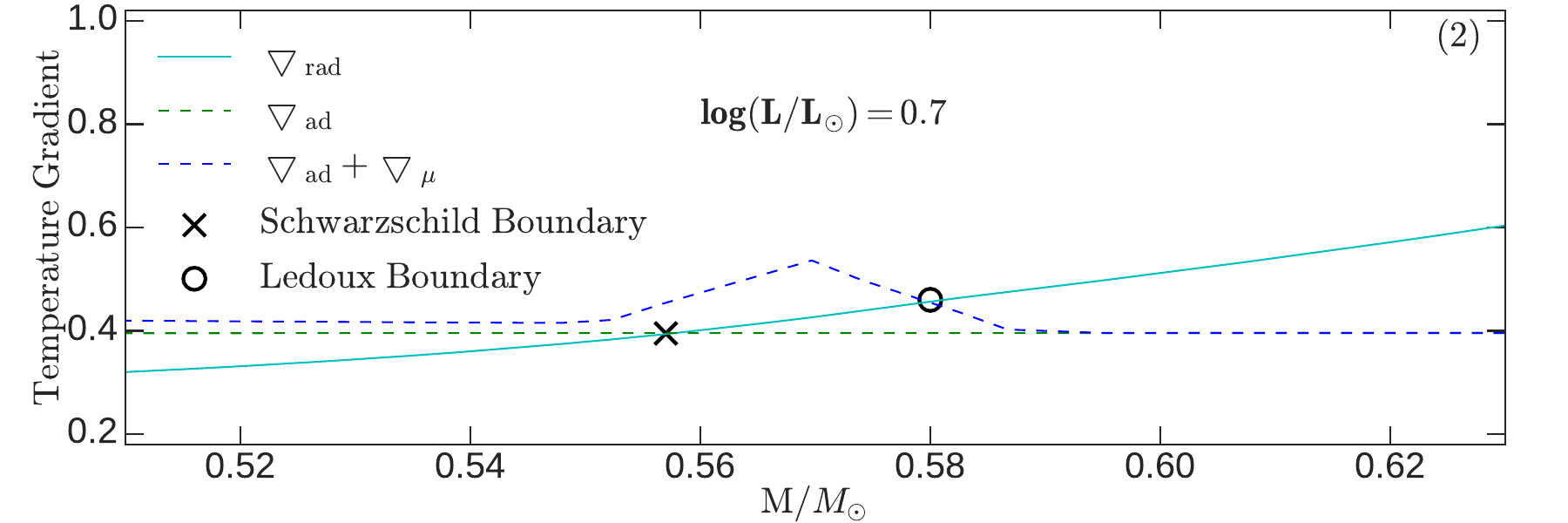}
        \includegraphics[scale=0.29]{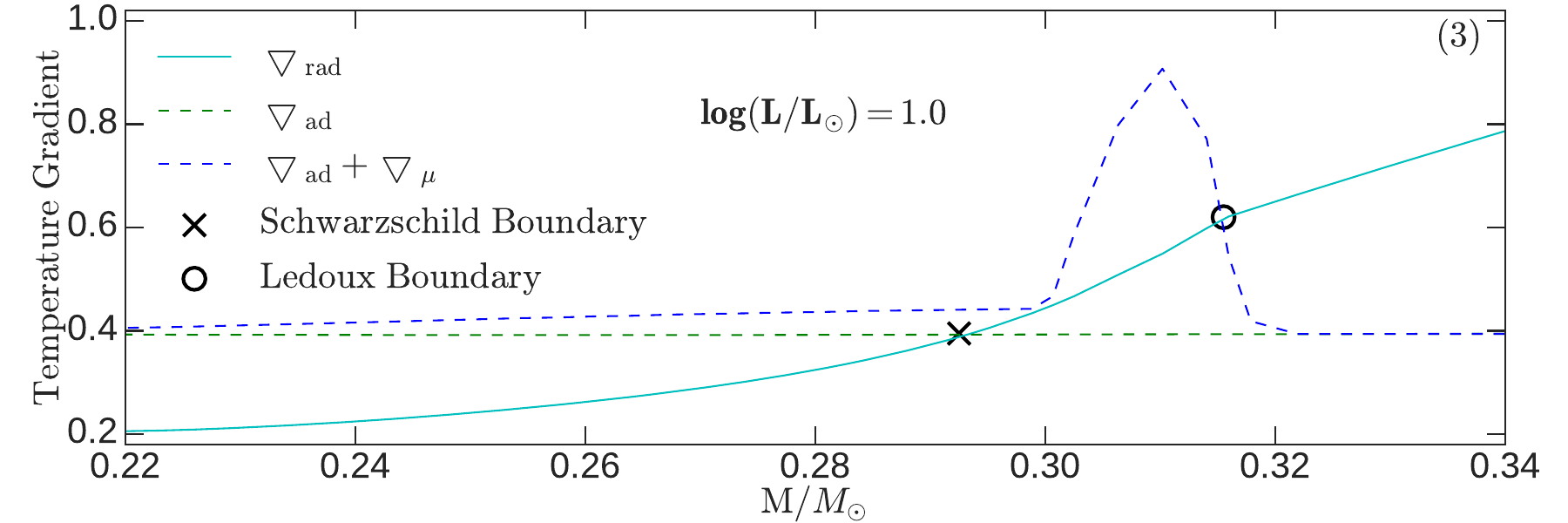}
        \includegraphics[scale=0.29]{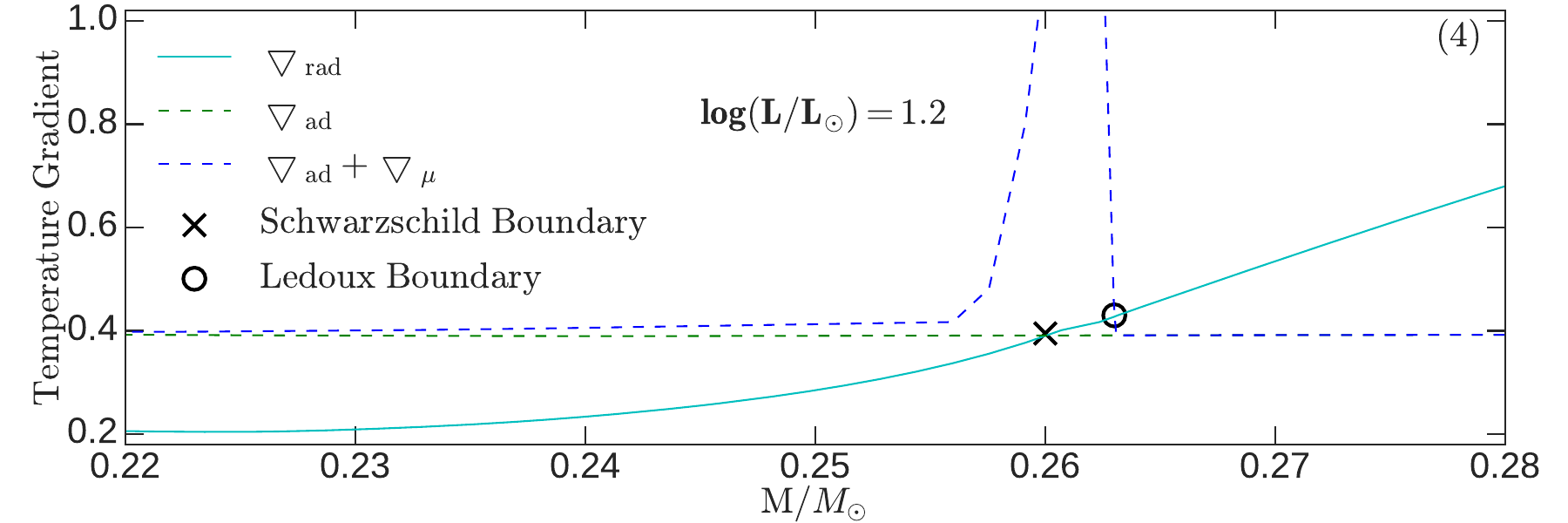}
        \includegraphics[scale=0.29]{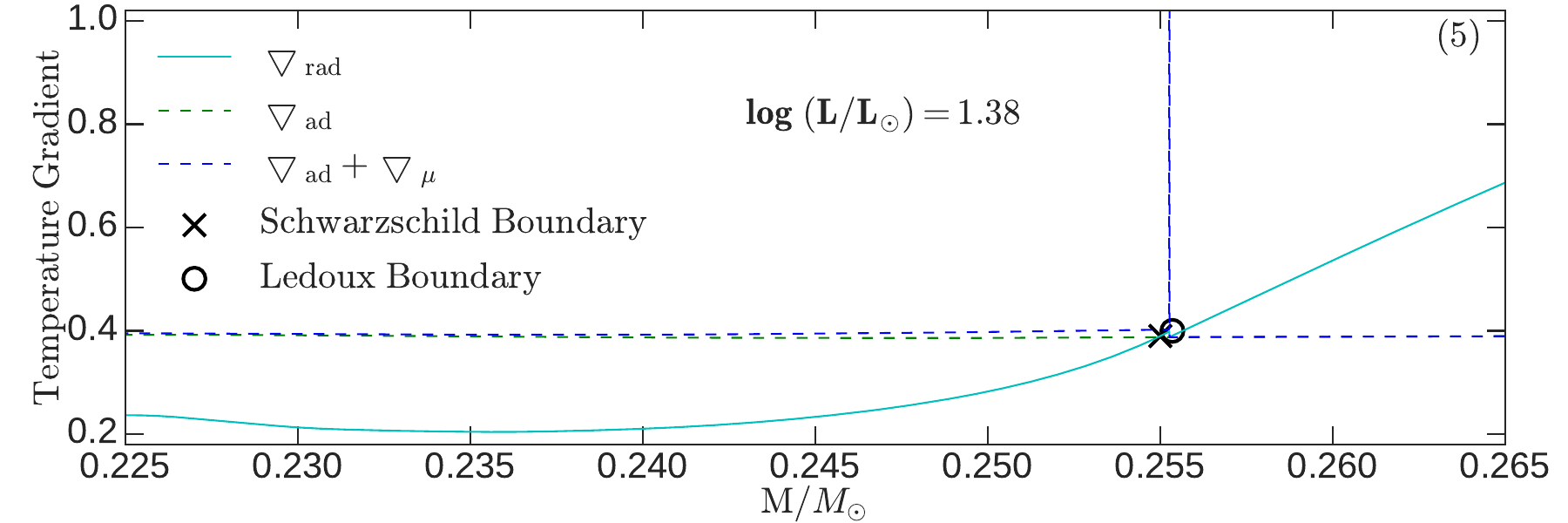}
        \includegraphics[scale=0.29]{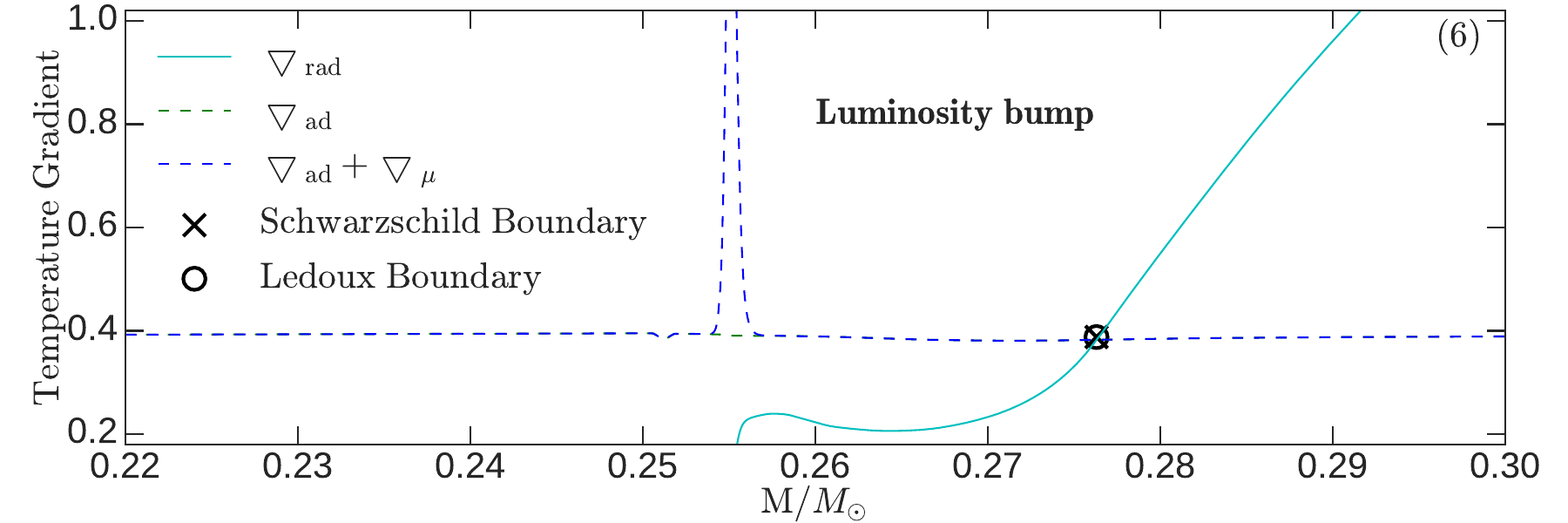}
        \includegraphics[scale=0.29]{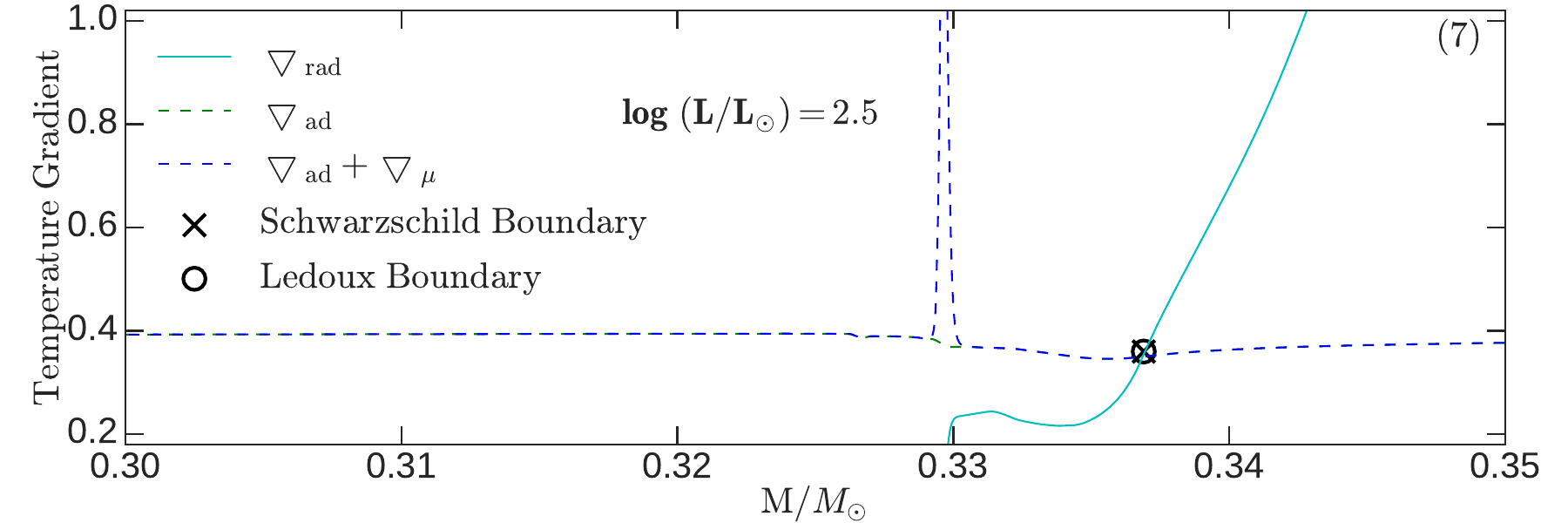}
        \includegraphics[scale=0.29]{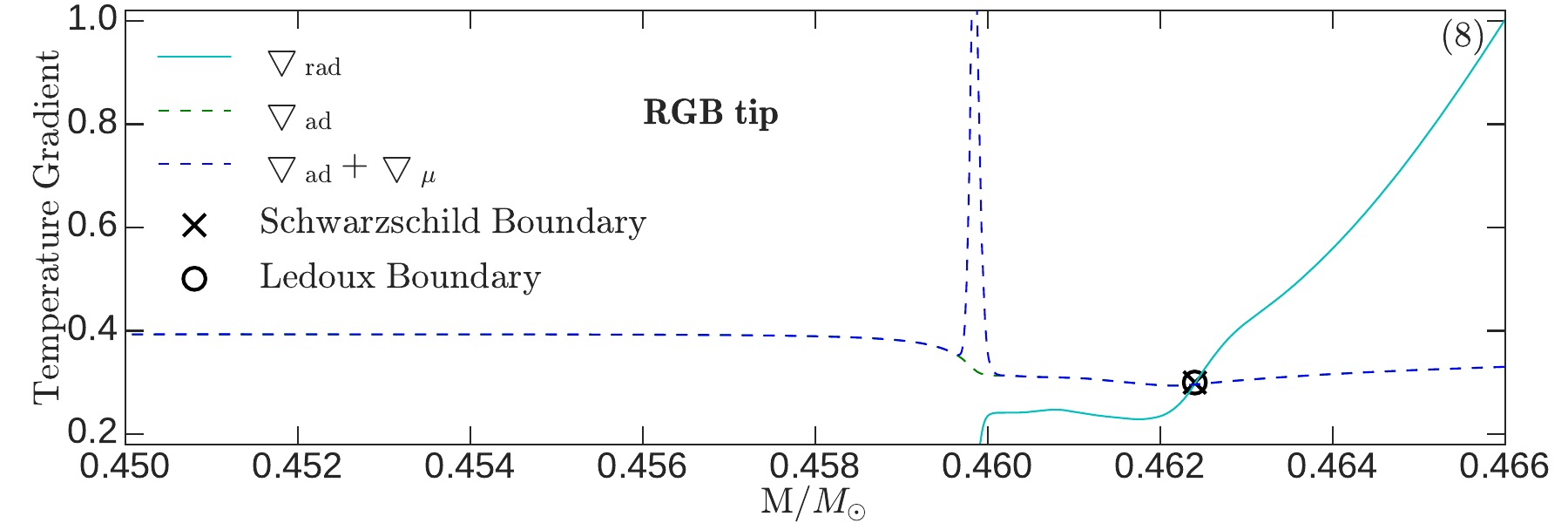}
	\caption{Similar to $\rm Fig.\,\ref{fig:boundary}\, (2)$. According to the evolution sequence of stars, we choose 8 time points from the MS turnoff to the RGB tip, which are the MS turnoff, $\rm log$$(L/L_{\odot})=0.7$, $\rm log$$(L/L_{\odot})=1.0$, $\rm log$$(L/L_{\odot})=1.2$, $\rm log$$(L/L_{\odot})=1.38$, the RGB bump, $\rm log$$(L/L_{\odot})=2.5$, and the RGB tip. We have highlighted it in bold in the figure. The Schwarzschild and Ledoux boundaries are represented by crosses and circles. The model parameter is $1.2\,M_{\odot}$, $Z=0.014$, and $A(\rm Li)_{ZMAS}=3.3\,dex$.} \label{fig:disc}
\end{figure*}

\begin{figure}
	\centering
	\includegraphics[scale=0.29]{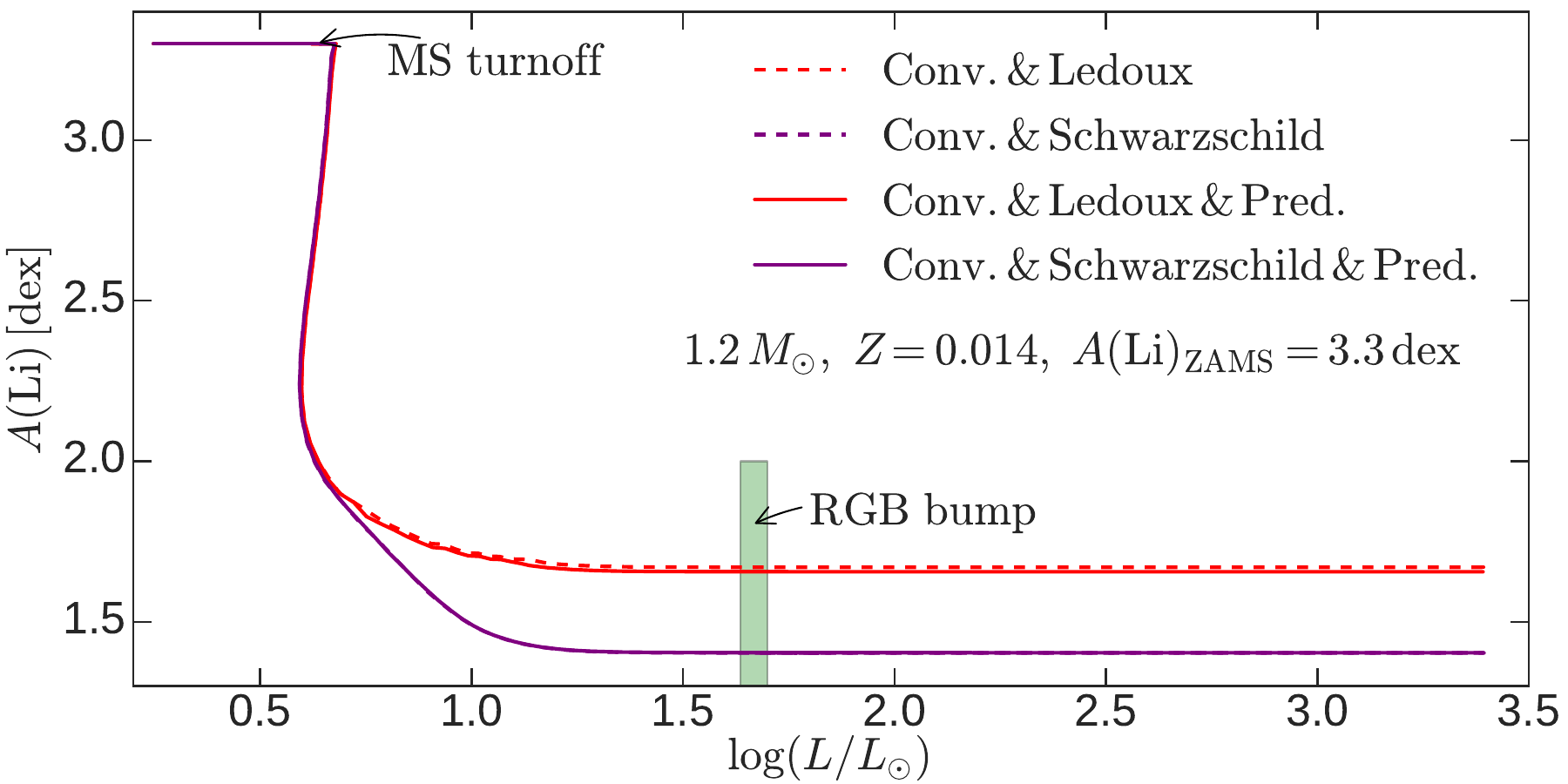}
	\caption{Similar to $\rm Fig.\,\ref{fig:boundary}\, (1)$. We consider only two convective convection models and supply corresponding models that involve predictive mixing. 
                $\rm Conv.\, \&\, Ledoux\, \&\,Pred.$: $\rm Conv.\, \&\, Ledoux$ + the predictive mixing. $\rm Conv.\, \&\, Schwarzschild\, \&\,Pred.$: $\rm Conv.\, \&\, Schwarzschild$ + the predictive mixing. The predictive mixing is only applied to the surface convective zone.} \label{fig:disc2}
\end{figure}

\begin{figure*}
	\centering
	\includegraphics[scale=0.29]{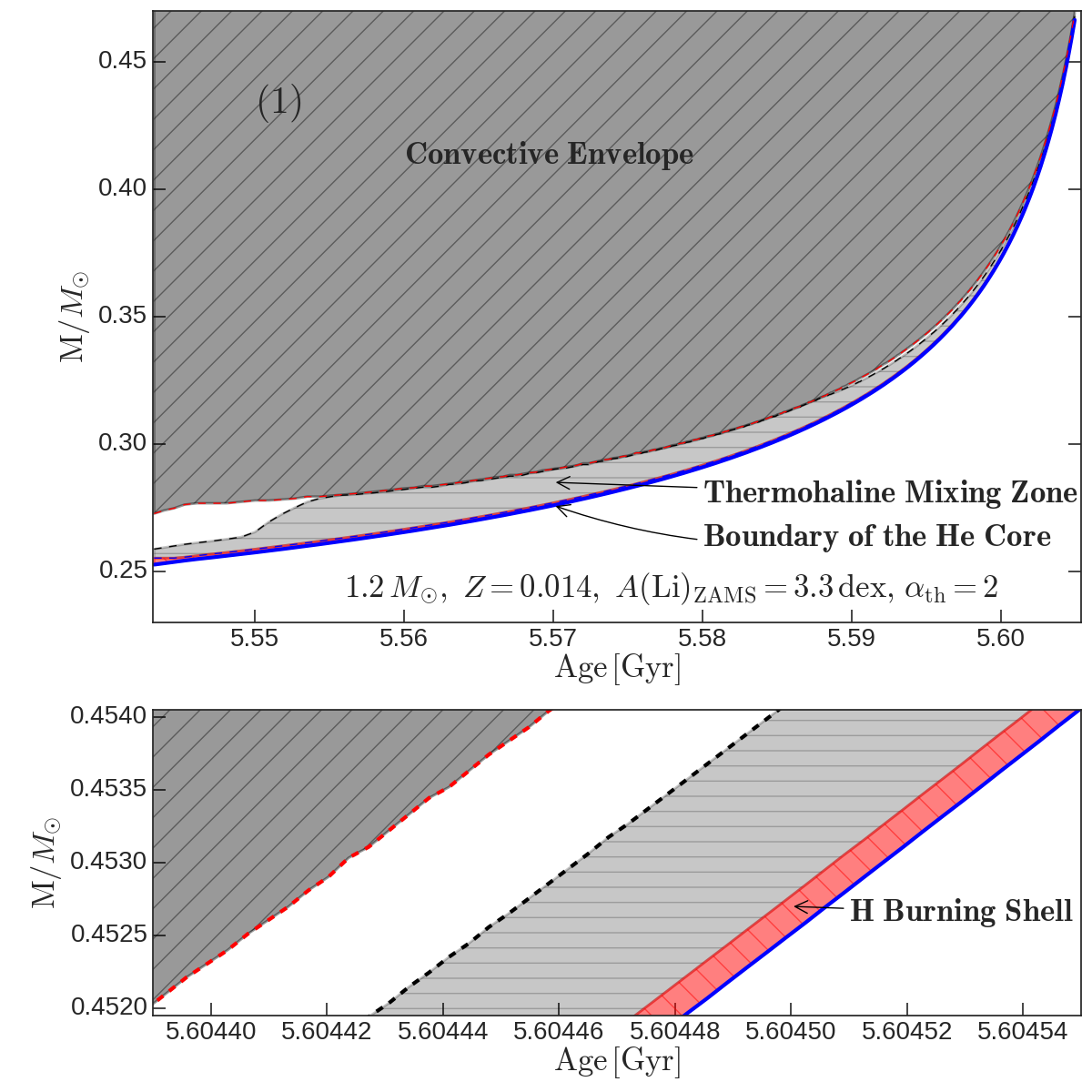}
        \includegraphics[scale=0.29]{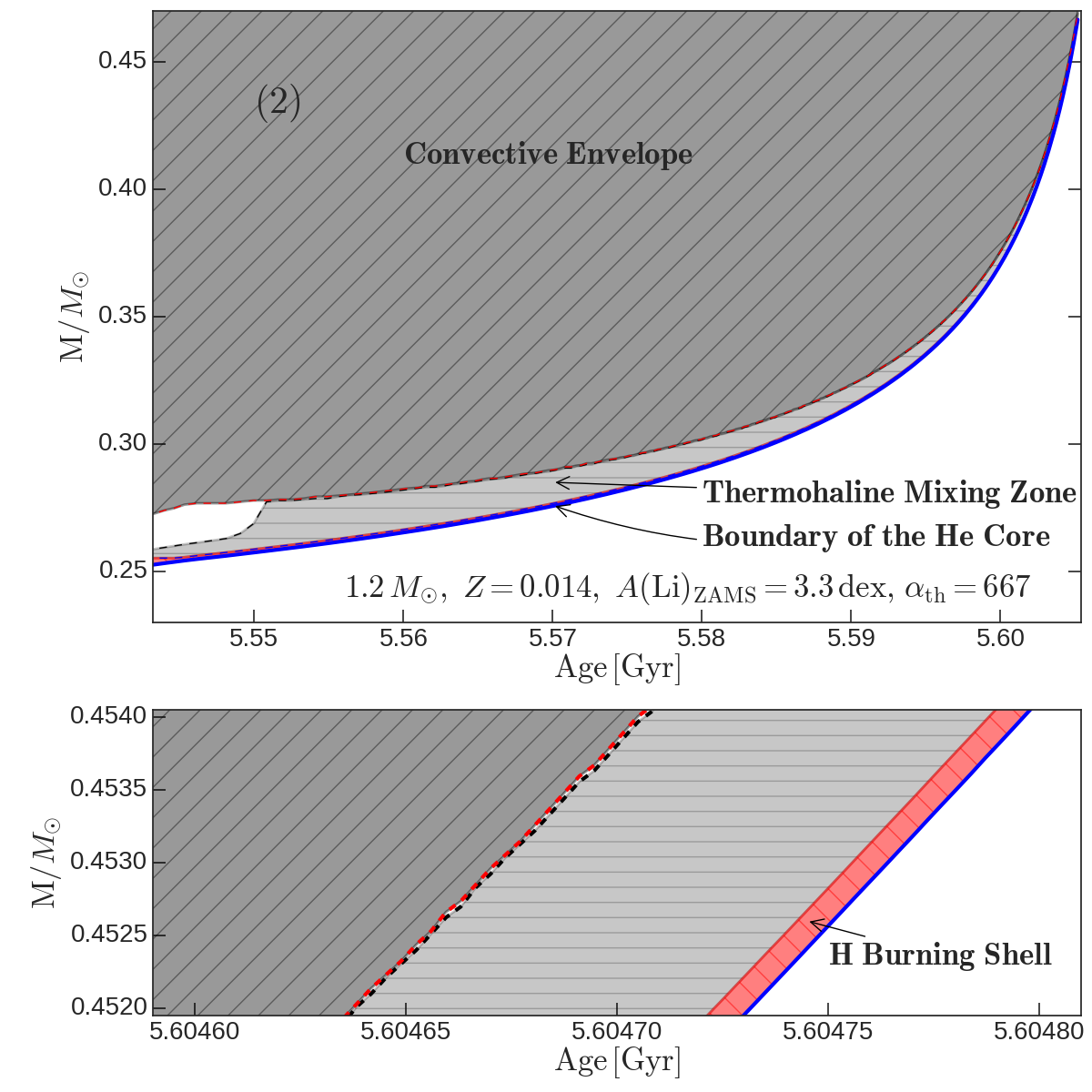}
	\caption{The Kippenhahn diagram. The evolutionary stage is from the RGB bump to the RGB tip. The thermohaline mixing coefficient $\alpha_{\rm th}$ for panels (1) and (2) are 2 and 667 respectively. Different colours represent different regions within the star. The bottom panels are  partial enlarged view of the area near the RGB tip.} \label{fig:kipp}
\end{figure*}

\subsection{Convective Boundary}\label{sec:4.1}
It can be seen from $\rm Fig.\,\ref{fig:boundary}\, (3)$ that the boundaries of two convection models coincide after the RGB bump, then the evolution of Li abundance after this stage is expected to be insensitive to the two convective boundary selection effects. We show the relative positions of the two boundaries in $\rm Fig.\,\ref{fig:disc}$. We choose eight points in time. As can be seen from $\rm Fig.\,\ref{fig:disc}$, the Schwarzschild and the Ledoux boundaries coincide near the MS turnoff (i.e., $\rm Fig.\,\ref{fig:disc}\, (1)$). When the stars enter the RGB, they begin to separate with the inward movement of the convective envelope, and then when the luminosity rises to $\sim 10^{1.4}\,L_{\odot}$ (i.e., $\rm Fig.\,\ref{fig:disc}\, (5)$), the two begin to merge again to maintain the RGB tip (i.e., $\rm Figs.\,\ref{fig:disc}\, (6), (7),\ and\ (8)$). The above behavior is closely related to the movement of the convective envelope, because at the beginning, the inward migration will enter the region with uneven chemical composition generated by the central hydrogen-burning, where $\bigtriangledown_\mu > 0$. After the first dredge-up, the convective envelope will rapidly retreat and its lower boundary will pass through the region covered by the previous convective envelope, and the chemical composition near the boundary is uniform (i.e., $\bigtriangledown_\mu = 0$). The Ledoux boundary is therefore the same as the Schwarzschild boundary. %The above behavior is the root cause of the evolutionary behavior of Li abundance in $\rm Fig.\,\ref{fig:boundary}\, (1)$.
As a result, our $\rm Conv. \&\, Ledoux$ model actually has an effect on the Li abundance only at the first dredge-up.

\citet{2022ApJ...928L..10A} conducted hydrodynamical simulations on the convective zone and its adjacent radiative zone, and pointed out that in the absence of the convective flows moving into the stably stratified region, the Schwarzschild and Ledoux boundaries are consistent. Throughout the entire stellar evolution phase for our $\rm Conv.\,\&\, Ledoux$ model, the use of the Ledoux boundary is indicated into three stages: 1) the MS; 2) the first dredge-up; and 3) after the RGB bump.

In stage 1), the Ledoux boundary is equivalent to the Schwarzschild boundary (e.g., $\rm Fig.\,\ref{fig:disc}\, (1)$). \citet{2014A&A...569A..63G} analyzed two incorrect methods for defining the convective boundaries, namely using convective instability conditions in the radiation zone and utilizing sign variations of discontinuous functions to select the convective boundary, and pointed out that interpolations or extrapolations from points within the convective zone for each iteration is a more appropriate method. From this point of view, the Schwarzschild and the Ledoux boundaries of during the MS are equivalent, as the evolution time scale far exceeds the convective overturn time scale \citep[e.g.,][]{2021A&A...650A.128G,2023Galax..11...56A}. Subsequent numerical simulations also confirm this point \citep{2022ApJ...928L..10A}. The convective boundary may not be uniquely deterministic in the case of utilizing sign variations of discontinuous functions to select convective boundary. The MESA code has this problem, which presents the possibility of non-convergence during running. The MESA provides a solution by adding the `predictive mixing' (see \citet{2018ApJS..234...34P} for more details). Our models do not take the predictive mixing into account because our results present the convergence, and reveal the equivalence of the Ledoux and the Schwarzschild boundaries except for the first dredge-up. During the first dredge-up, it can be seen from $\rm Figs.\,\ref{fig:disc}\, (2), (3),\ and\ (4)$ that the selection of the convective boundary obviously needs to place emphasis on $\bigtriangledown_\mu$, which is different from the MS and the stage after the RGB bump.

In stage 2), more detailed and appropriate handling of $\bigtriangledown_{\rm rad}=\bigtriangledown_{\rm ad}$ will not bring significant differences. \citet{2023Galax..11...75J} pointed out the Ledoux criterion can be simplified as the Schwarzschild criterion when the chemical composition tends towards homogenization (i.e., $\bigtriangledown_\mu \to 0$). Whereas, the mean molecular weight gradient near the convective boundary is significantly greater than the adiabatic temperature gradient during the first dredge-up (e.g., $\rm Figs.\,\ref{fig:disc}\, (2), (3),\ and\ (4)$), meaning that the Ledoux boundary and the Schwarzschild boundary are separated prominently. We have added the results of the predictive mixing in $\rm Fig.\,\ref{fig:disc2}$, and it can be seen that the modified convective boundary (i.e., the $\rm Conv.\, \&\, Ledoux\, \&\,Pred.$ and the $\rm Conv.\, \&\, Schwarzschild\, \&\,Pred.$ in $\rm Fig.\,\ref{fig:disc2}$) does not affect the discrepancy between the Ledoux and the Schwarzschild boundaries. In addition, the change of the modified boundary on Li abundance can be ignored.

In stage 3), due to the retreat of the convective envelope, the lower boundary is uniformly composed around it throughout the retreat process, so the two boundaries are equally equivalent. At this stage, the convective zone is far away from the hydrogen-burning shell, so the standardization of the convective boundaries does not affect the surface Li.

\subsection{Low-coefficient Thermohaline Mixing Model}\label{sec:4.2}

Recently, \citet{2022ApJ...935L..30T} used the low-coefficient thermohaline mixing (i.e., $\alpha_{\rm th}=2$) to explore the time difference in observed decay of Li and carbon near the RGB bump. Their simulation results indicate that the depletion of Li and carbon is simultaneous, which is inconsistent with observations.
Similar to their models, the mixing coefficient is also $\alpha_{\rm th}=2$ in our thermohaline mixing model, and differ from our models, they calculated more detailed spatial and temporal resolutions. Nevertheless, their simulation results show a slight Li depletion after the RGB bump due to the meet between the thermohaline mixing zone and the convective zone. The increment of surface $\rm log\,^7Li$ in their scale is about $-0.0012$. In our model, this increment corresponds to a Li depletion of about $\rm 0.003\, dex$. Our thermohaline mixing model has a  Li abundance of $\rm 1.67\, dex$ at the RGB bump, and the depletion of $\rm 0.003\, dex$ is negligible. Significant Li depletion after the RGB bump can exist in the thermohaline mixing model as long as the coefficient $\alpha_{\rm th}$ is appropriately increased \citep[e.g., $\alpha_{\rm th}=667$ for][]{2007A&A...467L..15C}.

On the other hand, their results indicate that the effect of thermohaline mixing alone is not sufficient to explain observations. If combined with our $\rm Conv.\&\ Ledoux$ model, it can be expected that the Ledoux criterion may not be appropriate because the thermohaline mixing has a homogenization effect, which may inhibit the formation of $\bigtriangledown_\mu > 0$.

\citet{2022ApJ...935L..30T} emphasized that they do not attempt to provide predictions for the surface Li abundances but to explore relative abundance variations. Similar to their views, our aim is not to use the low-coefficient thermohaline mixing to predict the Li abundances of giants, but only to discuss why does it produces the same result as the $\rm Conv.\,\&\,Ledoux$ models. We introduce the thermohaline mixing and the semiconvection because their implementation in the MESA code requires switching the Schwarzschild boundary to the Ledoux boundary.  Within the confines of the MESA code, they both have `\texttt{use\_Ledoux\_criterion = .true.}'. While the thermohaline mixing is driven by the mean molecular weight variation that is induced by the $\rm ^3He(^3He, 2p)^4He$ reaction, and it will form a mixing zone (see the $\rm section\,2$ of \cite{2007A&A...467L..15C}). Therefore, we will take the thermohaline mixing zone into account.

On the other hand, the low-coefficient thermohaline mixing model shows a nearly constant Li abundance beyond the RGB bump (see $\rm Fig.\,\ref{fig:structure}$), a result that differs from known observations \citep[e.g.,][]{2016ApJ...819..135K, 2021A&A...651A..84M, 2021A&A...655A..23M}. Therefore, we briefly discuss the difference in coefficient choice here, and $\rm Fig.\,\ref{fig:kipp}$ shows the Kippenhahn diagram considering the thermohaline mixing. The critical role of time step and spatial resolution in accurately modelling the mixing process of red giants has been much discussed in the recent literature, particularly with respect to the thermohaline mixing \citep[e.g.,][]{2015MNRAS.446.2673L, 2022ApJ...941..164F}. We show the test results in $\rm Figs.\,\ref{fig:appendix4}\ and\ \ref{fig:appendix5}$. It can be seen that increasing the time step does not significantly affect our results, and the thermohaline mixing zone still does not touch the convective envelope. However, increasing the spatial resolution can narrow the thermohaline mixing zone. It can be seen that for the low-coefficient thermohaline mixing model, increasing the time step and the spatial resolution affect the Li abundance on the order of $\rm 0.001\,dex$. However, the central focus of this paper is not on thermohaline mixing, but the modelling accuracy of thermohaline mixing is still very interesting and we will consider it further in the future.

For $\rm Fig.\,\ref{fig:kipp}$, we continue to use the MESA default setup. The low-coefficient thermohaline mixing does excite the mixing zone, but it does not reach the convective zone (see $\rm Fig.\,\ref{fig:kipp}\,(1)$), which is similar to the results of \citet{2011A&A...533A.139W}. In this case, the surface Li is not transferred into the interior of the star by the thermohaline mixing. Thus, the evolution behaviour of $A(\rm Li)$ is similar to the $\rm Conv.\,\&\,Ledoux$ models and  essentially unchanged beyond the RGB bump. The opposite is true for the high-coefficient model (see $\rm Fig.\,\ref{fig:kipp}\,(2)$), which can therefore cause the Li depletion after the RGB bump \citep[e.g.,][]{2015MNRAS.446.2673L}.

\subsection{The 1\% Puzzle}\label{sec:4.3}

The evolution of super Li-rich progenitors is a possible formation channel for Li-rich giants.
Inspecting the $\rm Conv.\,\&\,Ledoux$ models, the $1\%$ puzzle\footnote{We define as the $1\%$ puzzle that the currently unexplained observed fact of anomalous Li abundance for about $1\%$ of giants.} of giants Li abundance must be closely related to the Li abundance of their progenitors.
$\rm Fig.\,\ref{fig:mt}$ shows that most stars can naturally form the Li-rich giants through convective mixing, convective boundary defined by the Ledoux criterion, when initial Li abundance is around the meteoritic value. The reality, however, is that only about $1\%$ of giants are rich in Li, which means that more Li depletion is the order of the day for stellar objects. The Li abundance observations of giant progenitors also confirm the existence of Li depletion \citep[e.g.,][]{1993AJ....106.1080S, 1993ApJ...415..150T, 2008A&A...489..677P, 2010A&A...519A..87B, 2010A&A...515A..93T, 2016A&A...587A.100C, 2020NatAs...4.1059K, 2023MNRAS.522.3217M}. Except for a few stars that deplete a little Li before the MS turnoff, most cases are below the meteoritic value at the MS turnoff, and end up their giants life with a Li abundance of less than $\rm 1.5\,dex$. 

At present, rotation may be a good explanation for the MS Li depletion \citep{2010A&A...522A..10C, 2017AJ....153..128C, 2019AJ....158..163D}. The intense MS Li depletion is brought about by higher equatorial velocities \citep{2010A&A...522A..10C, 2020A&A...633A..34C}. Their low-velocity models predict slight Li depletion behavior during the MS, indicating that high Li abundance still exists in the MS turnoff even when considering possible the MS Li depletion processes, which can provide support for our models.

The physical process involved in our models is incomprehensive and limited. In stellar objects, however, there are many fascinating stories on the Li of the giant progenitors, including the  Li dip \citep{1986ApJ...302L..49B}, the Li plateau \citep{1982A&A...115..357S}, and the solar Li problem (e.g.,  \citet{1997ARA&A..35..557P} vs. \citet{1997AJ....113.1871K}). Therefore, regarding the setting of $A(\rm Li)_{\rm ZAMS}$, the real situation must be considered, that is, more often than not, it is less than the meteoritic value.  Taking into account some extra mixing (such as rotation, diffusion, overshooting, etc.), the probability of a star starting out with an initial value of $\rm 3.3\,dex$ forming a Li-rich giants will drop sharply. Perhaps the extra mixing is precisely the key to the $1\%$ puzzle.

Our models predict that the giant progenitors with higher Li abundances could naturally evolve into Li-rich giants without extra mixing. Thus, to put it further, we need to be combined with extra Li depletion to fit the $1\%$ Li-rich giants of observed fact. This requires the support of a large number of the MS Li abundance observed samples. In addition, the sample also helps us to understand its correlation with the formation of the Li abundance distribution of the giants.

\subsection{Li Enrichment}
\label{sec:4.4}
Li enrichment process may exist in giant progenitors. The extremely high Li of the giant progenitors might come from the accretion of circumstellar matter \citep{2005MNRAS.363L..81A, 2022ApJ...929L..14Y} and/or from some binary effects \citep{2011ApJ...738L..29K}. This further adds to the mystery of the Li abundances of the progenitors and improves the existence possibility of super Li-rich giant progenitors. 
Our models are completely unexplainable for super Li-rich giants, though they are a tiny minority of Li-rich giants \citep{2019ApJS..245...33G, 2022ApJ...931..136Z}. For those with a Li abundance of giants with more than $\rm 3.3\,dex$, inhibiting Li depletion is not enough. What is needed is some mixing processes that transports the inner beryllium to the stellar surface and/or Li contamination from the outer celestial objects.

On the other hand, there may also be a process of Li enrichment in the giant branch. \citet{2000A&A...359..563C} had analyzed the evolution behavior of Li on Hertzsprung-Russell diagram in detail, and two stages have caught our attention, one is the first dredge-up, the other is the RGB bump. For low-mass stars, their statistical results indicate the existence of Li-rich giants in the two phases described above. However, the sample of Li-rich stars during the first dredge-up is too small to determine whether there is Li enrichment process. From $\rm Fig.\,\ref{fig:boundary}\,(1)$, we can find that the Li abundance of stars during the whole first dredge-up is greater than $\rm 1.5\,dex$. 
The reason for the difference with \citet{2000A&A...359..563C} is that it has almost no sample stars, while the explore for the open cluster NGC 2506 shows many Li-rich subgiants \citep{2018AJ....155..138A}. The existence of a significantly Li-rich giant near the RGB bump \citep{2000A&A...359..563C}, with the Li abundance in the range of $\rm 2.5-3.0\,dex$, is obviously not explained by our convection models. Because for stars with masses in the $1.0-1.2\,M_{\odot}$ range, in the case of relying only on the dilution during the first dredge-up, the Li abundance can reach this range only if its $A(\rm Li)_{ZAMS}$ is greater than $\rm 4.5\,dex$, which can be inferred that there is a very high probability of Li enrichment processes taking place for this star.
For now, the research on the mechanism of Li enrichment is still in full swing.

\section{Conclusion} 
\label{sec:Conclusion}

In this paper, we investigate the impact of convective mixing on Li in giants. Some of the main conclusions are as follows:

1. From testing the convection model for the location of the convective boundary, we find that when the boundary is determined by the Ledoux criterion, the convection model can significantly damper the Li depletion during the first dredge-up, especially for lower masses and higher metallicities stars.

2. By examining the influence of convective mixing effects on Li abundance, we propose a possible channel for the formation of Li-rich giants: the evolution of their super Li-rich progenitors.

3. When the Li abundance input is more than the  meteoritic value, most low-mass stars will naturally form the Li-rich giants in our convection models.

4. When extra Li depletion processes are taken into account, the progenitor Li abundance usually falls below that of a meteorite at the MS turnoff, and they will naturally evolve into a normal giant.

\section*{Acknowledgements}
We thank the anonymous reviewers for their help in improving the manuscript.
Our research is supported by National Natural Science Foundation of China (grant Nos: 11973079, 12133011, 11833006, 11973052, 12022304, 12090040, 12090044, 12173080, 12273104, and 12373036) and the National Key R\&D Program of China Nos. 2019YFA0405502, 2021YFA1600400, 2021YFA1600402. This work is supported by a grant from National Basic Science Center Project of China (grant No. 12288102). X.-F. L. acknowledges supports from the Natural Science Foundation of Yunnan Province (No. 202201AT070158) and the Yunnan Fundamental Research Projects (Grant No. 202401AS070045). H.-L.Y. acknowledges the supports from Youth Innovation Promotion Association of CAS and the NAOC Nebula Talents Program. J.-H. Z. acknowledges support from NSFC grant No.12103063 and from China Postdoctoral Science Foundation funded project (grant No. 2020M680672).

%%%%%%%%%%%%%%%%%%%%%%%%%%%%%%%%%%%%%%%%%%%%%%%%%%
\section*{Data Availability}

The data involved in this paper are all the results of modeling, consulting the corresponding author to obtain relevant cases of the MESA code.



\bibliographystyle{mnras}
\bibliography{Li_references} % if your bibtex file is called example.bib
%\begin{thebibliography}{99}
%\end{thebibliography}

%%%%%%%%%%%%%%%%%%%%%%%%%%%%%%%%%%%%%%%%%%%%%%%%%%

%%%%%%%%%%%%%%%%% APPENDICES %%%%%%%%%%%%%%%%%%%%%

\appendix

\section{Additional Kippenhahn Diagram}

\begin{figure*}
	\centering
	\includegraphics[scale=0.4]{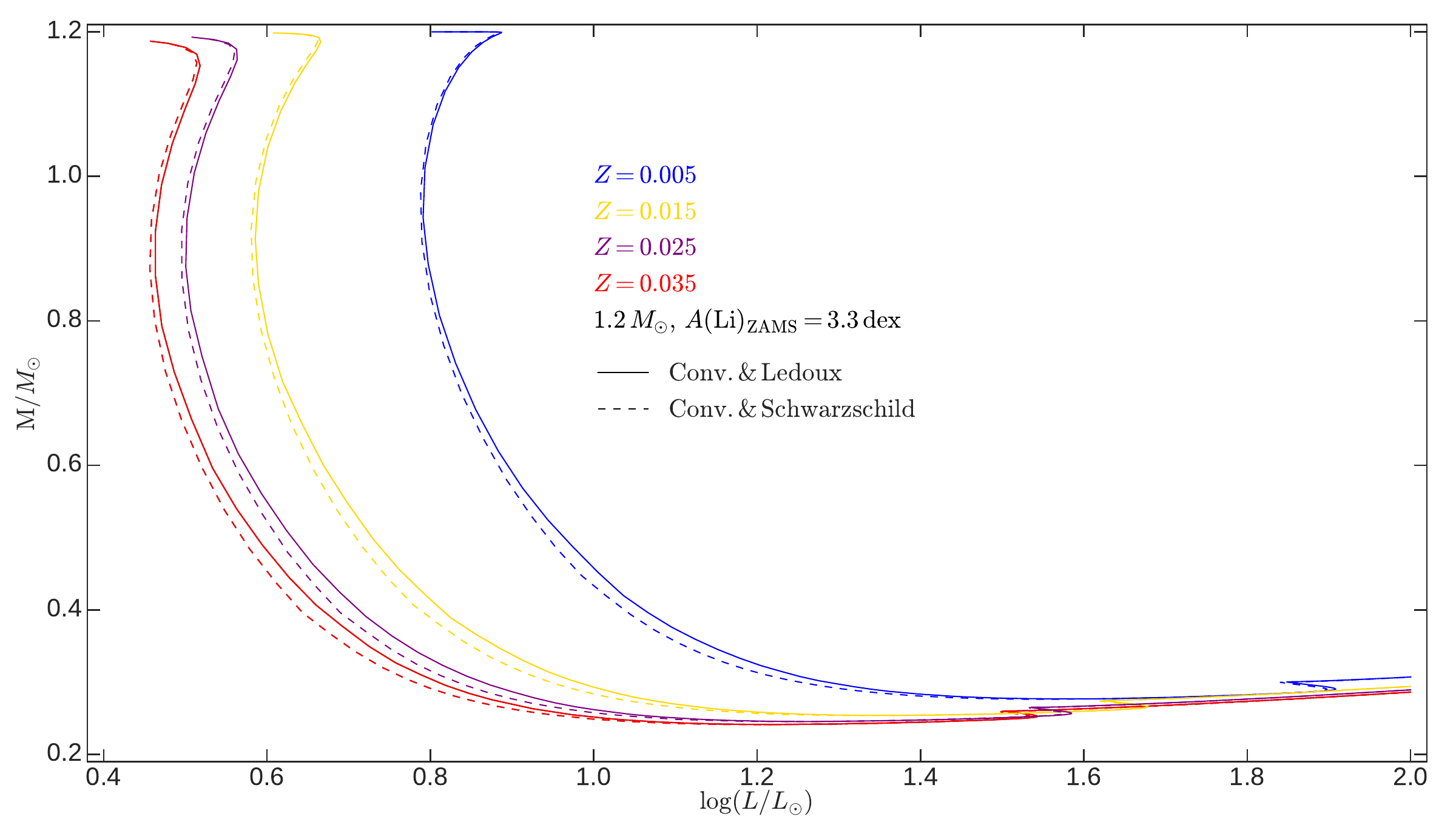}
 \includegraphics[scale=0.4]{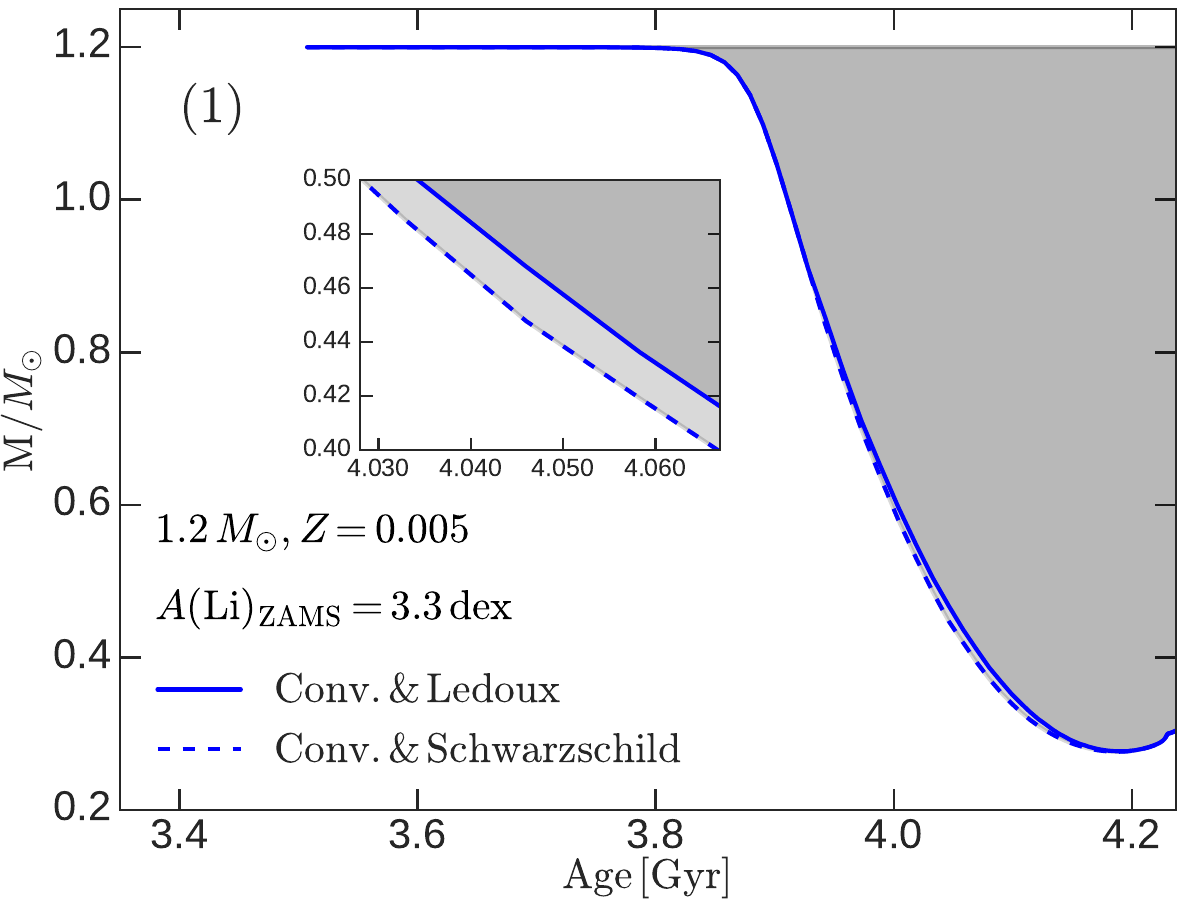}
 \includegraphics[scale=0.4]{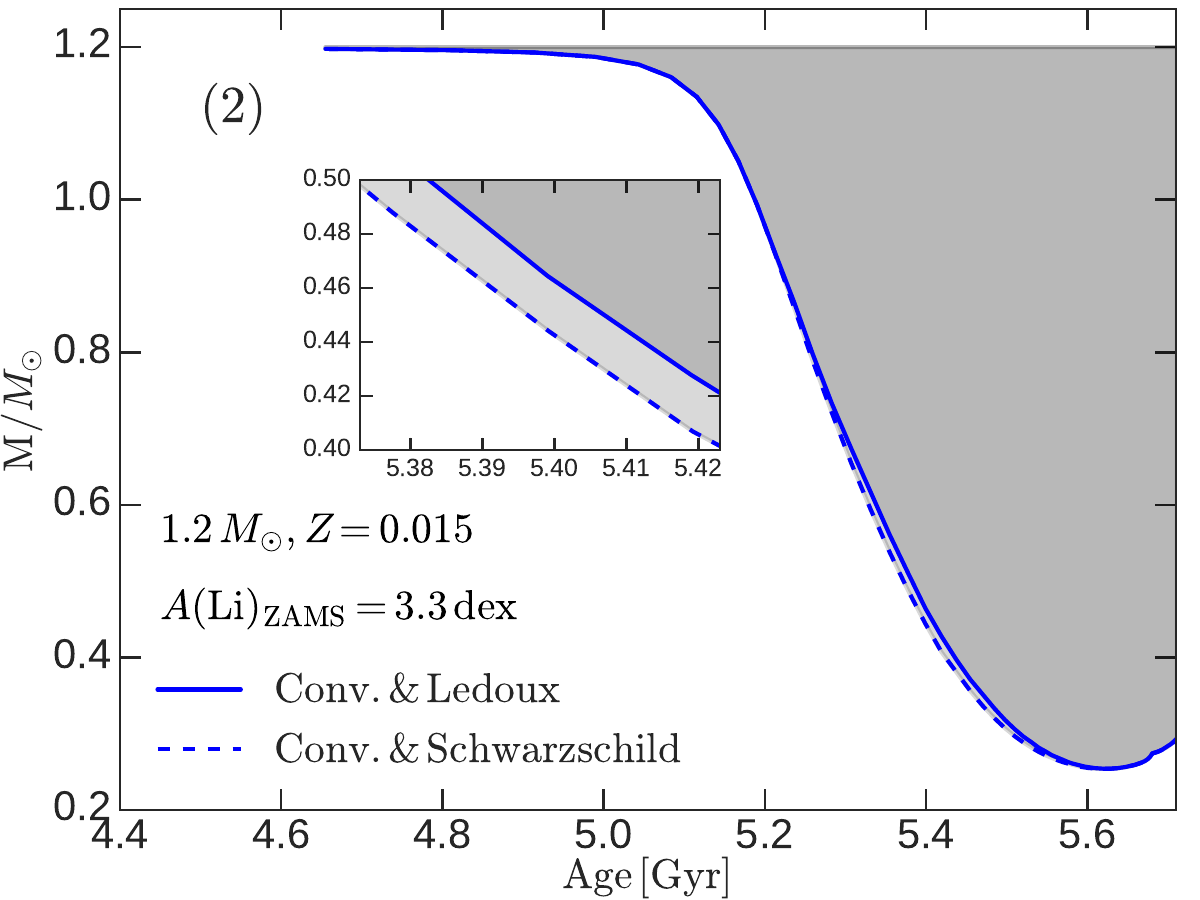}
 \includegraphics[scale=0.4]{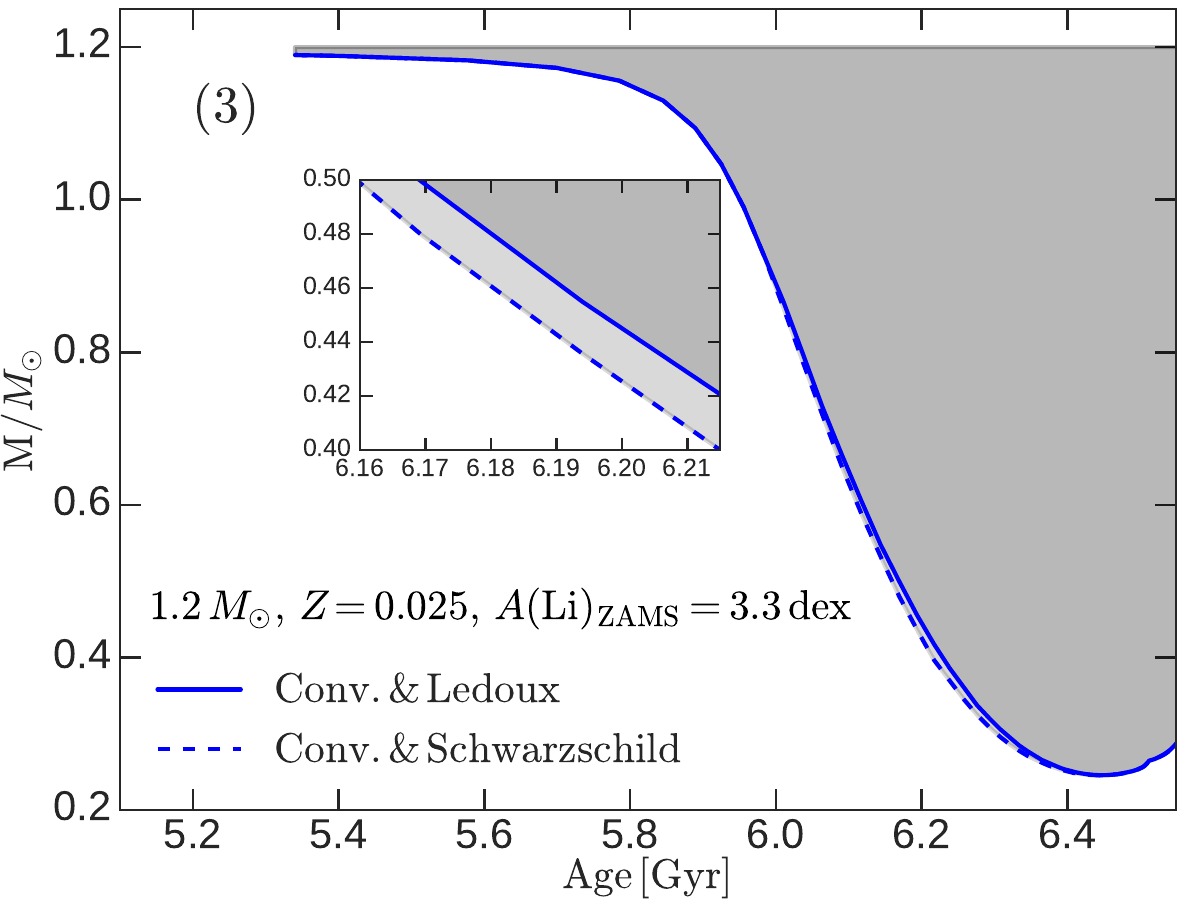}
 \includegraphics[scale=0.4]{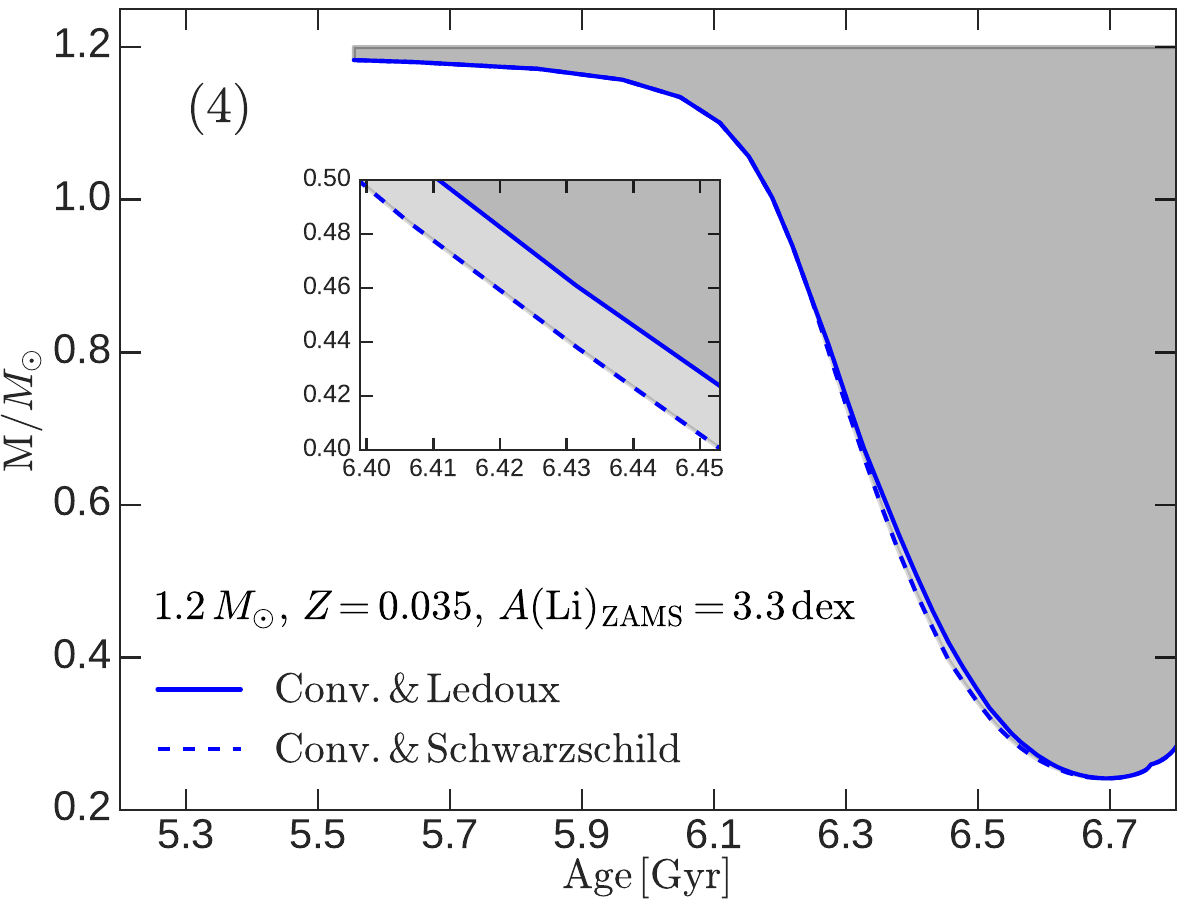}
	\caption{Structural information (similar to $\rm Fig.\,\ref{fig:boundary}\, (3)$) and the Kippenhahn diagram. We show information on the evolution of convective boundaries corresponding to the $\rm Conv.\, \&\, Schwarzschild$ and the $\rm Conv.\, \&\, Ledoux$ models in $\rm Fig.\,\ref{fig:structure}$ of the text. The small figures in the Kippenhahn diagrams are a partial enlargement. Shadows are the convective envelope.} \label{fig:appendix1}
\end{figure*}

\begin{figure*}
	\centering
	\includegraphics[scale=0.36]{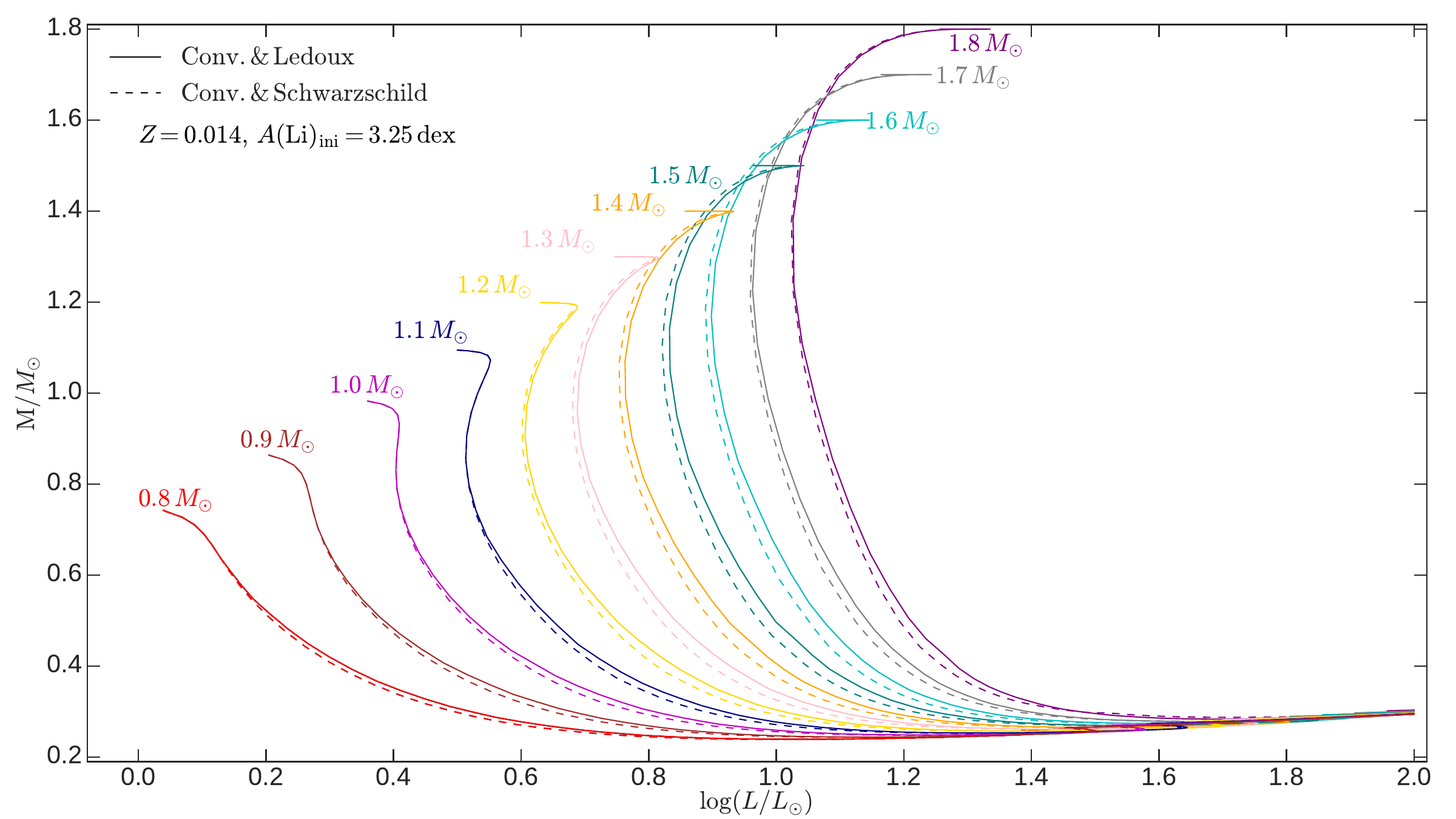}
 \includegraphics[scale=0.24]{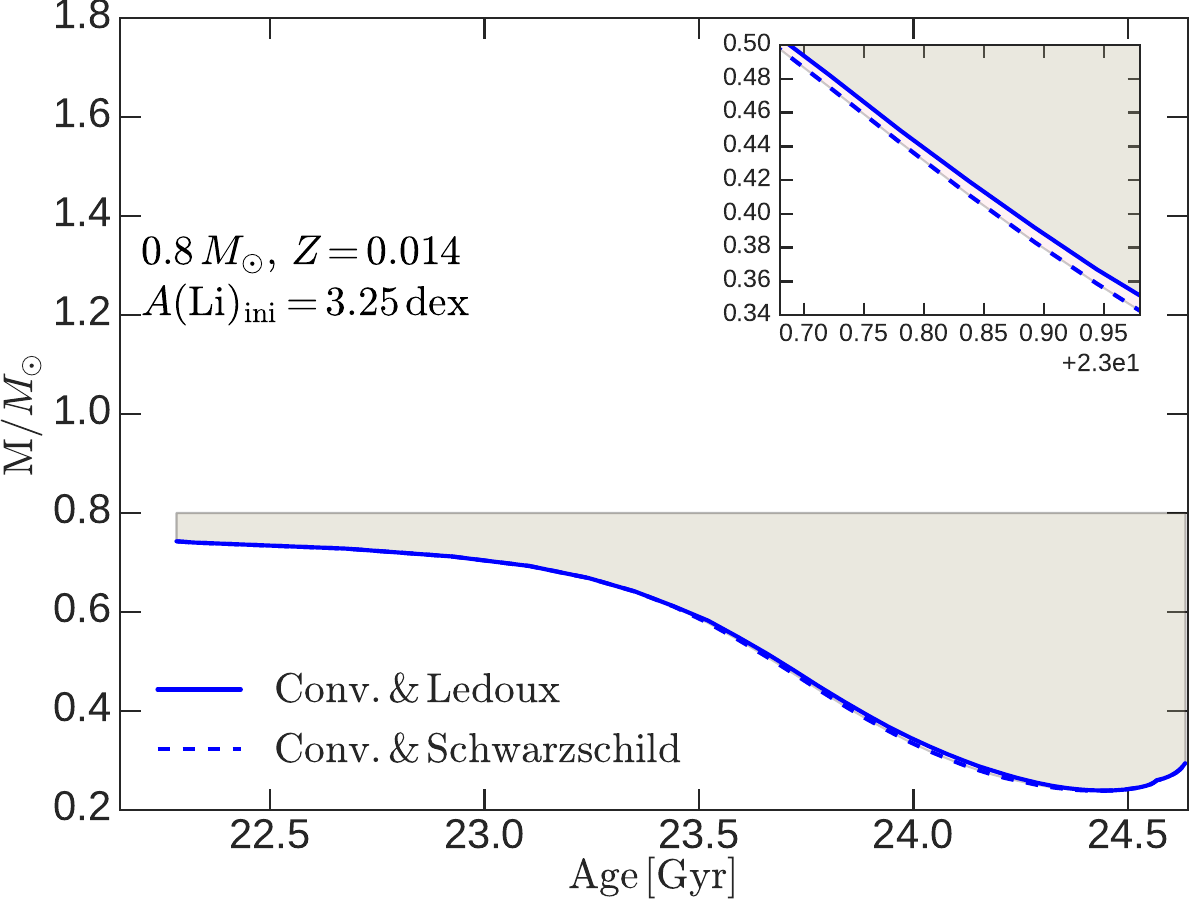}
 \includegraphics[scale=0.24]{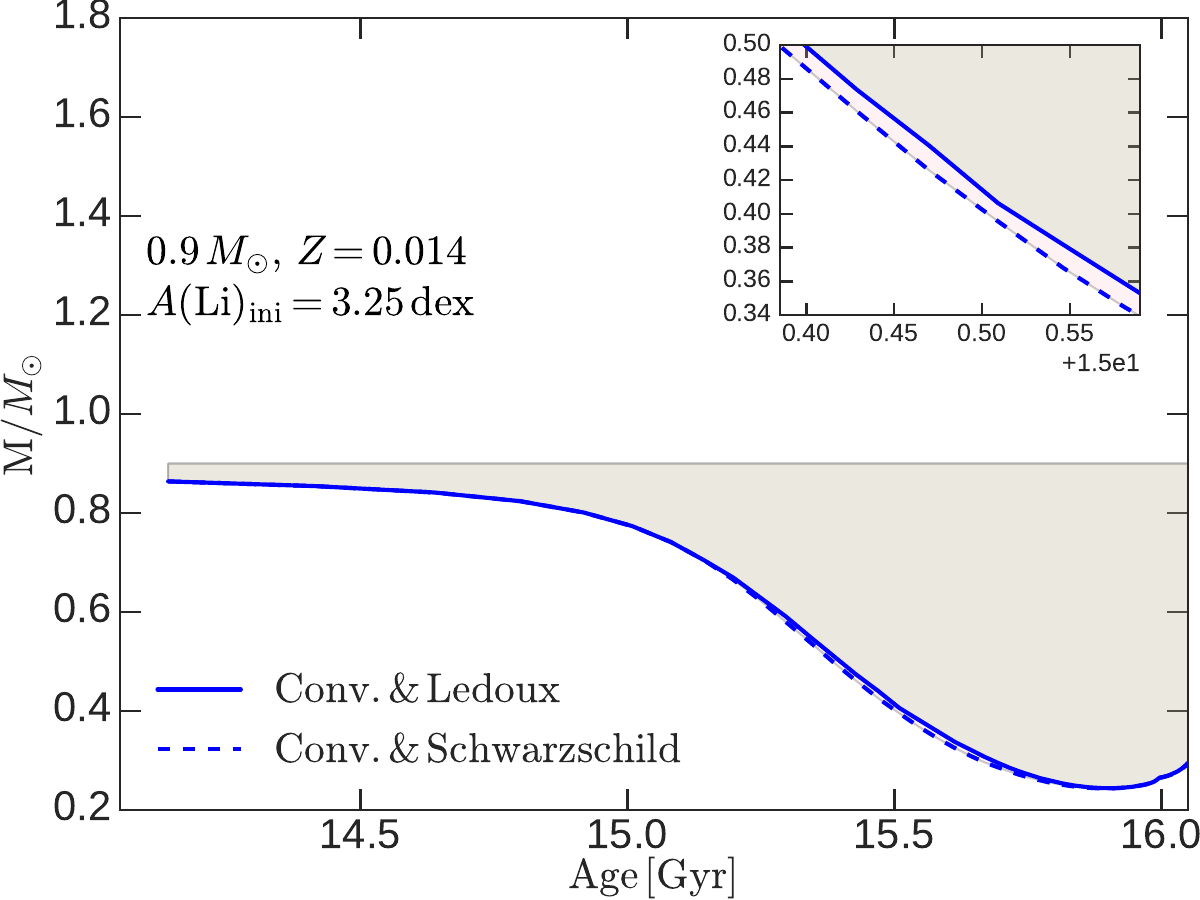}
 \includegraphics[scale=0.24]{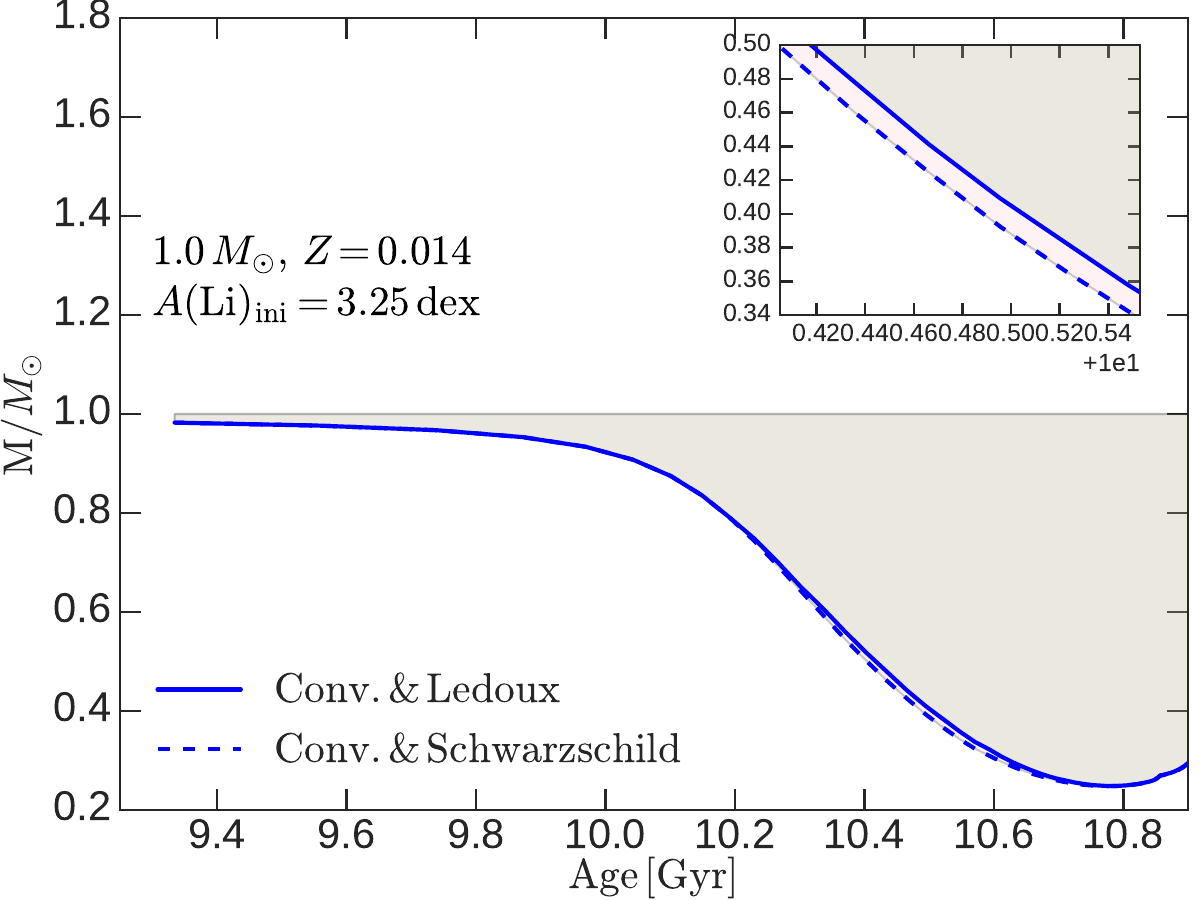}
 \includegraphics[scale=0.24]{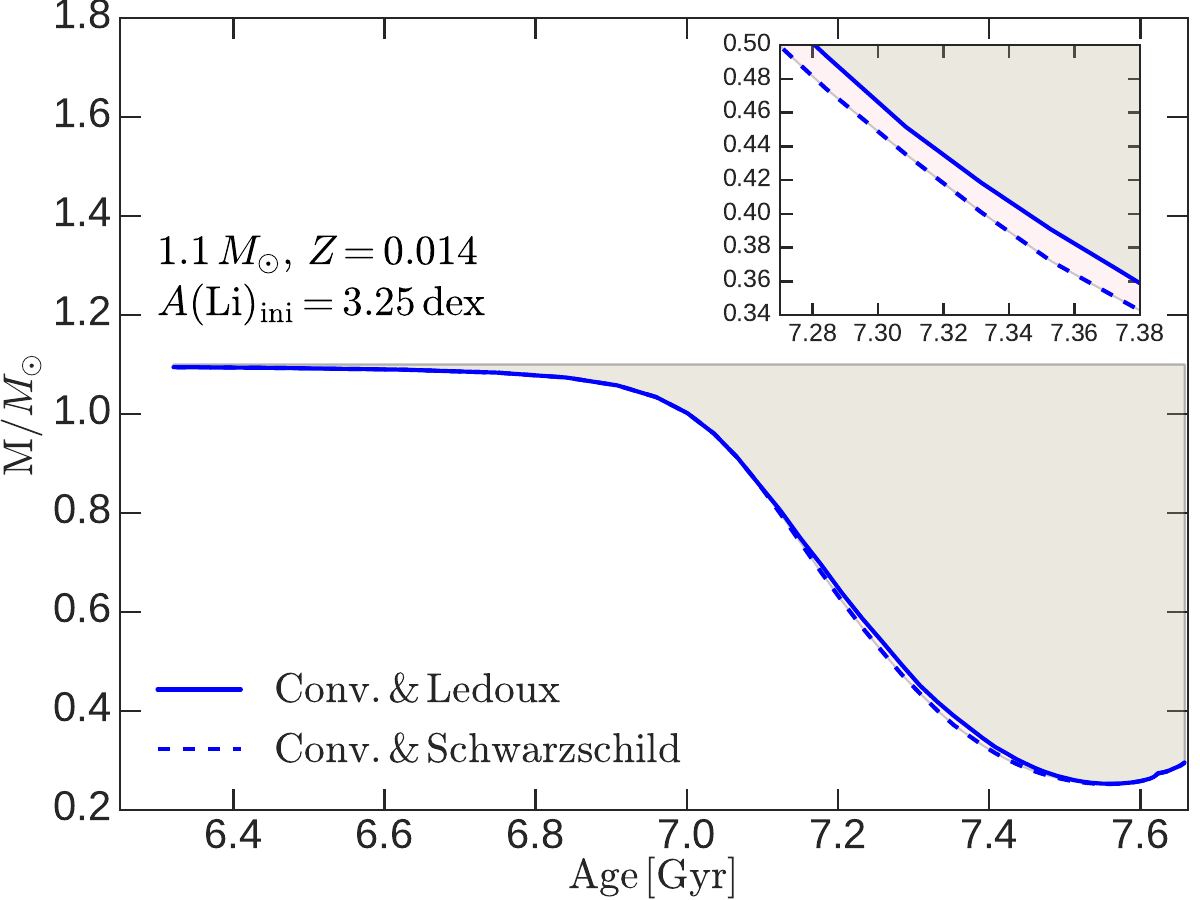}
 \includegraphics[scale=0.24]{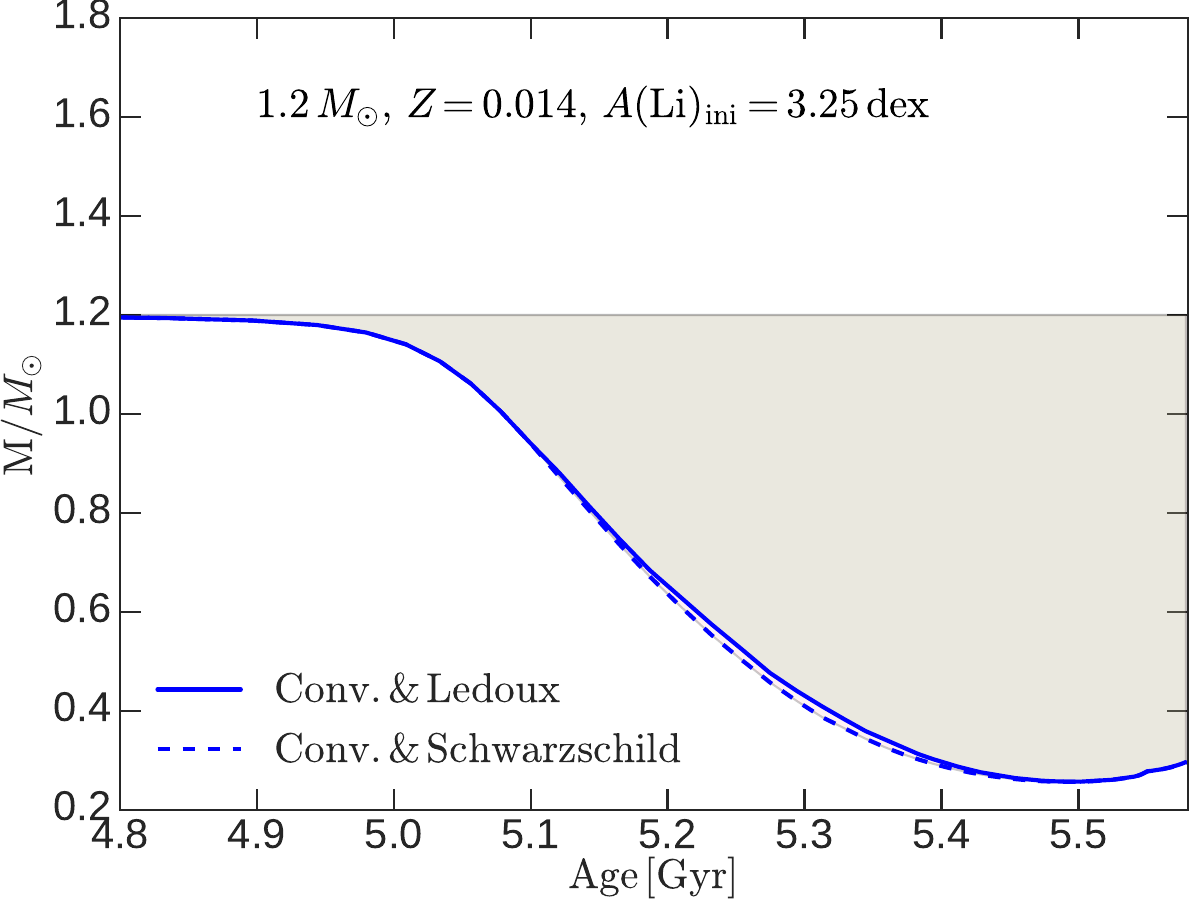}
 \includegraphics[scale=0.24]{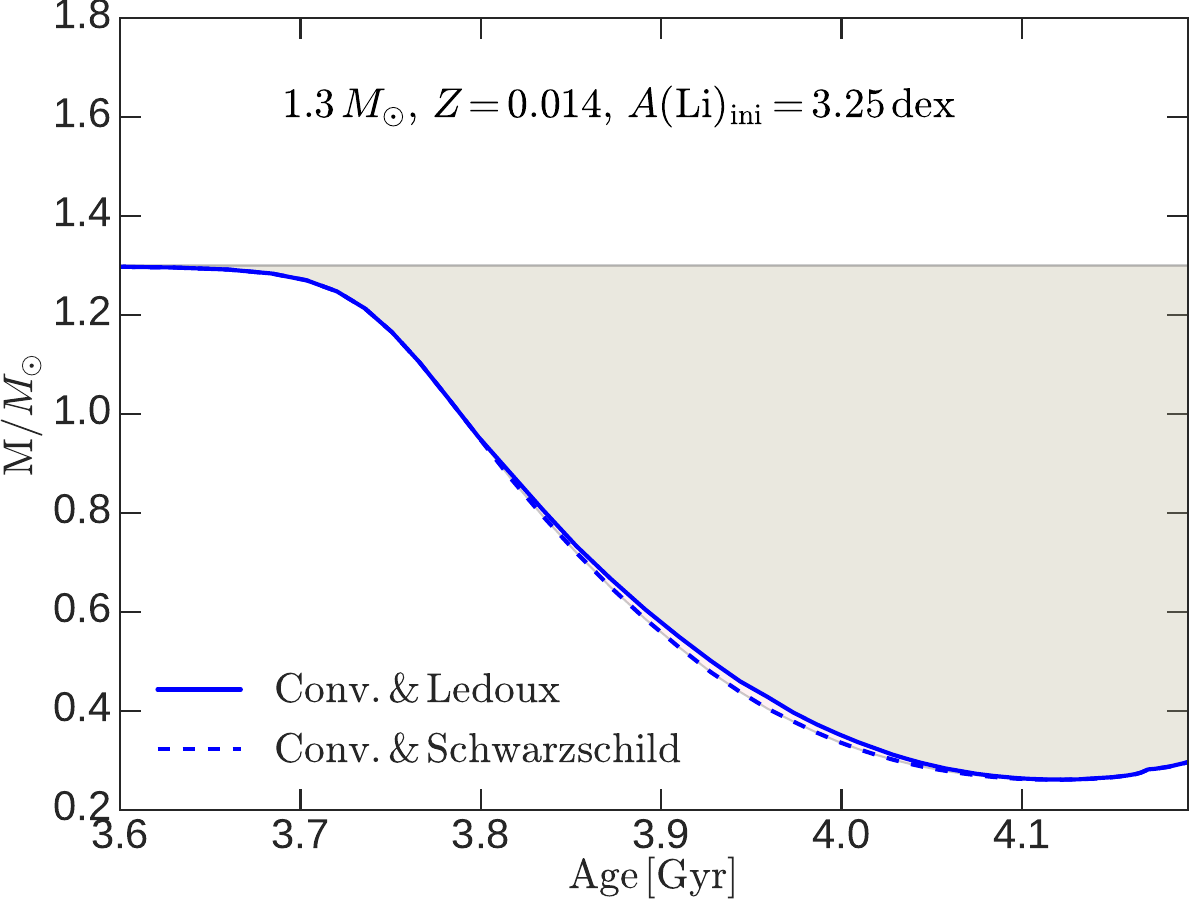}
 \includegraphics[scale=0.24]{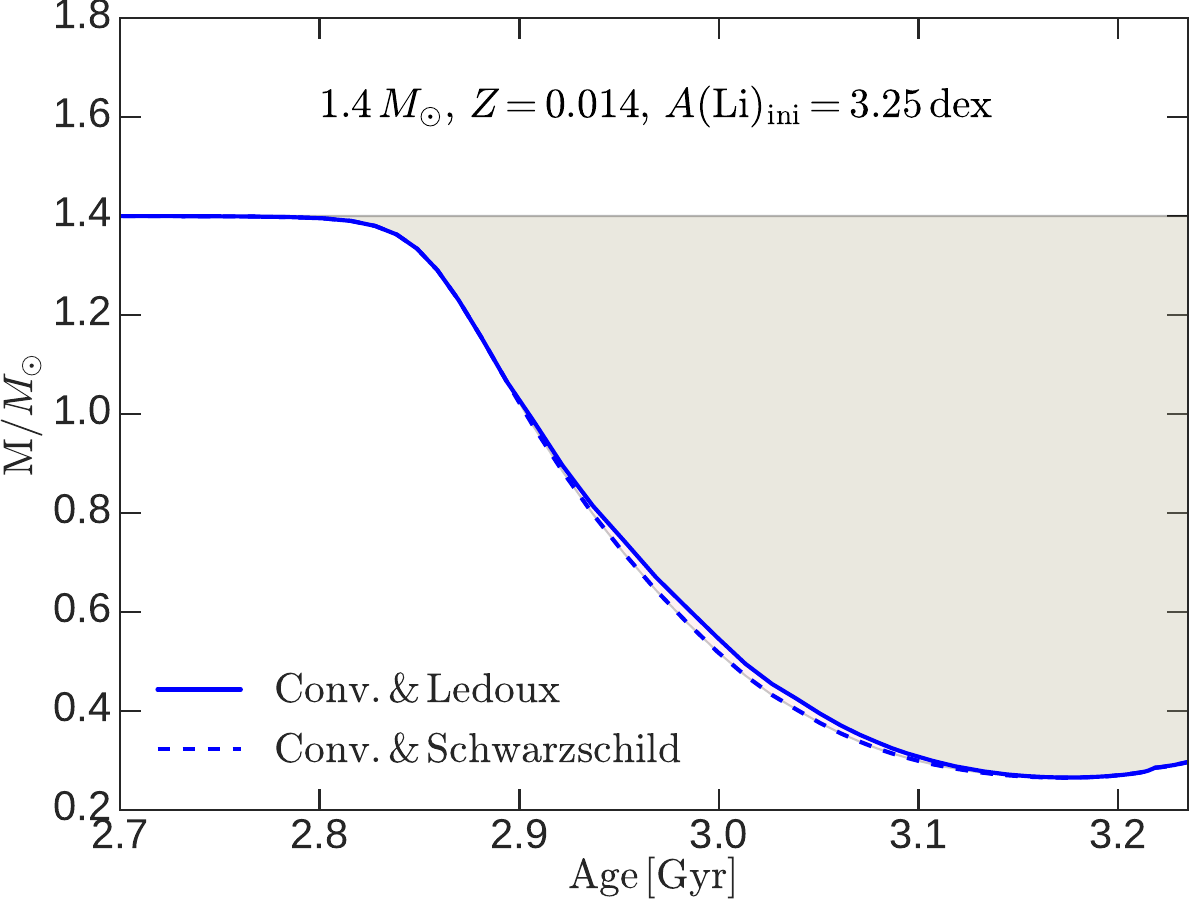}
 \includegraphics[scale=0.24]{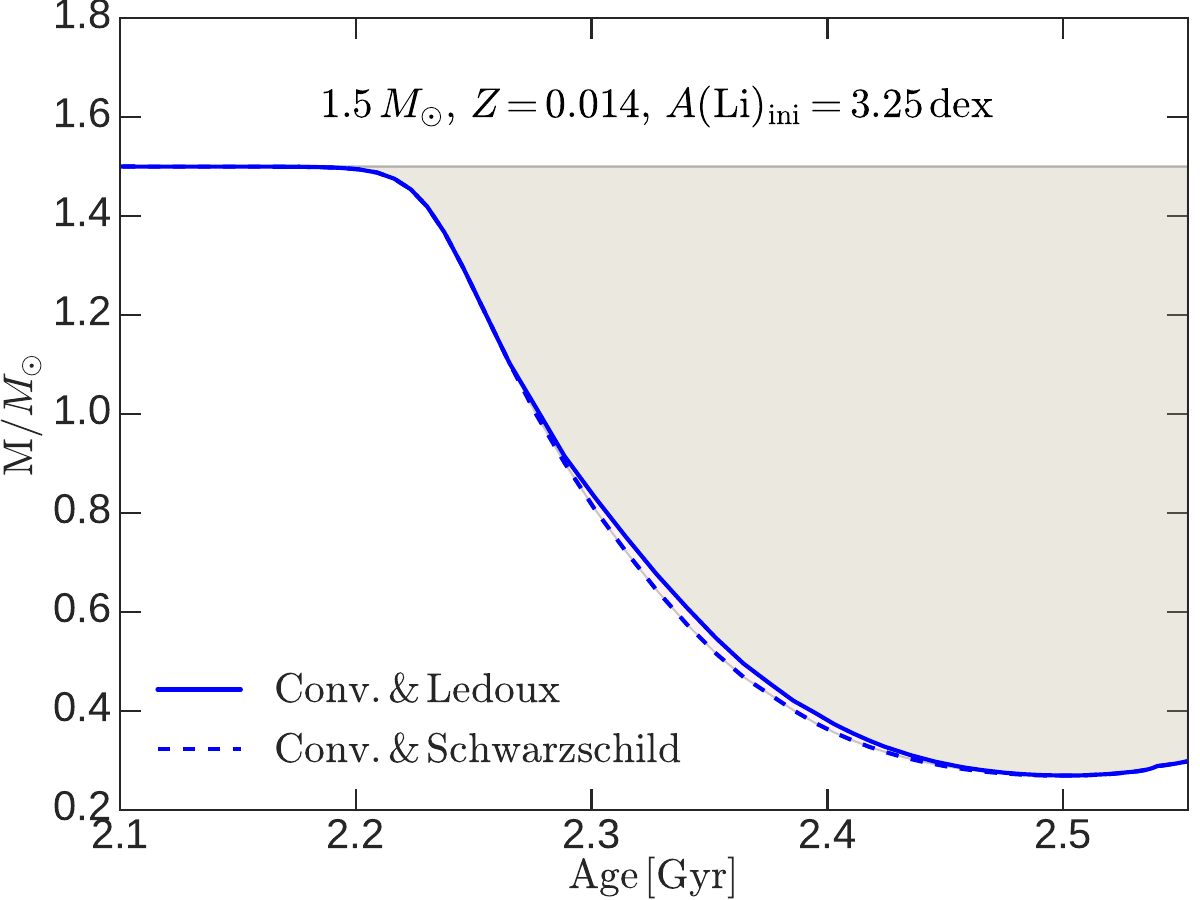}
 \includegraphics[scale=0.24]{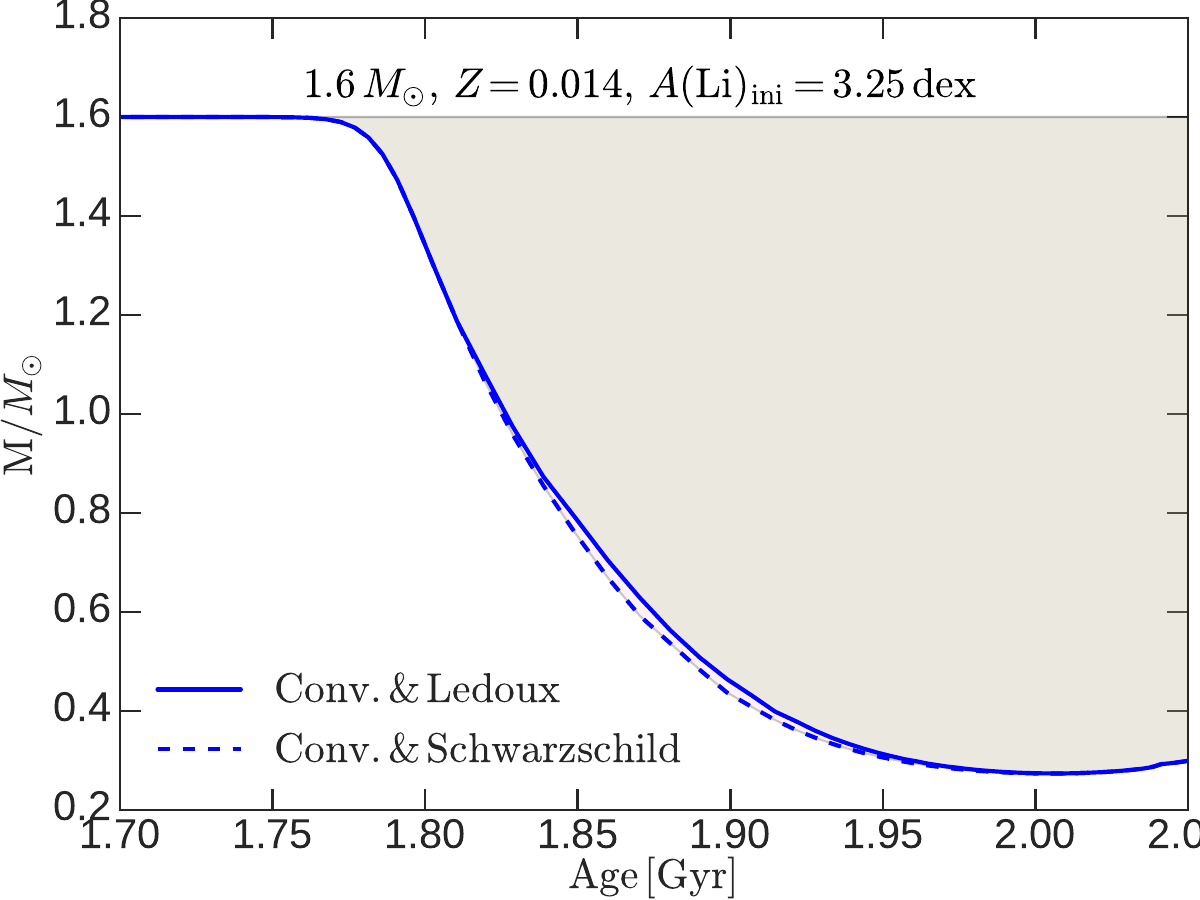}
 \includegraphics[scale=0.24]{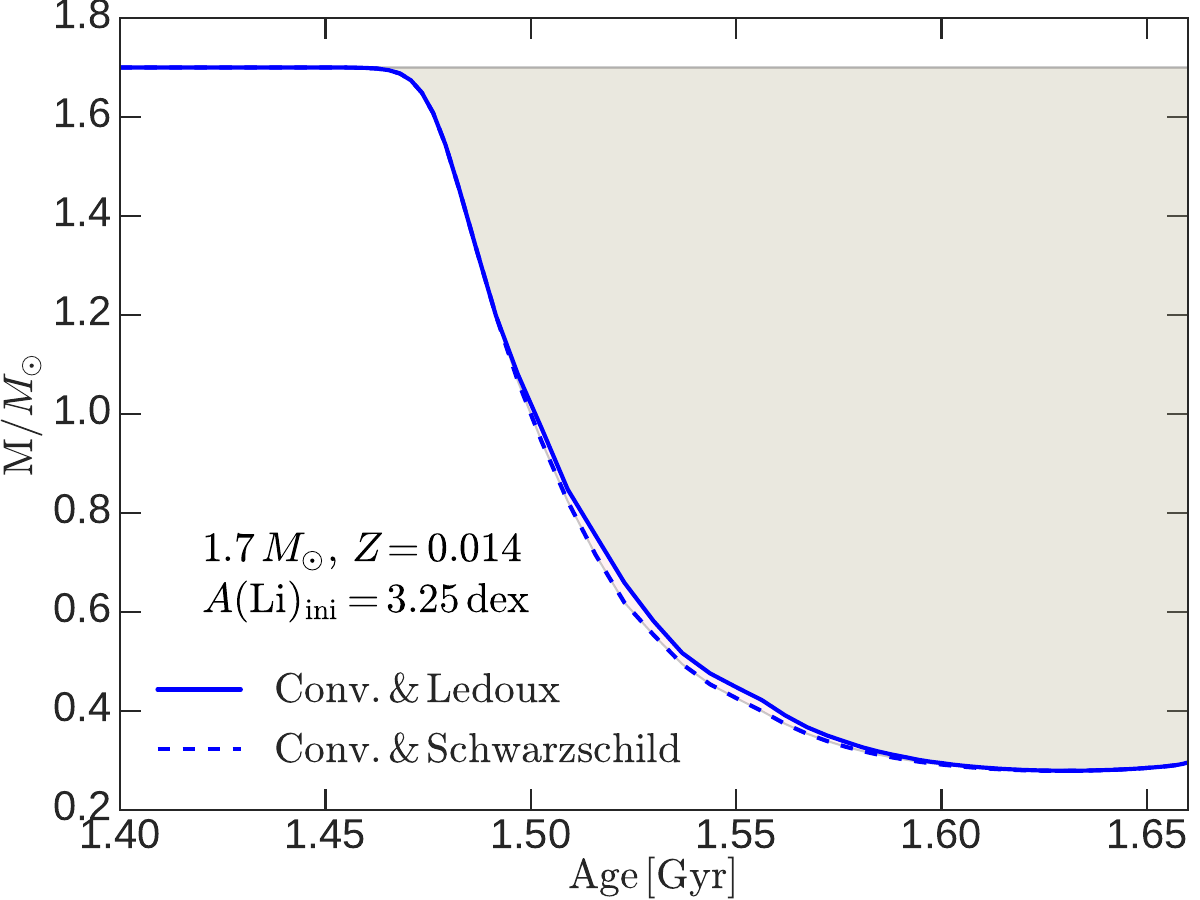}
 \includegraphics[scale=0.24]{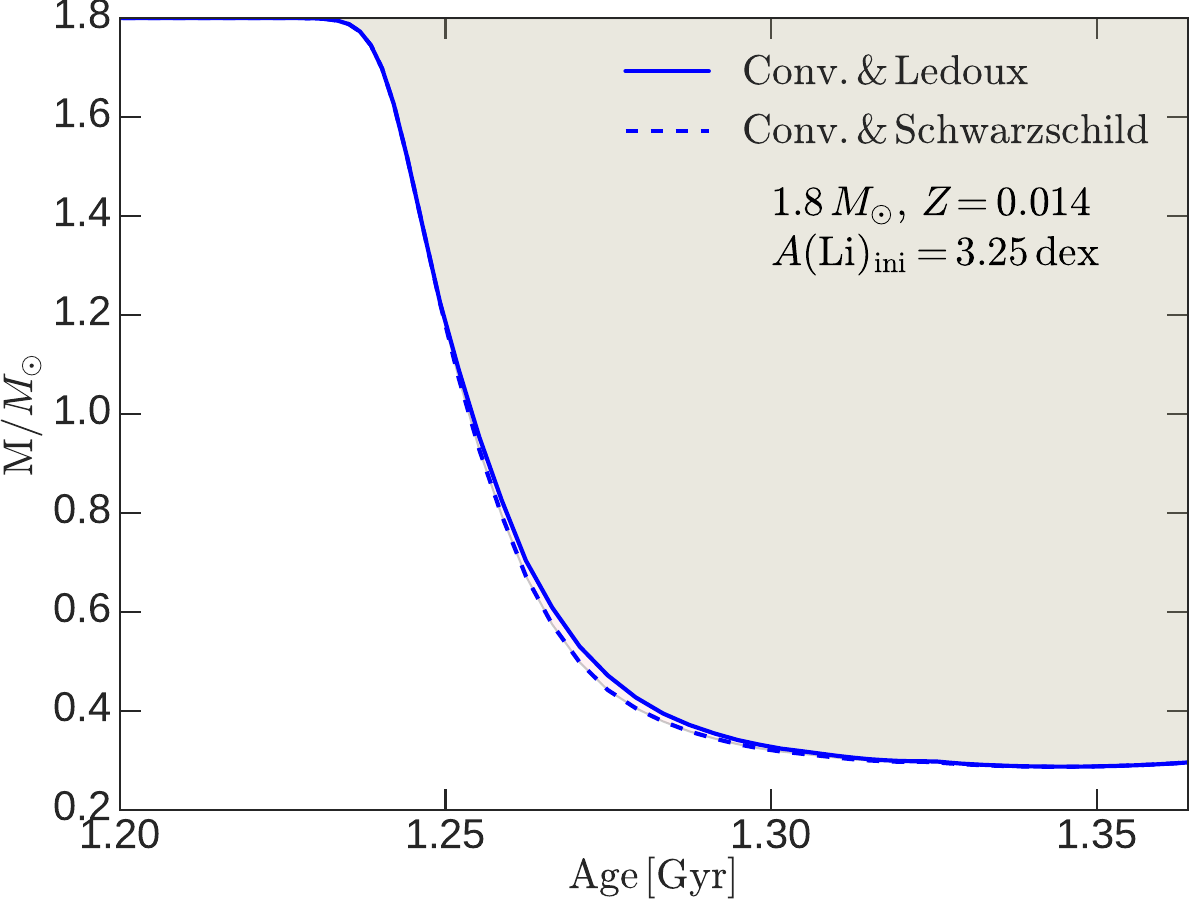}
	\caption{Structural information (similar to $\rm Fig.\,\ref{fig:boundary}\, (3)$) and the Kippenhahn diagram. We show information on the evolution of convective boundaries corresponding to the $\rm Conv.\, \&\, Schwarzschild$ and the $\rm Conv.\, \&\, Ledoux$ models in $\rm Fig.\,\ref{fig:convmass}\, (1)$ of the text. The small figures in the Kippenhahn diagrams are a partial enlargement. Shadows are the convective envelope.} \label{fig:appendix3}
\end{figure*}

\begin{figure*}
	\centering
	\includegraphics[scale=0.36]{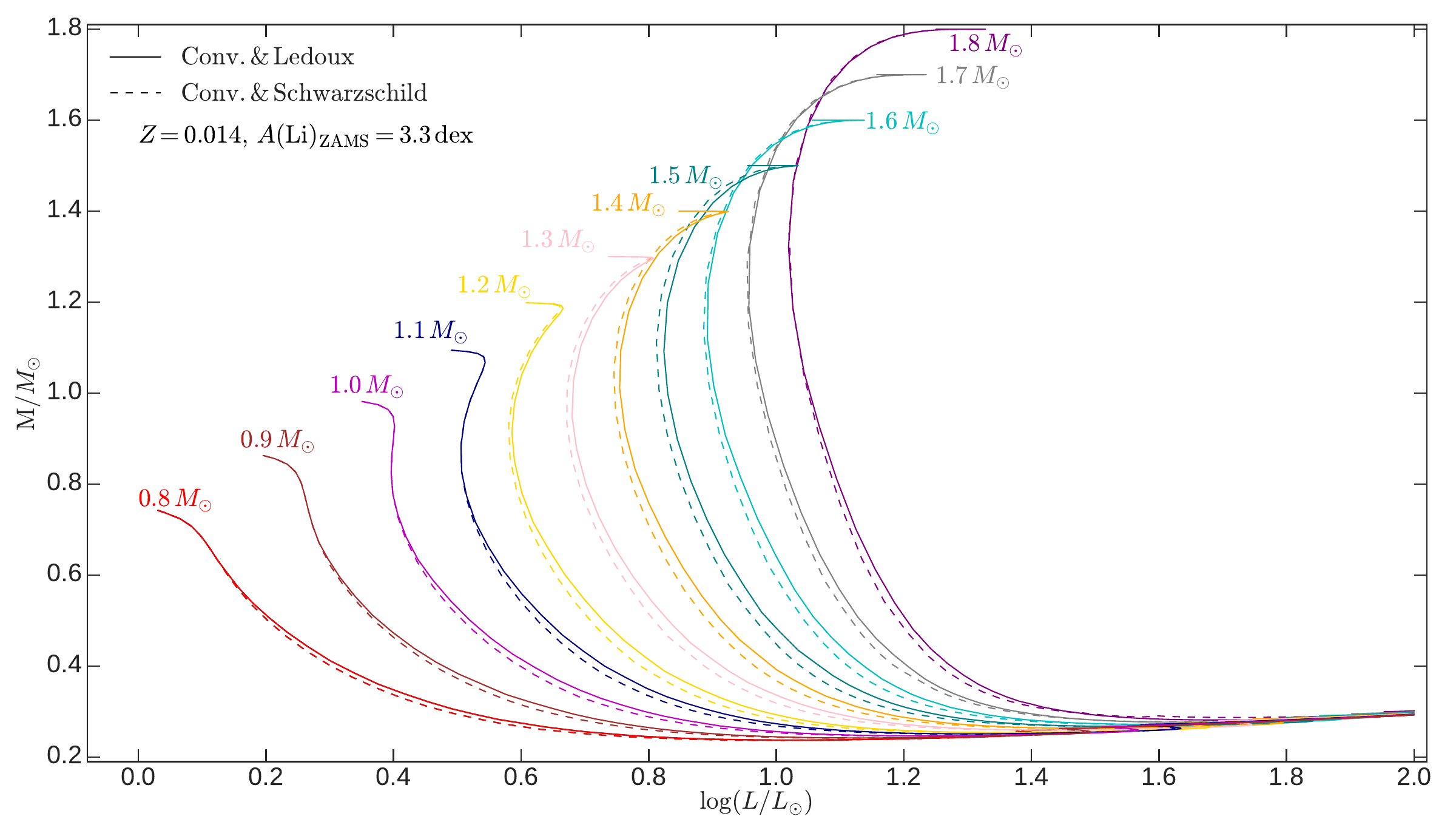}
 \includegraphics[scale=0.24]{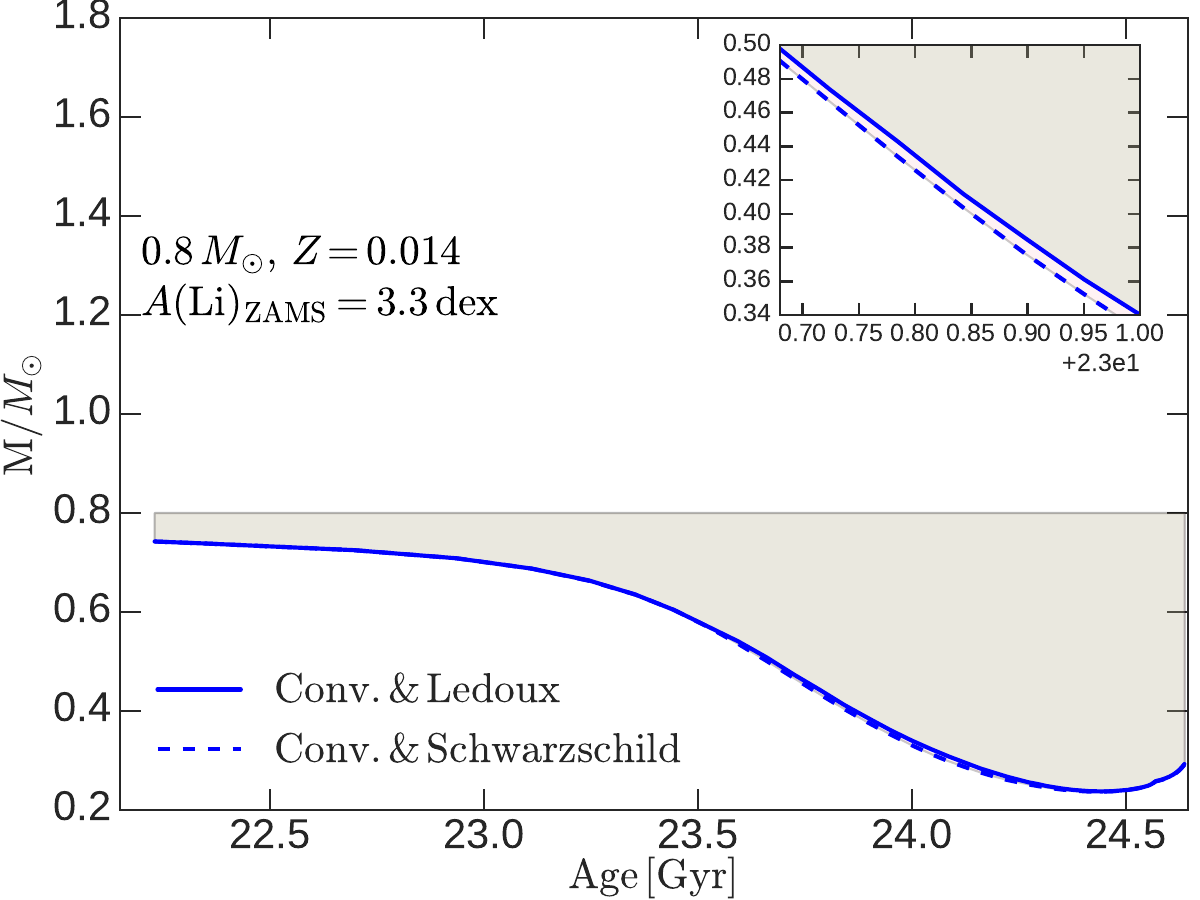}
 \includegraphics[scale=0.24]{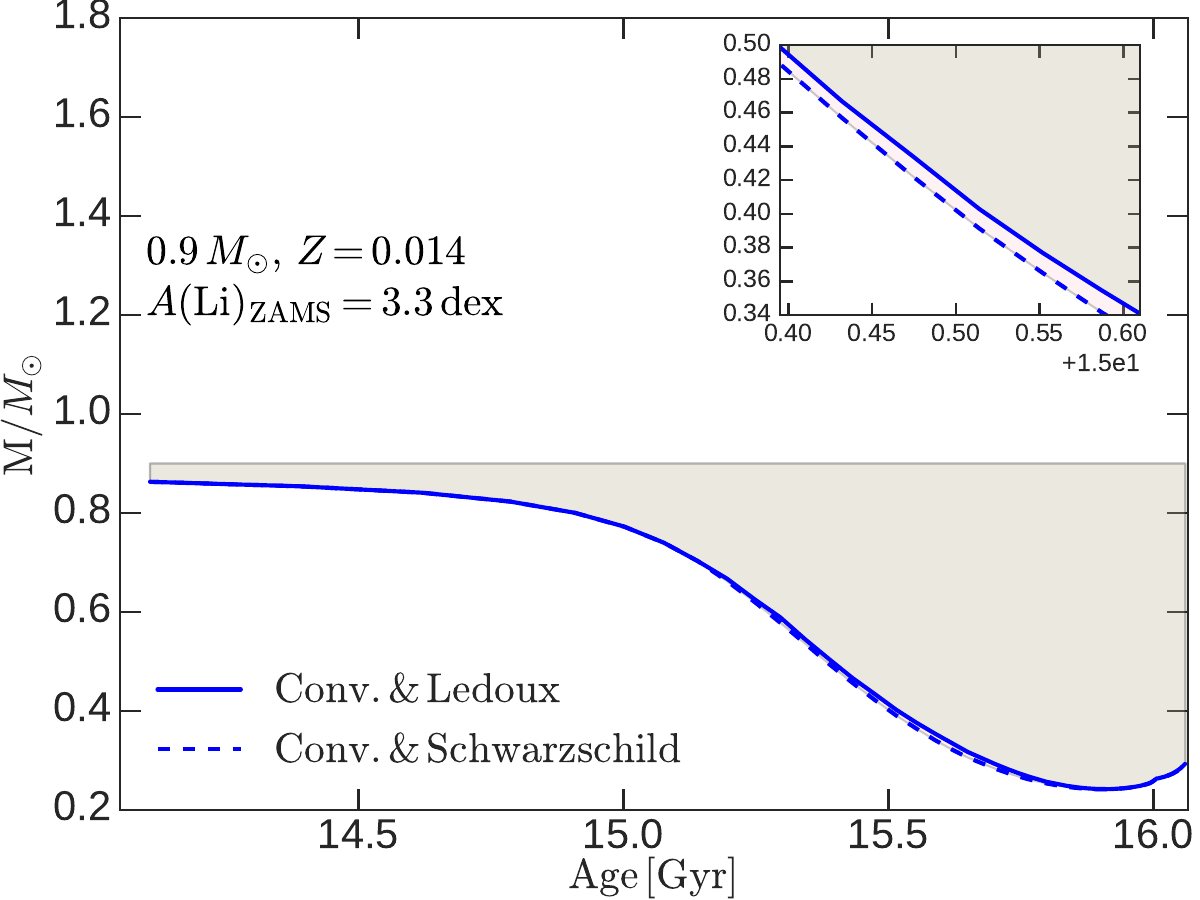}
 \includegraphics[scale=0.24]{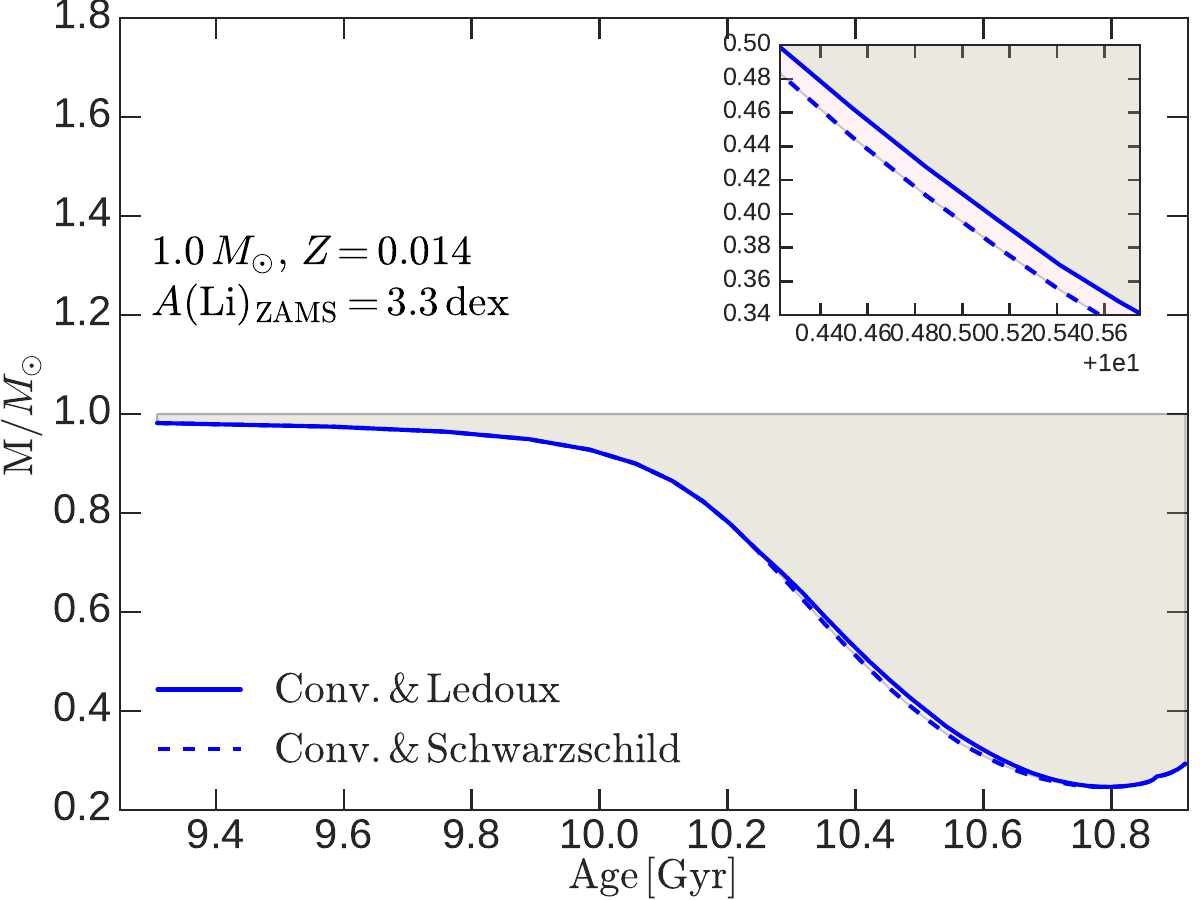}
 \includegraphics[scale=0.24]{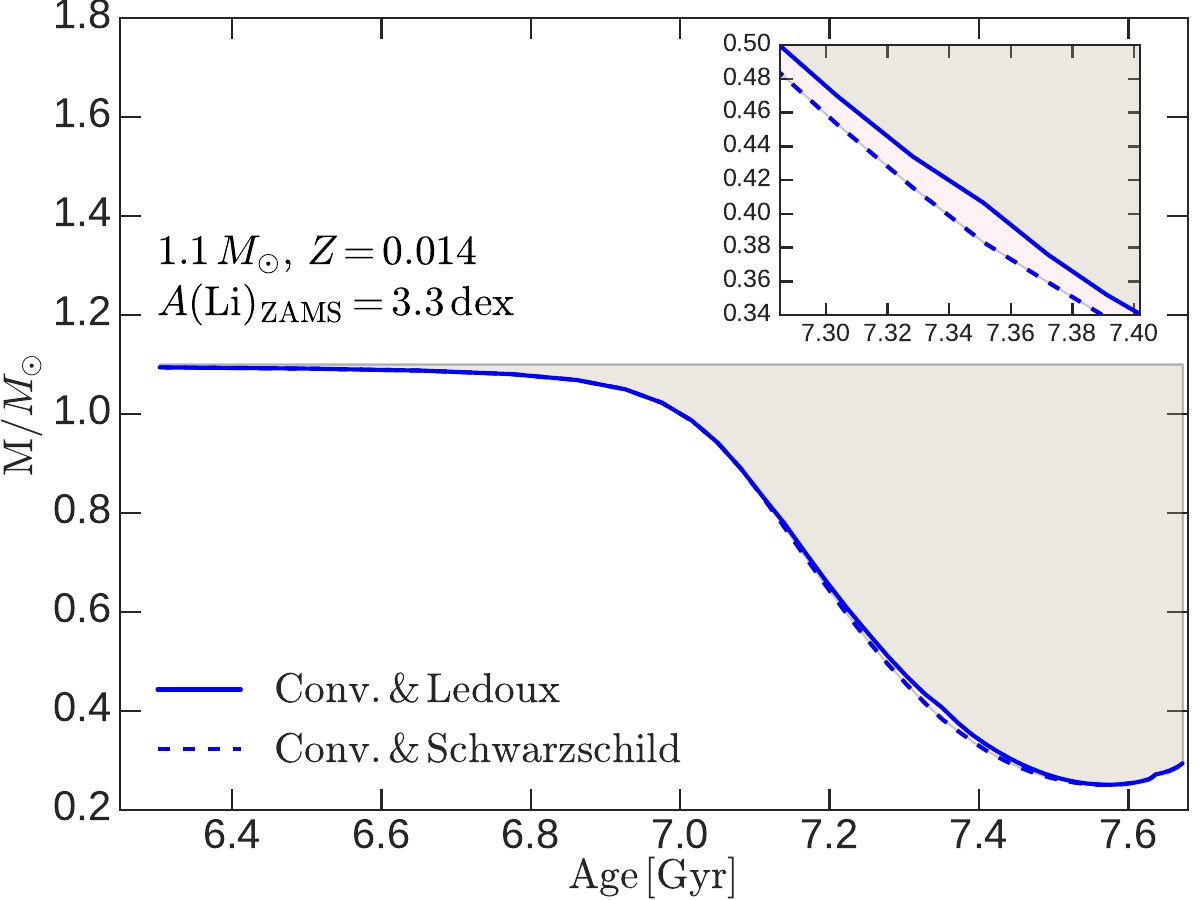}
 \includegraphics[scale=0.24]{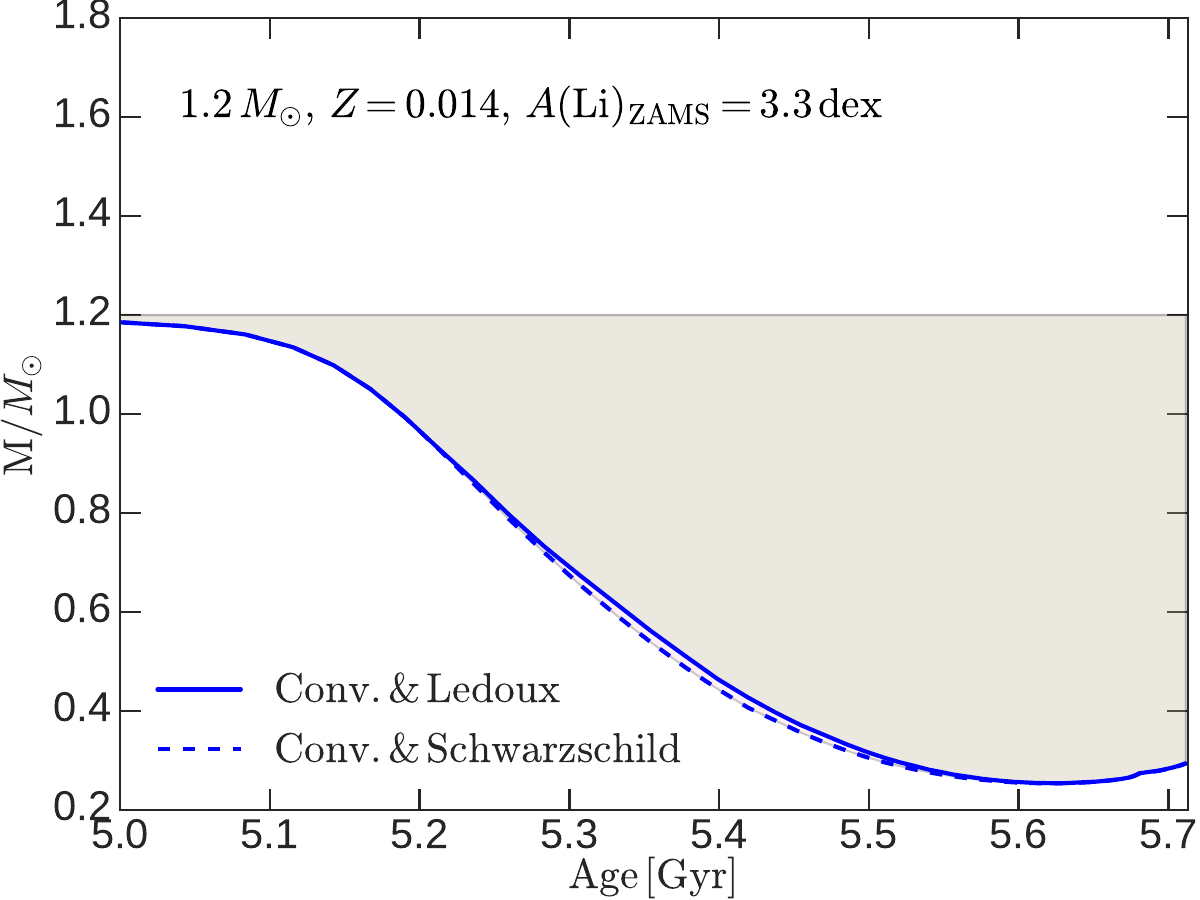}
 \includegraphics[scale=0.24]{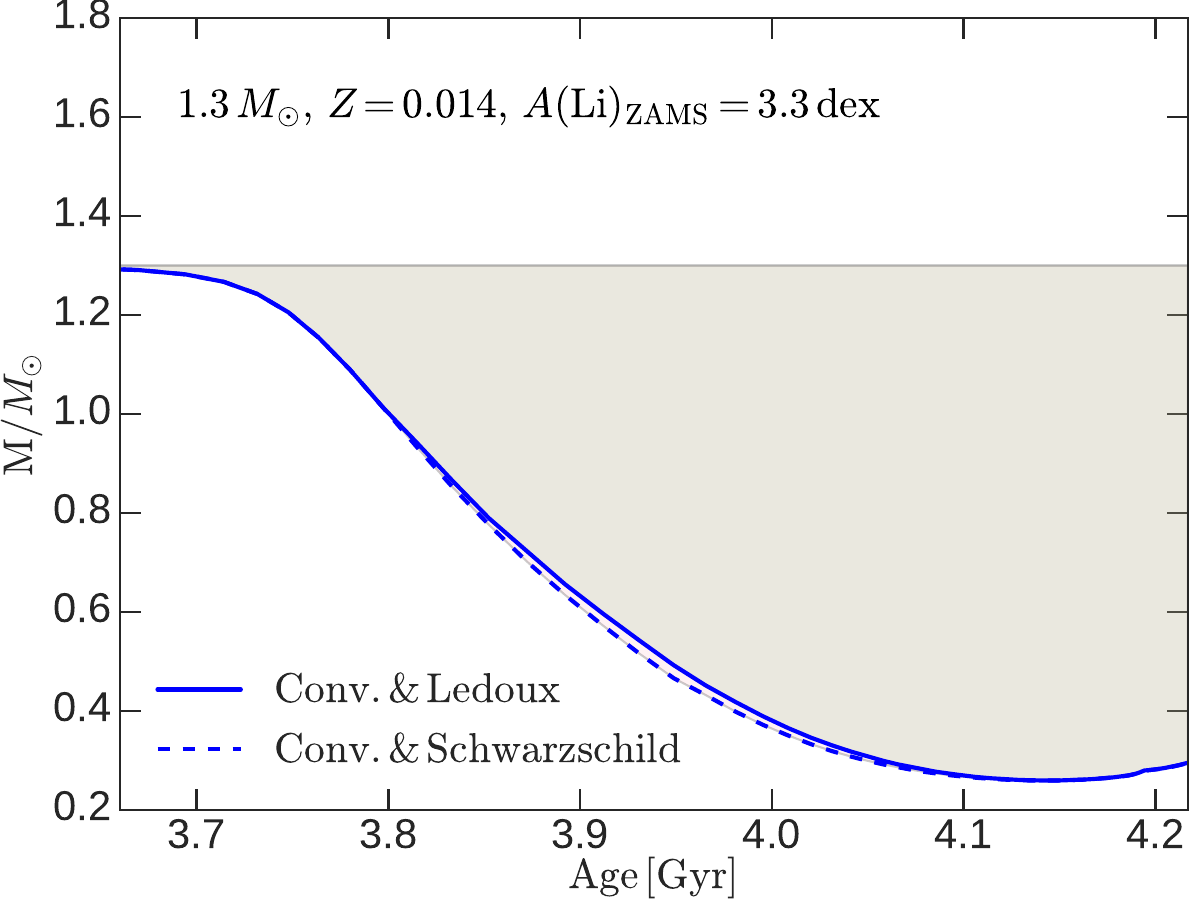}
 \includegraphics[scale=0.24]{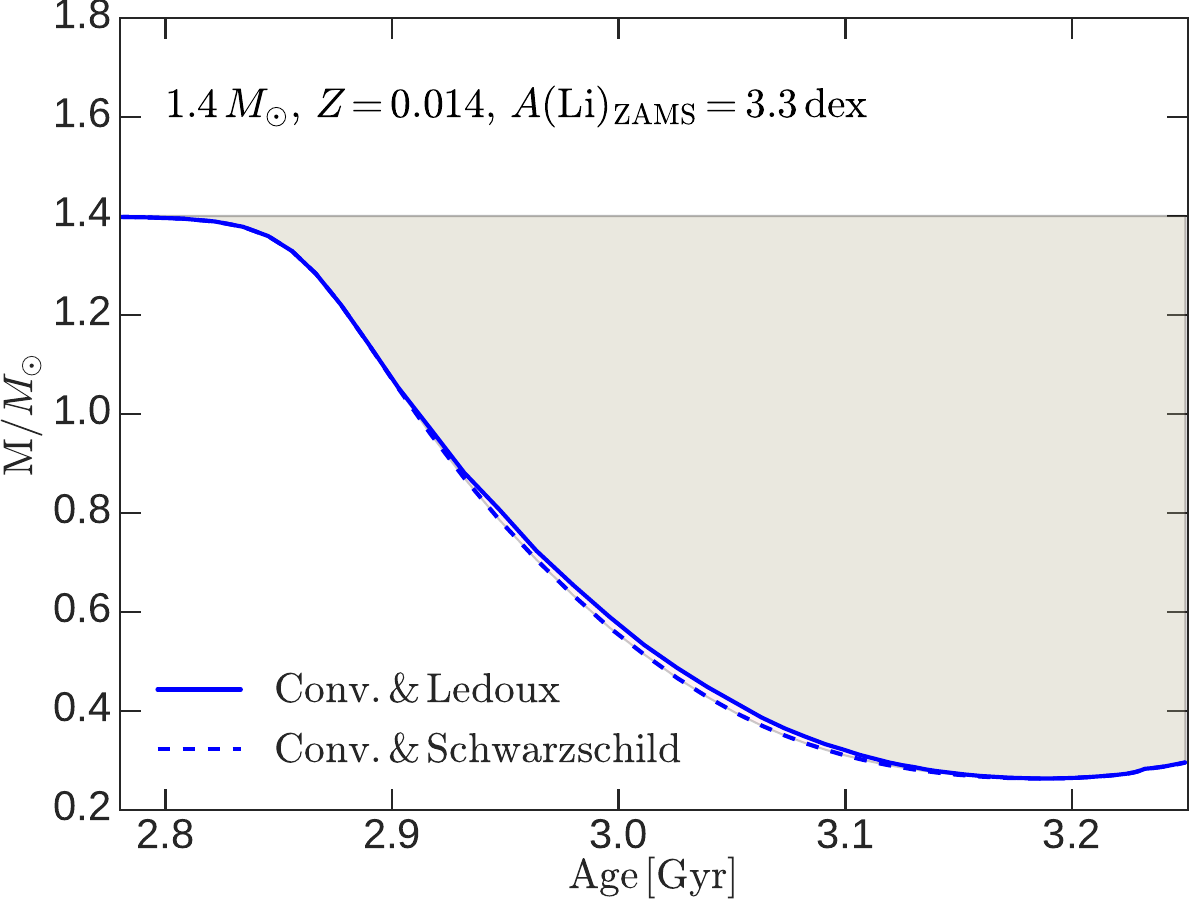}
 \includegraphics[scale=0.24]{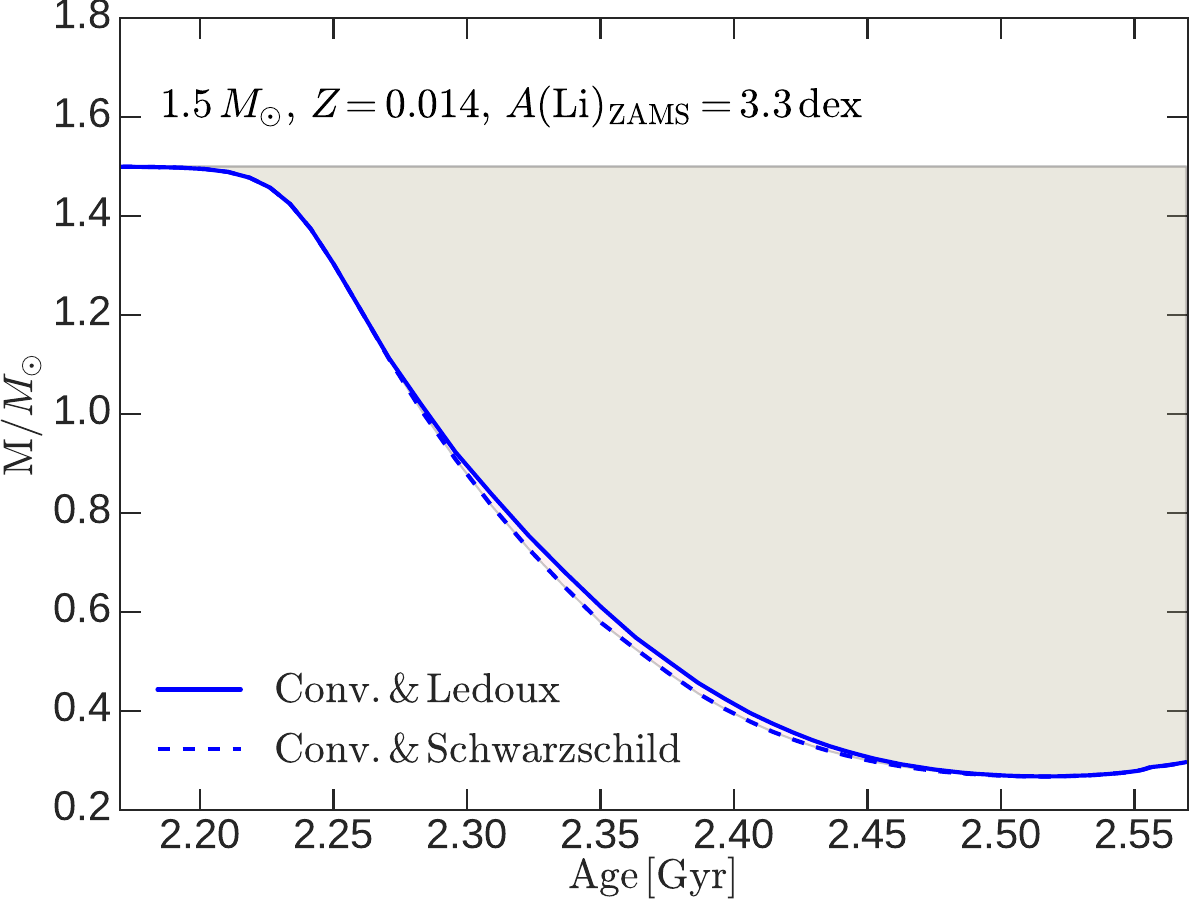}
 \includegraphics[scale=0.24]{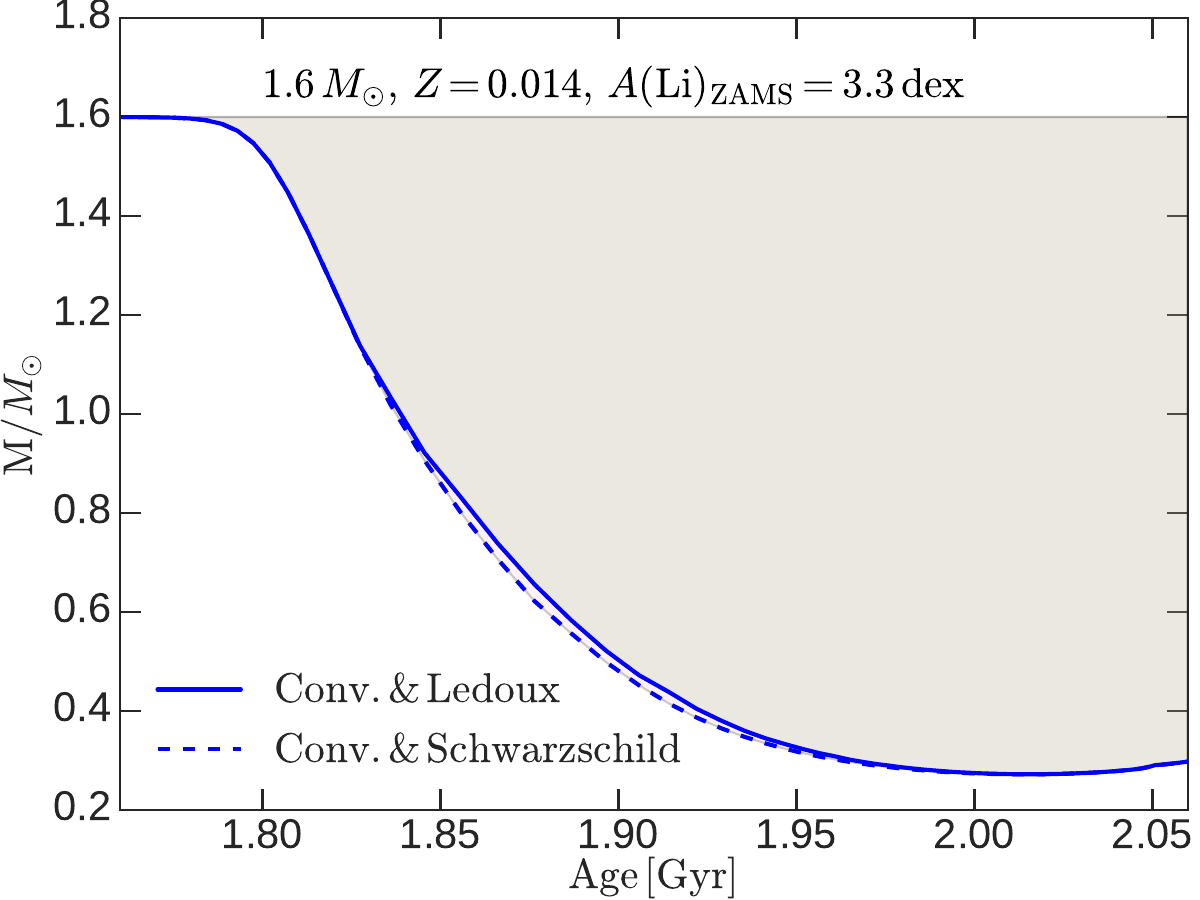}
 \includegraphics[scale=0.24]{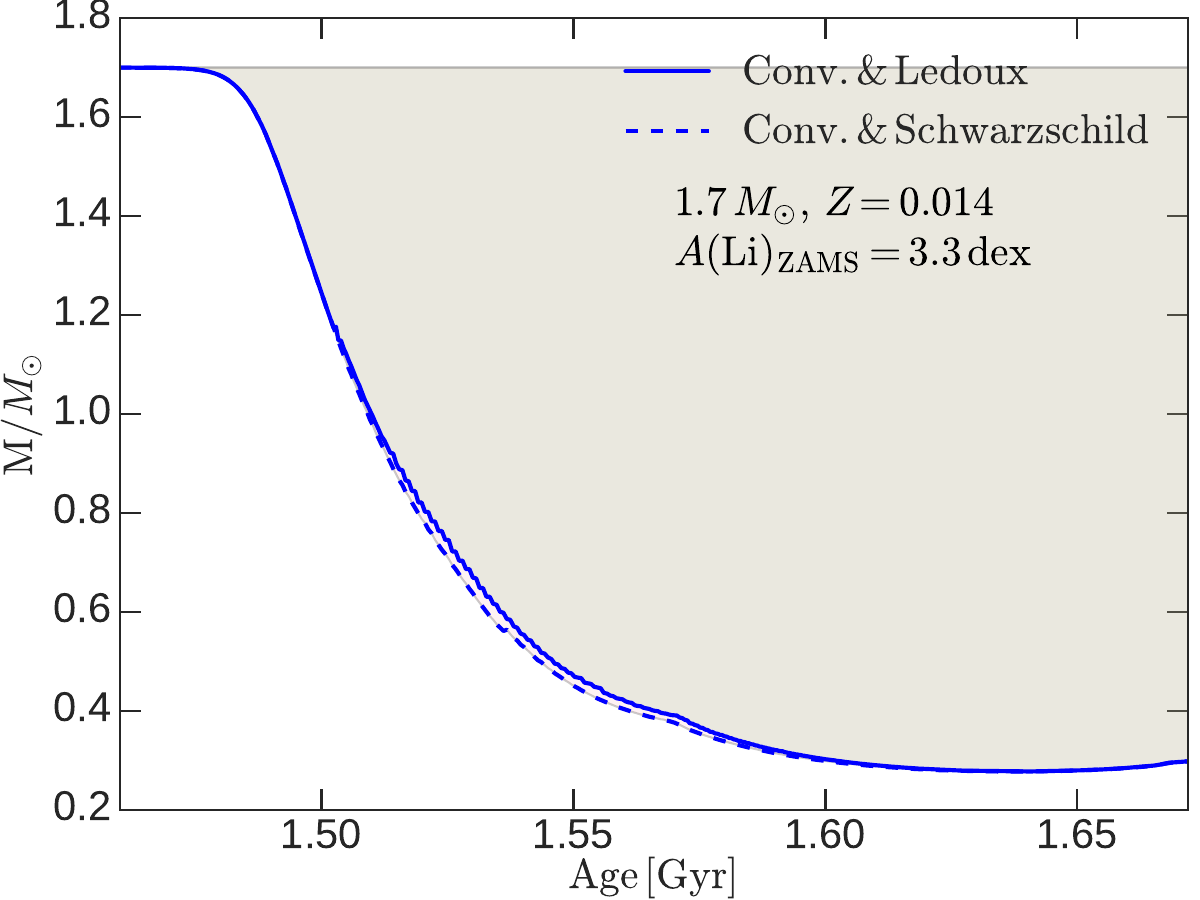}
 \includegraphics[scale=0.24]{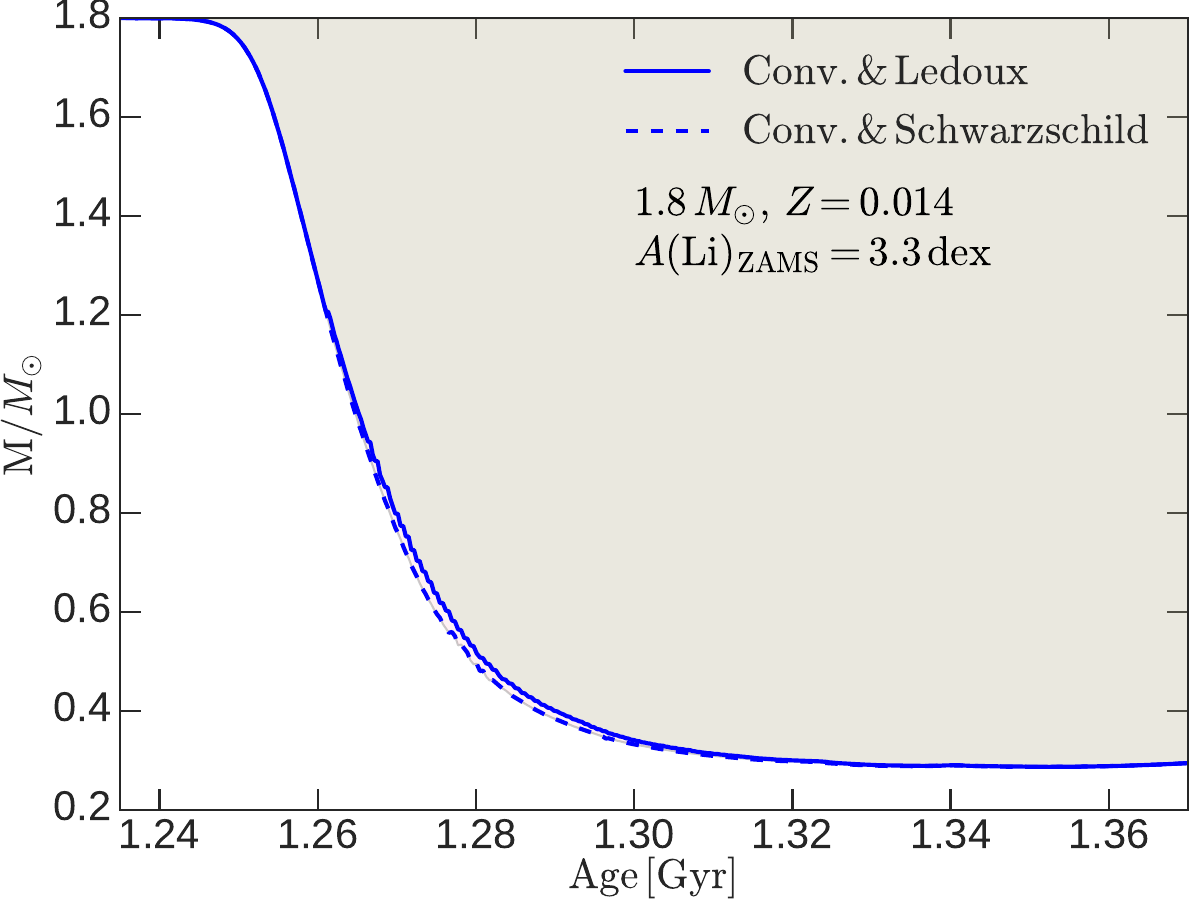}
	\caption{Structural information (similar to $\rm Fig.\,\ref{fig:boundary}\, (3)$) and the Kippenhahn diagram. We show information on the evolution of convective boundaries corresponding to the $\rm Conv.\, \&\, Schwarzschild$ and the $\rm Conv.\, \&\, Ledoux$ models in $\rm Fig.\,\ref{fig:convmass}\, (2)$ of the text. The small figures in the Kippenhahn diagrams are a partial enlargement. Shadows are the convective envelope.} \label{fig:appendix2}
\end{figure*}

We supplement this section with Kippenhahn diagrams of all the test models involving the comparison of the Ledoux and the Schwarzschild boundaries in the main text. As this section contains many structure plots and Kippenhahn diagrams, in order to show the difference between the two boundaries during the first dredge-up as much as possible. Similar to $\rm Fig.\,\ref{fig:structure}$, in $\rm Figs.\,\ref{fig:appendix1},\ \ref{fig:appendix3},\ and\ \ref{fig:appendix2}$ we show only the evolution from the MS turnoff to the luminosity increase to $100\,L_{\odot}$.

\begin{figure*}
	\centering
\includegraphics[scale=0.29]{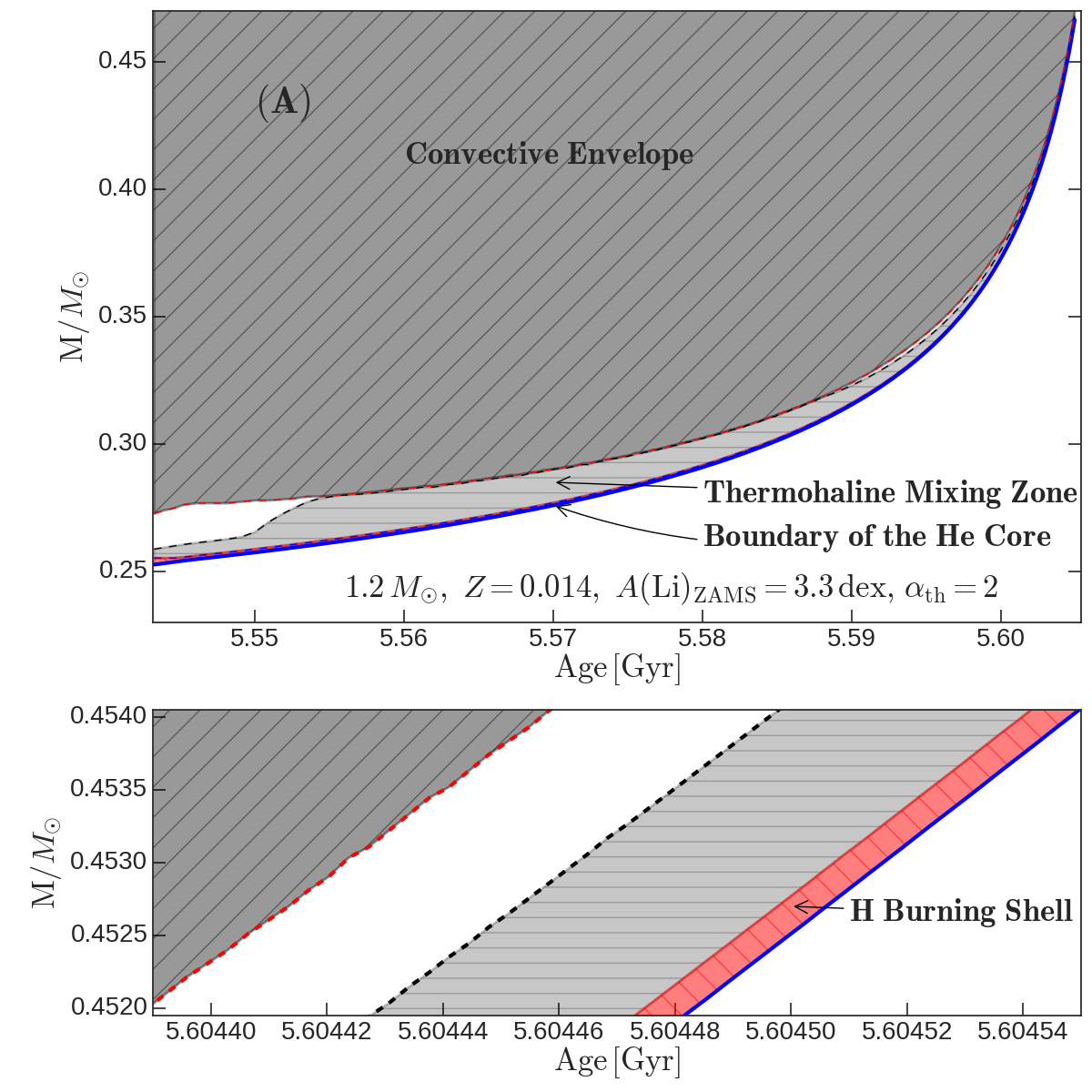}
\includegraphics[scale=0.29]{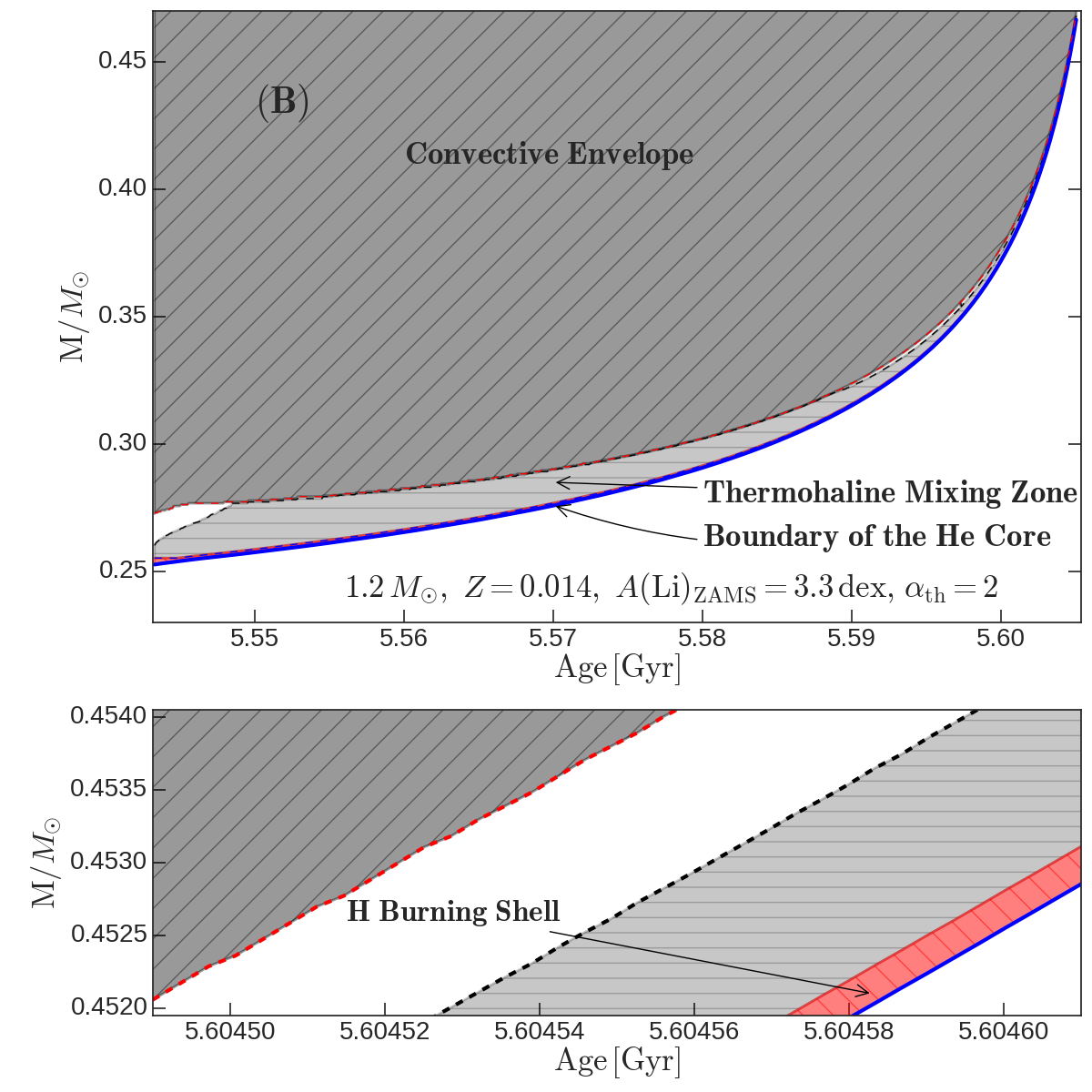}
\includegraphics[scale=0.29]{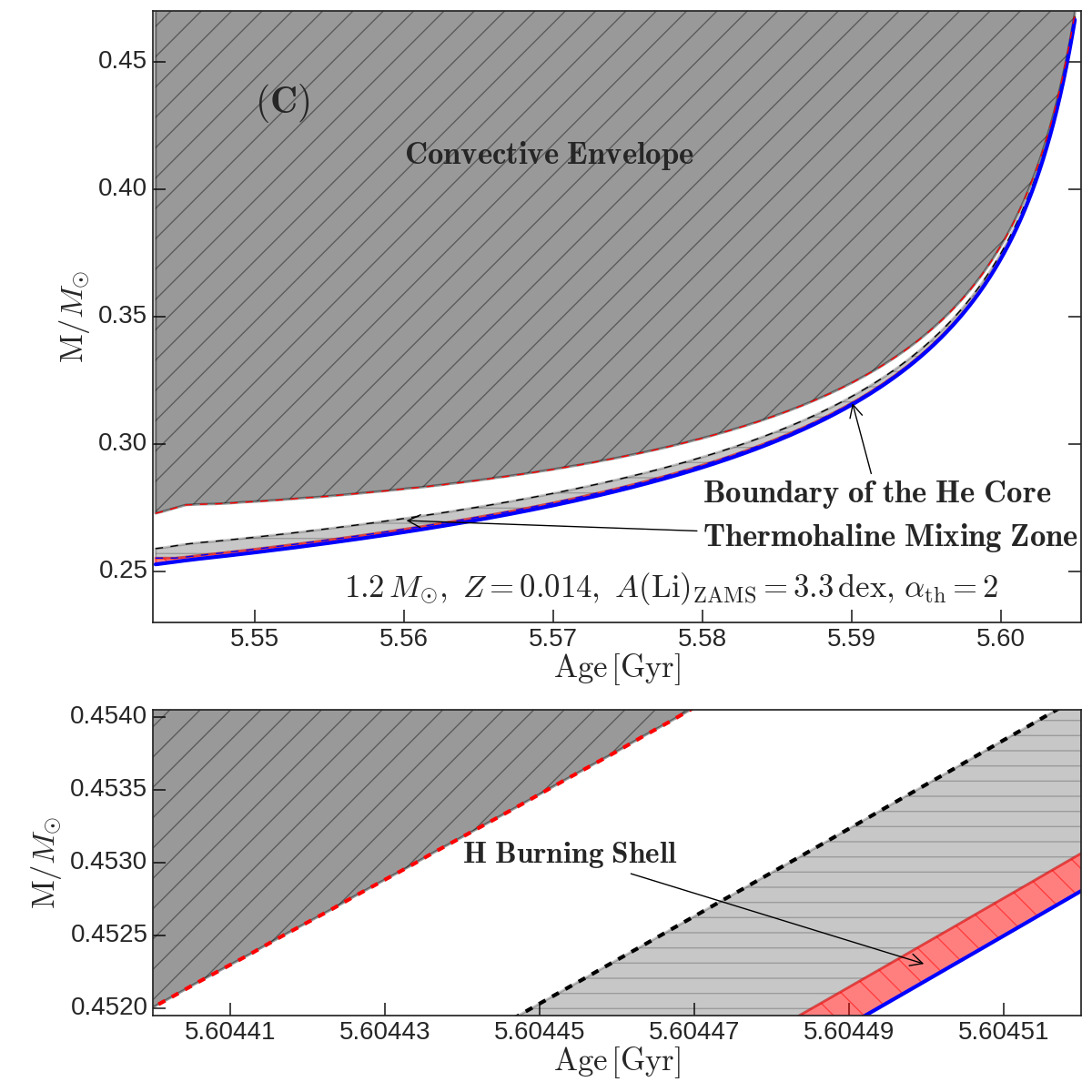}
	\caption{The Kippenhahn diagram containing the low-coefficient thermohaline mixing ($\alpha_{\rm th}=2$). Panel(A): same as $\rm Fig.\,\ref{fig:kipp}\,(1)$, and time step and spatial resolution are the default value for the MESA; (B): the time step is $\rm 10^4\,yr$ and the spatial resolution is the same as in panel (A); (C): five times better spatial resolution, same time step.} \label{fig:appendix4}
\end{figure*}

\begin{figure*}
	\centering
\includegraphics[scale=0.5]{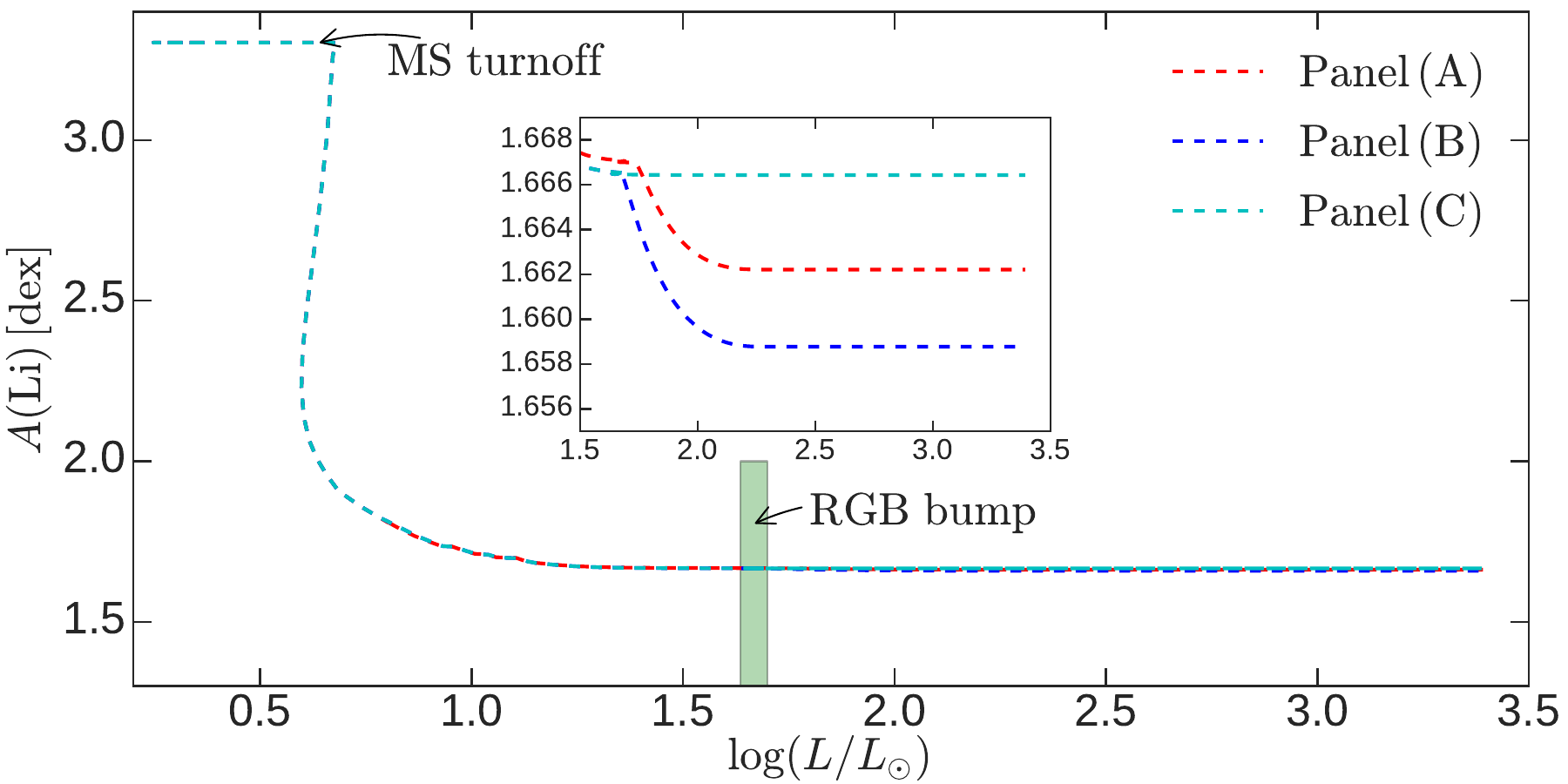}
	\caption{Similar to $\rm Fig.\,\ref{fig:boundary}\,(1)$. The three evolution lines correspond to the three models in $\rm Fig.\,\ref{fig:appendix4}$. The small image in the picture is a partial enlargement.} \label{fig:appendix5}
\end{figure*}
%%%%%%%%%%%%%%%%%%%%%%%%%%%%%%%%%%%%%%%%%%%%%%%%%%

% Don't change these lines
\bsp	% typesetting comment
\label{lastpage}
\end{document}